\documentclass[11pt]{article}
\usepackage{fullpage}
\usepackage{graphicx}
\usepackage{latexsym,amsmath,amssymb,amsthm,epic,eepic,multirow}
\usepackage{natbib}

\usepackage{multicol}
\usepackage{multirow}
\usepackage{subcaption} 
\usepackage{natbib}
\usepackage{algorithm}
\usepackage[noend]{algpseudocode}
\usepackage{mdframed}
\usepackage{tikz}
\usetikzlibrary{positioning}
\usepackage{hyperref}
\usepackage{listings} 
\usepackage{color} 
\usepackage{bbm}
\usepackage{relsize}
\usepackage{array}
\usepackage{mathtools}
\newcolumntype{P}[1]{>{\centering\arraybackslash}p{#1}}
\newcolumntype{R}[1]{>{\raggedleft\arraybackslash}p{#1}}

\definecolor{upmaroon}{rgb}{0.48, 0.07, 0.07}
\definecolor{royalazure}{rgb}{0.0, 0.22, 0.66}
\definecolor{pakistangreen}{rgb}{0.0, 0.4, 0.0}

\newcommand{\PP}{\mathbb{P}}
\newcommand{\EE}{\mathbb{E}}
\newcommand{\eps}{\epsilon}

\theoremstyle{definition}
\newtheorem{theo}{Theorem}

\newtheorem{lemm}{Lemma}
\newtheorem{corr}{Corollary}

\newtheoremstyle{dotless}{}{}{}{}{\bfseries}{}{ }{}
\theoremstyle{dotless}
\newtheorem{assa}{}

\newtheorem{assb}{}

\newtheorem{assc}{}

\usepackage{color}

\let\originalleft\left
\let\originalright\right
\renewcommand{\left}{\mathopen{}\mathclose\bgroup\originalleft}

\renewcommand{\right}{\aftergroup\egroup\originalright}
\makeatletter
\newcommand{\leqnomode}{\tagsleft@true}
\newcommand{\reqnomode}{\tagsleft@false}
 
\makeatother
\begin{document}
\title{Higher-order least squares: assessing partial goodness of fit of linear causal models} 
 
\author{Christoph Schultheiss, Peter B\"uhlmann and Ming Yuan\\
Seminar for Statistics, ETH Z\"urich and Department of Statistics, Columbia University}

\maketitle

\begin{abstract}
We introduce a simple diagnostic test for assessing the overall or partial goodness of fit of a linear causal model with errors being independent of the covariates.
In particular, we consider situations where hidden confounding is potentially present.
We develop a method and discuss its capability to distinguish between covariates that are confounded with the response by latent variables and those that are not.
Thus, we provide a test and methodology for \emph{partial} goodness of fit. The test is based on comparing a novel higher-order least squares principle with ordinary least squares.
In spite of its simplicity, the proposed method
is extremely general and is also proven to be valid for high-dimensional settings.
\end{abstract}

\noindent%

\section{Introduction} \label{intro}
Linear models are the most commonly used statistical tools to study the relationship between a response and a set of covariates. 
The regression coefficient corresponding to a particular covariate is usually interpreted as its net effect on the response
variable when all else is held fixed. Such an interpretation is essential in many applications and yet could be rather misleading when the linear model assumptions are in question,
in particular, when there are hidden confounders.

In this work, we develop a simple but powerful approach to goodness of fit tests for potentially high-dimensional linear causal models, including also tests for partial goodness of fit of single predictor variables. 
While hidden confounding is the primary alternative in mind, different nonlinear deviations from the linear model assumption are also in scope.
Tests for goodness of fit tests are essential to statistical modeling \cite[e.g.,][]{lehmann2005testing} and the concept is also very popular in econometrics where it is referred to as specification tests. For an overview of such methods, see, e.g., \cite{godfrey1991misspecification} or \cite{maddala2009introduction}.

Another set of related works is \cite{buja2019modelsI,buja2019modelsII}, which elaborately discusses deviations from the (linear) model and how distributional robustness, i.e., robustenss against shifts in the covariates' distribution, links to correctly specified models. For this, they introduce the definition of ``well-specified'' statistical functionals. Distributional robustness, implied by well-specification, is also related to the causal interpretation of a linear model as discussed in \cite{peters2016causal}.

We consider here the question when and which causal effects can be inferred from the ordinary least squares estimator or a debiased Lasso procedure for the high-dimensional setting, even when there is hidden confounding. We address this by partial goodness of fit testing: if the data speaks against a linear causal model, we are able to specify which components of the least squares estimator should be rejected to be linear causal effects and which not. In the case of a joint Gaussian distribution, one cannot detect anything: this corresponds to a well-known unidentifiability result in causality \citep{hyvarinen2000independent, peters2014causal}. But, in certain models, we are able to identify some causal relations. Of particular importance are non-Gaussian linear structural equation models, as used in \cite{shimizu2006linear} or \cite{wang2020high} amongst others. The latter constructs the causal graph from observational data in a stepwise procedure using a test statistic similar to the one we suggest.

Our strategy has a very different focus than other approaches which do not rely on the least squares principle any longer to deal with the issue of hidden confounding. Most prominent, particularly in econometrics, is the framework of instrumental variables regression: assuming valid instruments, one can identify all causal effects,
see, e.g., \cite{angrist1996identification} or the books by \cite{bowden1990instrumental} and \cite{imbens2015causal}. 
The popular Durbin-Wu-Hausman test \citep{hausman1978specification} for validity of instruments bears a relation to our methodology, namely that we are also looking at the difference of two estimators to test goodness of fit.

Our automated partial goodness of fit methodology is easy to be used as it is based on ordinary (or high-dimensional adaptions of) least squares and its novel higher-order version: we believe that this simplicity is attractive for statistical practice.

\subsection{Our contribution}
We propose a novel method with a corresponding test, called higher-order least squares (HOLS). The test statistic is based on the residuals from an ordinary least squares or Lasso fit. In that regard, it is related to \cite{shah2018goodness} who use ``residual prediction'' to test for deviation from the linear model. However, our approach does neither assume Gaussian errors nor does it rely on sample splitting, and our novel test statistic has a $\sqrt{n}$ convergence rate (with $n$ denoting the sample size).

In addition to presenting a ``global'' goodness of fit test for
the entire model, we also develop a local interpretation that allows detecting which among the covariates are giving evidence for hidden
confounding or nonlinear relations. Thus, we strongly increase the amount of extracted information compared to a global goodness of fit test. 
In particular, in the case of localized (partial) confounding in linear structural equation models, we are able to recover the unconfounded regression parameters for a subset of predictors. This is a 
setting where techniques assuming dense (essentially global) confounding, as in \cite{cevid2020spectral} or \cite{guo2020doubly}, fail.

The work by \cite{buja2019modelsI,buja2019modelsII}, especially the second paper, shows how to detect deviations in a linear model using reweighting of the data. Our HOLS technique can be seen as a special way of reweighting. In contrast to their work, we provide a simple test statistic that tests for well-specification without requiring any resampling. Furthermore, we provide guarantees for a local interpretation under suitable modeling assumptions while as their per-covariate view remains rather exploratory.

\subsection{Outline}
The remainder of this paper will be structured as follows. We conclude this section with the necessary notation. In Section \ref{low-dim}, we present the main idea of HOLS and the according global null hypothesis. For illustrative purposes, we first discuss univariate regression. Then, we consider multivariate regression and extend our theory to high-dimensional problems incorporating the (debiased) Lasso. In Section \ref{lin-conf}, we present the local interpretation when the global null does not hold true alongside with theoretical guarantees. Models for which this local interpretation is most suitable are discussed in Section \ref{ex-models}. Section \ref{sachs} contains a real data analysis. We conclude with a summarizing discussion in Section \ref{diss}.

\subsection{Notation}
We present some notation that is used throughout this work.
Vectors and matrices are written in boldface, while scalars have the usual lettering. This holds for both random and fixed quantities.
We use upper case letters to denote a random variable, e.g., $\mathbf{X}$ or $Y$. 
We use lower case letters to denote i.i.d.\ copies of a random variable, e.g., $\mathbf{x}$. If $\mathbf{X} \in \mathbb{R}^{p}$, then $\mathbf{x} \in \mathbb{R}^{n \times p}$.
With a slight abuse of notation, $\mathbf{x}$ can either denote the copies or realizations thereof.
We write $\mathbf{x}_j$ to denote the j-th column of matrix $\mathbf{x}$ and $\mathbf{x}_{-j}$ to denote all but the j-th column.
We write $\overset{H_0}{=} $ to state that equality holds under $H_0$. With $\leftarrow$, we emphasize that an equality between random variables is induced by a causal mechanism.
We use $\odot$ to denote elementwise multiplication of two vectors, e.g., $\mathbf{x} \odot \mathbf{y}$. Similarly, potencies of vectors are also to be understood in an elementwise fashion, e.g., $\mathbf{x}^2=\mathbf{x}\odot\mathbf{x}$.
$\boldsymbol{I}_n$ is the $n$-dimensional identity matrix.
$\boldsymbol{P}_{-j}$ denotes the orthogonal projection onto $\mathbf{x}_{-j}$ and $\boldsymbol{P}^{\perp}_{-j} = \boldsymbol{I}_n - \boldsymbol{P}_{-j}$ denotes the orthogonal projection onto its complement.
For some random vector $\mathbf{X}$, we have the moment matrix $\boldsymbol{\Sigma}^{\mathbf{X}} \coloneqq \EE\left[\mathbf{X}\mathbf{X}^\top\right]$. Note that this equals the covariance matrix for centered $\mathbf{X}$.
We denote statistical independence by $\perp$. We write $\mathbf{e}$ to denote a vector for which every entry is $1$ and $\mathbf{e}_j$ to denote the unit vector in the direction of the $j$-th coordinate axis.

\section{Higher-order least squares (HOLS)}\label{low-dim}
We develop here the main idea of higher-order least squares (HOLS)
estimation.

\subsection{Univariate regression as a motivating case}\label{1D}
It is instructive to begin with the case of simple linear regression where we have a pair of random variables $X$ and $Y$. We consider the causal linear model
\begin{equation}\label{eq:1D-model}
Y \leftarrow X\beta + \mathcal{E}, \quad \text{where} \quad X \perp \mathcal{E}, \quad \EE\left[\mathcal{E}\right]=0 \quad \text{and} \quad \EE\left[\mathcal{E}^2\right]=\sigma^2 < \infty. \end{equation}
We formulate a null hypothesis that the model in \eqref{eq:1D-model} is correct and we
denote such a hypothesis by $H_0$. This model is of interest as $\beta$ describes the effect of a unit change if we were to intervene on covariate $X$ without intervening on the independent $\mathcal{E}$. Such model, or its multivariate extension, is often assumed in causal discovery, see, e.g., \cite{shimizu2006linear} or \cite{hoyer2008causal}. Therefore, we aim to provide a test for its well-specification.

Estimation of the regression parameter is typically done by the least
squares principle
\begin{equation*}
\beta^{OLS} \coloneqq \underset{b \in \mathbb{R}}{\text{argmin}}\EE\left[\left(Y-Xb\right)^2\right] = \dfrac{\EE\left[ XY\right]}{\EE\left[ X^2 \right]} \overset{H_0}{=} \beta,
\end{equation*}
where we use the superscript OLS to denote ordinary least squares. Alternatively, we can pre-multiply the linear model \eqref{eq:1D-model} with $X$: the
parameter minimizing the expected squared error loss is then
\begin{equation*}
\beta^{HOLS} \coloneqq \underset{b \in \mathbb{R}}{\text{argmin}}\EE\left[\left(X Y-X^2b\right)^2\right] = \dfrac{\EE\left[ X^3Y\right]}{\EE\left[ X^4 \right]} \overset{H_0}{=}\dfrac{\EE\left[ X^4 \beta\right]}{\EE\left[ X^4 \right]}=\beta.
\end{equation*}
More generally, $\beta^{HOLS} =\beta^{OLS} = \beta$, if $\EE\left[Y \vert X\right] = X\beta$. Using the definition from \cite{buja2019modelsII}, this means that the OLS parameter is well-specified. The estimation principle is called higher-order least squares, or HOLS for short, as it involves higher-order moments of $X$.
One could also multiply the linear model with a factor other than $X$, which may have implications on the power to detect deviations from \eqref{eq:1D-model}. We shall focus here on the specific choice to fix ideas.

The motivation to look at HOLS is when $H_0$ is violated, in terms of a
hidden confounding variable: let $H$ be a hidden confounder leading to a model
\begin{equation*}
X \leftarrow \mathcal{E}_X + H \rho, \quad Y \leftarrow X \beta + H \alpha+ \mathcal{E},
\end{equation*}
where $\mathcal{E}_X$, $H$, and $\mathcal{E}$ are all independent and $\alpha$ and $\rho$ define additional model parameters. In particular, we can compute under such a confounding model that
\begin{equation} \label{eq:1D-diff}
\beta^{HOLS} - \beta^{OLS} = \rho \alpha \left(\dfrac{3 \EE\left[\mathcal{E}_X^2\right]\EE\left[H^2\right] + \rho^2 \EE\left[H^4\right]}{\EE\left[\mathcal{E}_X^4\right] + 6 \rho^2 \EE\left[\mathcal{E}_X^2\right] \EE\left[H^2\right] + \rho^4 \EE\left[H^4\right]} - \dfrac{\EE\left[H^2\right]}{\EE\left[\mathcal{E}_X^2\right] + \rho^2 \EE\left[H^2\right]} \right).
\end{equation}
For simplicity, we assumed here $\EE\left[\mathcal{E}_X\right] = \EE\left[H\right] = \EE\left[\mathcal{E}\right] =0$. In practice, one can get rid of this assumption by including an intercept in the model. If either $\alpha$ or $\rho$ equals to $0$, we see that the difference in \eqref{eq:1D-diff} is $0$. This is not surprising as there is no confounding effect when either $X$ or $Y$ is unaffected. However, this is not the only possibility how the difference can be $0$. Namely,
\begin{equation*}
\EE\left[H^2\right] \left(\EE\left[\mathcal{E}_X^4\right]-3 \EE\left[\mathcal{E}_X^2\right]^2\right) = \rho^2 \EE\left[\mathcal{E}_X^2\right] \left(\EE\left[H^4\right]- 3\EE\left[H^2\right]^2\right) \Rightarrow \beta^{HOLS} - \beta^{OLS} = 0.
\end{equation*}
Especially, if neither $\mathcal{E}_X$ nor $H$ have excess kurtosis, the difference is $0$ for any $\rho$. This can be intuitively explained as it corresponds to Gaussian data (up to the moments we consider). For Gaussian $\mathcal{E}_X$ and $H$, one can always write
\begin{equation*}
Y = X \beta^{OLS} + \tilde{\mathcal{E}} \quad \text{where} \quad X \perp \tilde{\mathcal{E}}, 
\end{equation*}
which cannot be distinguished from the null model \eqref{eq:1D-model}. Or in other words $\EE\left[Y \vert X\right] = X\beta^{OLS}$, i.e., the OLS parameter is well-specified although it is not the parameter $\beta$ that we would like to recover. For other data generating distributions, one
should be able to distinguish $H_0$ from certain deviations when hidden
confounding is present. We discuss the implications of this in the general multivariate setting in Section \ref{inter-U}. Similar behaviour occurs for a violation of
$H_0$ in terms of a nonlinear model $Y = f(X,\eps)$ which then (typically)
leads to $\beta^{HOLS} - \beta^{OLS} \neq 0$. 

One can construct a test based on the sample estimates of
$\beta^{HOLS}$ and $\beta^{OLS}$. We consider the centered data 
\begin{equation*}
\tilde{\mathbf{x}} = \mathbf{x} - \bar{x}\mathbf{e},\quad \tilde{\mathbf{y}} = \mathbf{y} - \bar{y}\mathbf{e} \quad \text{and} \quad \tilde{\boldsymbol{\eps}} = \boldsymbol{\eps} - \bar{\eps}\mathbf{e} = \left(\boldsymbol{I}_n - \dfrac{1}{n}\mathbf{e}\mathbf{e}^{\top}\right) \boldsymbol{\eps},
\end{equation*}
where we use the upper bar to denote sample means. We can derive
\begin{equation*}
\tilde{\mathbf{y}} = \mathbf{y} - \bar{y}\mathbf{e} = \mathbf{x} \beta - \bar{x}\mathbf{e}\beta + \boldsymbol{\eps} - \bar{\eps} \mathbf{e} =\tilde{\mathbf{x}}\beta + \tilde{\boldsymbol{\eps}}.
\end{equation*}
We now obtain $\hat{\beta}^{OLS}$ from regression through the origin of $\tilde{\mathbf{y}}$ versus $\tilde{\mathbf{x}}$ with an error term of $\tilde{\boldsymbol{\eps}}$ and $\hat{\beta}^{HOLS}$ from regression through the origin of $\tilde{\mathbf{x}} \odot \tilde{\mathbf{y}}$ versus $\tilde{\mathbf{x}}^2$ with an error term of $\tilde{\mathbf{x}}\odot \tilde{\boldsymbol{\eps}}$. More precisely, we define
\begin{equation*}
\hat{\beta}^{OLS} \coloneqq \dfrac{\tilde{\mathbf{x}}^{\top} \tilde{\mathbf{y}}}{\tilde{\mathbf{x}}^{\top}\tilde{\mathbf{x}}} \quad \text{and} \quad \hat{\beta}^{HOLS} \coloneqq \dfrac{\left(\tilde{\mathbf{x}}^2\right)^{\top}\left(\tilde{\mathbf{x}}\odot \tilde{\mathbf{y}}\right)}{\left(\tilde{\mathbf{x}}^2\right)^{\top}\left(\tilde{\mathbf{x}}^2\right)} = \dfrac{\left(\tilde{\mathbf{x}}^3\right)^{\top}\left( \tilde{\mathbf{y}}\right)}{\left(\tilde{\mathbf{x}}^2\right)^{\top}\left(\tilde{\mathbf{x}}^2\right)}.
\end{equation*}
Under $H_0$, one can see that $\left(\hat{\beta}^{HOLS}-\hat{\beta}^{OLS}\right)$ given $\mathbf{x}$ is some known linear combination of $\boldsymbol{\eps}$. Assuming further Gaussianity of $\boldsymbol{\eps}$, it is conditionally Gaussian. We find
\begin{equation}\label{eq:1D-variance}
\left(\hat{\beta}^{HOLS}-\hat{\beta}^{OLS}\right)\Big \vert \mathbf{x} \overset{H_0}{\sim} \mathcal{N}\left(0, \sigma^2 \left(\dfrac{\left(\tilde{\mathbf{x}}^3\right)^{\top}\left(\boldsymbol{I}_n - \dfrac{1}{n}\mathbf{e}\mathbf{e}^{\top}\right)\left(\tilde{\mathbf{x}}^3\right)}{\left(\left(\tilde{\mathbf{x}}^2\right)^{\top}\left(\tilde{\mathbf{x}}^2\right)\right)^2} - \dfrac{1}{\left(\tilde{\mathbf{x}}^{\top}\tilde{\mathbf{x}}\right)} \right)\right).
\end{equation}
We can calculate this variance except for $\sigma^2$. Further, we can consistently estimate $\sigma^2$, for example, with the standard formula
\begin{equation*}
\hat{\sigma}^2 = \dfrac{\left\Vert \tilde{\mathbf{y}} - \tilde{\mathbf{x}} \hat{\beta}^{OLS} \right\Vert_2^2 }{n-2}.
\end{equation*}
Thus, we receive asymptotically valid z-tests for the null-hypothesis $H_0$ that the model \eqref{eq:1D-model} holds. We treat the extension to non-Gaussian $\boldsymbol{\eps}$ in Section \ref{mv} (for the multivariate case directly). As discussed above, in the presence of confounding, we can have that
$\beta^{HOLS} \neq \beta^{OLS}$. In such situations, a test assuming \eqref{eq:1D-variance} will have asymptotic power equal to 1 for
correctly rejecting $H_0$ under some conditions. These asymptotic results are discussed in Section \ref{conv-par} and Section \ref{recov-U}.

\subsection{Multivariate regression}\label{mv}
We typically want to examine the goodness of fit of
a linear model with $p > 1$ covariates. We consider $p$ to be fixed in this section and discuss the case where $p$ is allowed to diverge with $n$ in Section \ref{high-dim}.

We consider the causal model
\begin{equation}\label{eq:mvmodel}
\quad Y \leftarrow \mathbf{X}^{\top}\boldsymbol{\beta} + \mathcal{E}, \quad \text{where} \quad \mathbf{X} \perp \mathcal{E}, \quad \EE\left[\mathcal{E}\right] = 0 \quad \text{and} \quad \EE\left[\mathcal{E}^2\right]=\sigma^2 < \infty
\end{equation}
with $\mathbf{X} \in \mathbb{R}^p$ and $\boldsymbol{\beta}\in \mathbb{R}^p$. Note that $\EE\left[\mathcal{E}\right] = 0$ can always be enforced by including an intercept in the set of predictors. We assume the according moment matrix $\boldsymbol{\Sigma}^\mathbf{X}$ to be invertible. Then, the principal submatrices $\boldsymbol{\Sigma}^\mathbf{X}_{-j,-j} \coloneqq \EE\left[\mathbf{X}_{-j}\mathbf{X}_{-j}^\top\right]$ are also invertible. We formulate a global null hypothesis that the model in \eqref{eq:mvmodel} is correct and we denote it by $H_0$. To make use of the test described for the univariate case, we consider every component $j \in \{1,\ldots ,p\}$ separately
and work with partial regression, see, e.g., \cite{belsley2005regression}. For the population version, we define

\begin{equation} \label{eq:z-w-def}
 \begin{aligned}
Z_j & \coloneqq X_j - \mathbf{X}_{-j}^{\top}\boldsymbol{\gamma}_j, \quad &&\text{where}\quad \boldsymbol{\gamma}_j \coloneqq \underset{\mathbf{b} \in \mathbb{R}^{p-1}}{\text{argmin}}\EE\left[\left(X_j-\mathbf{X}_{-j}^\top \mathbf{b}\right)^2\right] = \left(\boldsymbol{\Sigma}^\mathbf{X}_{-j,-j}\right)^{-1}\EE\left[\mathbf{X}_{-j}X_j\right]\\
W_j & \coloneqq Y - \mathbf{X}_{-j}^{\top}\boldsymbol{\zeta}_j, \quad &&\text{where}\quad \boldsymbol{\zeta}_j \coloneqq \underset{\mathbf{b} \in \mathbb{R}^{p-1}}{\text{argmin}}\EE\left[\left(Y-\mathbf{X}_{-j}^\top \mathbf{b}\right)^2\right] = \left(\boldsymbol{\Sigma}^\mathbf{X}_{-j,-j}\right)^{-1}\EE\left[\mathbf{X}_{-j}Y\right].
 \end{aligned}
\end{equation}
Under $H_0$, it holds that $W_j = Z_j \beta_j + \mathcal{E}$. For $\boldsymbol{\beta}^{OLS} \coloneqq \left(\boldsymbol{\Sigma}^\mathbf{X}\right)^{-1}\EE\left[\mathbf{X}Y\right]$, we find
\begin{equation*}
\beta_j^{OLS} = \dfrac{\EE\left[Z_j W_j\right]}{\EE\left[Z_j^2 \right]}\overset{H_0}{=} \beta_j.
\end{equation*}
The first equality is a well-known application of the Frish-Waugh theorem, see, e.g., \cite{greene2003econometric}.
We define the according HOLS parameter by partial regression for every component $j$ separately, namely 
\begin{equation*}
 \beta_j^{HOLS} \coloneqq \dfrac{\EE\left[Z_j^3 W_j\right]}{\EE\left[Z_j^4\right]}\overset{H_0}{=} \beta_j.
\end{equation*}
We define a local, i.e., per-covariate null hypothesis $H_{0,j}: \ \beta^{OLS}_j =
\beta^{HOLS}_j$.
The difference $\beta^{OLS}_j -
\beta^{HOLS}_j$ can detect certain local alternatives from the
null hypothesis $H_0$.  Here, local refers to the covariate $X_j$ whose effect on $Y$ is
potentially confounded or involves a nonlinearity. Under model \eqref{eq:mvmodel}, $H_{0,j}$ holds true for every $j$. We discuss in Sections \ref{lin-conf} and \ref{ex-models} some concrete examples, where it is insightful to consider tests for individual $H_{0,j}$. 

We turn to sample estimates of the parameters. The residuals are estimated by
\begin{align*}
\hat{\mathbf{z}}_j &= \mathbf{x}_j - \boldsymbol{P}_{-j} \mathbf{x}_j = \boldsymbol{P}^{\perp}_{-j} \mathbf{x}_j \quad \text{and} \\
\hat{\mathbf{w}}_j & = \mathbf{y} - \boldsymbol{P}_{-j} \mathbf{y} = \boldsymbol{P}^{\perp}_{-j} \mathbf{y} \overset{H_0}{=} \boldsymbol{P}^{\perp}_{-j} \left(\mathbf{x}\boldsymbol{\beta} + \boldsymbol{\eps}\right) = \hat{\mathbf{z}}_j \beta_j + \boldsymbol{P}^{\perp}_{-j} \boldsymbol{\eps}.
\end{align*}
With ordinary least squares, we receive $\hat{\beta}_j^{OLS}$ from regression of $\hat{\mathbf{w}}_j$ versus $\hat{\mathbf{z}}_j$, where the error term is $\boldsymbol{P}^{\perp}_{-j} \boldsymbol{\eps}$. Accordingly, we calculate $\hat{\beta}_j^{HOLS}$ from regression of $\hat{\mathbf{z}}_j \odot \hat{\mathbf{w}}_j$ versus $\hat{\mathbf{z}}_j ^2$ with an error term $\hat{\mathbf{z}}_j \odot \boldsymbol{P}^{\perp}_{-j} \boldsymbol{\eps}$. Thus, we define
\begin{equation}\label{eq:bhat-def}
\hat{\beta}_j^{OLS} \coloneqq \dfrac{\hat{\mathbf{z}}_j^{\top}\hat{\mathbf{w}}_j}{\hat{\mathbf{z}}_j ^{\top}\hat{\mathbf{z}}_j }\quad \text{and} \quad \hat{\beta}_j^{HOLS} \coloneqq \dfrac{\left(\hat{\mathbf{z}}_j^2\right)^{\top}\left(\hat{\mathbf{z}}_j\odot \hat{\mathbf{w}}_j\right)}{\left(\hat{\mathbf{z}}_j^2\right)^{\top}\left(\hat{\mathbf{z}}_j^2\right)} = \dfrac{\left(\hat{\mathbf{z}}_j^3\right)^{\top}\hat{\mathbf{w}}_j}{\left(\hat{\mathbf{z}}_j^3\right)^{\top}\hat{\mathbf{z}}_j}.
\end{equation}

This is analogous to the univariate case, where we have $\tilde{\mathbf{y}}$ instead of $\hat{\mathbf{w}}_j$, $\tilde{\mathbf{x}}$ instead of $\hat{\mathbf{z}}_j$ and $\left(\boldsymbol{I}_n - \dfrac{1}{n}\mathbf{e}\mathbf{e}^{\top}\right)$ instead of $\boldsymbol{P}^{\perp}_{-j}$, and $\left(\boldsymbol{I}_n - \dfrac{1}{n}\mathbf{e}\mathbf{e}^{\top}\right)$ can be thought of as orthogonal projection onto $\mathbf{e}$'s complement, which completes the analogy. Again, we see that given $\mathbf{x}$, $\left(\hat{\beta}_j^{HOLS}-\hat{\beta}_j^{OLS}\right)$  is some known linear combination of $\boldsymbol{\eps}$, thus, it is conditionally Gaussian for Gaussian $\boldsymbol{\eps}$. The same holds for $\left(\hat{\boldsymbol{\beta}}^{HOLS}-\hat{\boldsymbol{\beta}}^{OLS}\right)$.

Naturally, Gaussian $\mathcal{E}$ is an overly strong assumption. Therefore, we consider additional assumptions such that the central limit theorem can be invoked.
\begin{assa} \label{ass:invertible}
The moment matrix $\boldsymbol{\Sigma}^{\mathbf{X}}$ has positive smallest eigenvalue.
\end{assa}
\begin{assa} \label{ass:mom6}
 $\EE\left[X_j ^6 \right] < \infty$ and $\EE\left[Z_j ^6 \right] < \infty \ \forall j.$
\end{assa}
Further, let
\begin{equation}\label{eq:ztilde-def}
\tilde{Z}_j^3 \coloneqq Z_j^3- \mathbf{X}_{-j}^{\top}\tilde{\boldsymbol{\gamma}}_j, \ \text{where}\ \tilde{\boldsymbol{\gamma}}_j \coloneqq \underset{\mathbf{b} \in \mathbb{R}^{p-1}}{\text{argmin}}\EE\left[\left(Z_j^3-\mathbf{X}_{-j}^\top \mathbf{b}\right)^2\right] = \left(\boldsymbol{\Sigma}^\mathbf{X}_{-j,-j}\right)^{-1}\EE\left[\mathbf{X}_{-j}Z_j^3\right].
\end{equation}
Note that $\EE\left[\left(\tilde{Z}_j^3\right)^2\right] \leq \EE\left[Z_j^6\right] < \infty$.
\begin{theo}\label{theo:non-gauss}
Assume that the data follows the model \eqref{eq:mvmodel} and that \ref{ass:invertible} - \ref{ass:mom6} hold. Let $p$ be fixed and $n \rightarrow \infty$. Then,
\begin{align*}
\sqrt{n} \left(\hat{\boldsymbol{\beta}}^{HOLS}-\hat{\boldsymbol{\beta}}^{OLS}\right) & \overset{\mathbb{D}}{\to} \mathcal{N}\left(\mathbf{0}, \sigma^2 \EE\left[\mathbf{V}\mathbf{V}^\top\right]\right)\\
\dfrac{1}{n} \hat{\mathbf{v}} ^\top \hat{\mathbf{v}} & \overset{\PP}{\to} \EE\left[\mathbf{V}\mathbf{V}^\top\right],
\end{align*}
where $\hat{\mathbf{v}}_j =\dfrac{\boldsymbol{P}_{-j}^\perp \left( \hat{\mathbf{z}}_j^3\right)}{\dfrac{1}{n}\left(\hat{\mathbf{z}}_j^2\right)^\top\left(\hat{\mathbf{z}}_j^2\right)} - \dfrac{\hat{\mathbf{z}}_j}{\dfrac{1}{n}\hat{\mathbf{z}}_j^\top\hat{\mathbf{z}}_j}$ and $V_j = \dfrac{\tilde{Z}_j^3}{\EE\left[Z_j^4\right]}- \dfrac{Z_j}{\EE\left[Z_j^2\right]}$.
\end{theo}
Note that 
\begin{equation*}
\left(\hat{\boldsymbol{\beta}}^{HOLS}-\hat{\boldsymbol{\beta}}^{OLS}\right)\overset{H_0}{=} \dfrac{1}{n}\hat{\mathbf{v}}^\top\boldsymbol{\eps}, \quad \text{and, in analogy to \eqref{eq:1D-variance},} \quad
\dfrac{1}{n^2}\hat{\mathbf{v}}_j ^\top \hat{\mathbf{v}}_j = \dfrac{\left( \hat{\mathbf{z}}_j^3\right)^\top\boldsymbol{P}_{-j}^\perp \left( \hat{\mathbf{z}}_j^3\right)}{\left(\left(\hat{\mathbf{z}}_j^2\right)^\top\left(\hat{\mathbf{z}}_j^2\right)\right)^2} - \dfrac{1}{\hat{\mathbf{z}}_j^\top\hat{\mathbf{z}}_j}.
\end{equation*}
Following Theorem \ref{theo:non-gauss}, we can test the null hypothesis $H_0$ with a consistent estimate for $\sigma^2$. Such an estimate can be obtained, e.g., using the standard formula
\begin{equation*}
\hat{\sigma}^2 = \dfrac{\big\Vert \mathbf{y} - \mathbf{x} \hat{\boldsymbol{\beta}}^{OLS}\big\Vert_2 ^2 }{n-p}.
\end{equation*}
We define for later reference
\begin{equation}\label{eq:var-hat}
\widehat{\text{Var}}\left(\hat{\beta}_j^{HOLS}-\hat{\beta}_j^{OLS}\right) \coloneqq \hat{\sigma}^2 \dfrac{1}{n^2} \hat{\mathbf{v}}_j^\top\hat{\mathbf{v}}_j.
\end{equation}
To test $H_{0,j}$, we can compare $\left(\hat{\beta}_j^{HOLS}-\hat{\beta}_j^{OLS}\right)$ to the quantiles of the univariate normal distribution with the according variance.
The joint distribution leads to a global test that controls the type I error.
Namely, one can look at the maximum test statistic $T = \underset{k}{\text{max}} \left\vert \hat{\beta}_k^{HOLS} - \hat{\beta}_k^{OLS} \right\vert \overset{H_0}{\sim} \underset{k}{\text{max}} \left\vert S_k\right\vert$
, where $\mathbf{S} \sim \mathcal{N}\left(\boldsymbol{0}, \hat{\sigma}^2 \hat{\mathbf{v}} ^\top \hat{\mathbf{v}}/n^2\right)$ can be easily simulated. Further, one receives multiplicity corrected individual p-values for $H_{0,j}$ by comparing each $\left\vert \hat{\beta}_j^{HOLS} - \hat{\beta}_j^{OLS} \right\vert$ to the distribution of $\underset{k}{\text{max}} \left\vert S_k\right\vert$. This is in analogy to the multiplicity correction suggested by \cite{buhlmann2013statistical}. Naturally, other multiplicity correction techniques such as Bonferroni-Holm are valid as well.

Algorithm \ref{alg:HOLS} summarizes how to find both raw and multiplicity corrected p-values for each component $j$ corresponding to the $j$th covariate, $p_j$ and $P_j$ respectively.
Then, one would reject the global null hypothesis $H_0$ that the model \eqref{eq:mvmodel} holds if $\underset{j}{\text{min}} \ P_j \leq \alpha$, and such a decision procedure provides control of the type I error at level $\alpha$. Note that this means that we have strong control of the FWER for testing all $H_{0,j}$.
\begin{algorithm}
\caption{HOLS check}\label{alg:HOLS}
\begin{algorithmic}[1]
\For{$j=1$ to $p$}
\State $\boldsymbol{P}_{-j}^\perp = \boldsymbol{I}_n - \mathbf{x}_{-j}\left(\mathbf{x}_{-j}^\top \mathbf{x}_{-j}\right)\mathbf{x}_{-j}^\top$
\State Regress $\mathbf{x}_j$ versus $\mathbf{x}_{-j}$ via least squares, denote the residual by $\hat{\mathbf{z}}_j=\boldsymbol{P}_{-j}^\perp\mathbf{x}_j$
\State Regress $\mathbf{y}$ versus $\mathbf{x}_{-j}$ via least squares, denote the residual by $\hat{\mathbf{w}}_j=\boldsymbol{P}_{-j}^\perp\mathbf{y}$
\State $\hat{\beta}_j^{OLS} = \dfrac{\hat{\mathbf{z}}_j^\top \hat{\mathbf{w}}_j}{\hat{\mathbf{z}}_j^\top \hat{\mathbf{z}}_j}$, $\hat{\beta}_j^{HOLS} = \dfrac{\left(\hat{\mathbf{z}}_j^3\right)^\top \hat{\mathbf{w}}_j}{\hat{\mathbf{z}}_j^\top \hat{\mathbf{z}}_j}$ and $\hat{\mathbf{v}}_j =\dfrac{\boldsymbol{P}_{-j}^\perp \left( \hat{\mathbf{z}}_j^3\right)}{\dfrac{1}{n}\left(\hat{\mathbf{z}}_j^2\right)^\top\left(\hat{\mathbf{z}}_j^2\right)} - \dfrac{\hat{\mathbf{z}}_j}{\dfrac{1}{n}\hat{\mathbf{z}}_j^\top\hat{\mathbf{z}}_j}$
\EndFor
\State $\hat{\sigma}^2 = \dfrac{\big\Vert \mathbf{y} - \mathbf{x} \hat{\boldsymbol{\beta}}^{OLS}\big\Vert_2 ^2 }{n-p}$
\State Create $n_{sim}$ i.i.d\ copies of $\mathbf{S} \sim \mathcal{N}\left(\boldsymbol{0}, \hat{\sigma}^2 \hat{\mathbf{v}} ^\top \hat{\mathbf{v}}/n^2\right)$, say, $\mathbf{s}^1$ to $\mathbf{s}^{n_{sim}}$
\For{$j=1$ to $p$}
\State $p_j = 2 \left(1 - \Phi\left(\dfrac{\left\vert \hat{\beta}_j^{HOLS} - \hat{\beta}_j^{OLS}\right\vert}{\hat{\sigma}\dfrac{1}{n}\left\Vert \hat{\mathbf{v}}_j \right\Vert_2}\right)\right)$
\State $P_j = \dfrac{1}{n_{sim}}\sum_{i=1}^{n_{sim}}\mathbbm{1}\left(\left\vert \hat{\beta}_j^{HOLS} - \hat{\beta}_j^{OLS}\right\vert > \left\Vert \mathbf{s}^i\right\Vert_\infty\right)$
\EndFor 
\end{algorithmic}
\end{algorithm}
\begin{corr}\label{corr:alpha-H0}
Assume the conditions of Theorem \ref{theo:non-gauss}. Consider the decision rule to reject $H_0$ iff $\min_j P_j \le \alpha$, where $P_j$ is as in Step 10 of Algorithm \ref{alg:HOLS}. Then, the type I error
is asymptotically controlled at $\alpha$. Furthermore, the FWER is asymptotically controlled at level $\alpha$ for testing all local hypotheses $\{H_{0,j};\ j=1,\ldots ,p\}$ with the decision rule to reject $H_{0,j}$ iff $P_j \le \alpha$.
\end{corr}
We provide simulation results supporting this theory in Section \ref{app:sim} of the supplemental material.

\subsection{High-dimensional data}\label{high-dim}
We now turn to high-dimensional situations. We assume the global null hypothesis \eqref{eq:mvmodel} but allow for $p$ to diverge with and even exceed $n$ such that ordinary least squares regression is not applicable. Instead, we apply the debiased Lasso introduced in \cite{zhang2014confidence} and further discussed in \cite{van2014asymptotically}. We denote the estimator again by $\hat{\boldsymbol{\beta}}^{OLS}$ since it converges under certain conditions  to the population parameter $\boldsymbol{\beta}^{OLS}$.

\begin{algorithm}[t!]
\caption{HOLS check for $p>n$}\label{alg:HOLS-HD-alt}
\begin{algorithmic}[1]
\State Regress $\mathbf{y}$ versus $\mathbf{x}$ via Lasso with a penalty parameter $\lambda$, denote the estimated regression coefficients by $\hat{\boldsymbol{\beta}}$
\For{$j=1$ to $p$}
\State Regress $\mathbf{x}_j$ versus $\mathbf{x}_{-j}$ via Lasso with a penalty parameter $\lambda_j$, denote the residual by $\hat{\mathbf{z}}_j$
\State Regress $\hat{\mathbf{z}}_j ^3$ versus $\mathbf{x}_{-j}$ via Lasso with a penalty parameter $\tilde{\lambda}_j$, denote the residual by $\hat{\tilde{\mathbf{z}}}_j^3$
\State $\hat{\mathbf{w}}_j=\mathbf{y}- \mathbf{x}_{-j}\hat{\boldsymbol{\beta}}_{-j}$
\State $\hat{\beta}_j^{OLS} = \dfrac{\hat{\mathbf{z}}_j^\top \hat{\mathbf{w}}_j}{\hat{\mathbf{z}}_j^\top \mathbf{x}_j}$, $\hat{\beta}_j^{HOLS} = \dfrac{\left(\hat{\tilde{\mathbf{z}}}_j^3\right)^\top \hat{\mathbf{w}}_j}{\left(\hat{\tilde{\mathbf{z}}}_j^3\right)^\top \mathbf{x}_j}$ and $\hat{\mathbf{v}}_j =\dfrac{\left( \hat{\tilde{\mathbf{z}}}_j^3\right)}{\dfrac{1}{n}\left(\hat{\tilde{\mathbf{z}}}_j^3\right)^\top\mathbf{x}_j} - \dfrac{\hat{\mathbf{z}}_j}{\dfrac{1}{n}\hat{\mathbf{z}}_j^\top\mathbf{x}_j}$
\EndFor
\State $\hat{\sigma}^2 = \dfrac{\big\Vert \mathbf{y} - \mathbf{x} \hat{\boldsymbol{\beta}}\big\Vert_2 ^2 }{n-\left\vert \hat{\boldsymbol{\beta}} \right\vert_0}$ (or any other reasonable variance estimator)
\State Create $n_{sim}$ i.i.d\ copies of $\mathbf{S} \sim \mathcal{N}\left(\boldsymbol{0}, \hat{\sigma}^2 \dfrac{1}{n^2} \hat{\mathbf{v}} ^\top \hat{\mathbf{v}}\right)$, say, $\mathbf{s}^1$ to $\mathbf{s}^{n_{sim}}$
\For{$j=1$ to $p$}
\State $p_j = 2 \left(1 - \Phi\left(\dfrac{\left\vert \hat{\beta}_j^{HOLS} - \hat{\beta}_j^{OLS}\right\vert}{\hat{\sigma} \dfrac{1}{n} \left\Vert \hat{\mathbf{v}}_j \right\Vert_2}\right)\right)$
\State $P_j = \dfrac{1}{n_{sim}}\sum_{i=1}^{n_{sim}}\mathbbm{1}\left(\left\vert \hat{\beta}_j^{HOLS} - \hat{\beta}_j^{OLS}\right\vert > \left\Vert \mathbf{s}^i\right\Vert_\infty\right)$
\EndFor 
\end{algorithmic}
\end{algorithm}

From the debiased Lasso, we receive $\hat{\mathbf{z}}_j = \mathbf{x}_j - \mathbf{x}_{-j}\hat{\boldsymbol{\gamma}}_j$, where $\hat{\boldsymbol{\gamma}}_j$ is obtained by regressing $\mathbf{x}_j$ versus $\mathbf{x}_{-j}$ using the Lasso, and $\hat{\mathbf{w}}_j = \mathbf{y}-\mathbf{x}_{-j}\hat{\boldsymbol{\beta}}_{-j}$ with $\hat{\boldsymbol{\beta}}$ coming from the Lasso fit of $\mathbf{y}$ versus $\mathbf{x}$. Since $\hat{\beta}^{OLS}_j = \hat{\mathbf{z}}_j^\top \hat{\mathbf{w}}_j / \hat{\mathbf{z}}_j^\top\mathbf{x}_j$, one might want to use $\left(\hat{\mathbf{z}}_j^3\right)^\top \hat{\mathbf{w}}_j / \left(\hat{\mathbf{z}}_j^3\right)^\top\mathbf{x}_j$ for HOLS.
However, this leads in general to an uncontrollable approximation error since $\EE\left[Z_j^3 \mathbf{X}_{-j} \right]\neq \mathbf{0}$. As a remedy, we suggest a second level of orthogonalization based on $\tilde{Z}_j^3$ and $\boldsymbol{\tilde{\gamma}}_j$ as defined in \eqref{eq:ztilde-def}. Naturally, we have $\tilde{Z}_j^3 =Z_j^3$ iff $\EE\left[Z_j^3 \mathbf{X}_{-j}\right]=\mathbf{0}$ and always $\EE\left[\tilde{Z}_j^3 \mathbf{X}_{-j}\right]=\mathbf{0}$. To approximate $\tilde{\mathbf{z}}_j^3$ we use the Lasso for the regression $\hat{\mathbf{z}}_j^3$ versus $\mathbf{x}_{-j}$ leading to $\hat{\tilde{\mathbf{z}}}_j^3 = \hat{\mathbf{z}}_j^3 - \mathbf{x}_{-j} \hat{\tilde{\boldsymbol{\gamma}}}_j $. We define $\hat{\boldsymbol{\beta}}^{HOLS}$ as
\begin{equation*} 
\hat{\beta}^{HOLS}_j \coloneqq \dfrac{\left(\hat{\tilde{\mathbf{z}}}_j^3\right)^\top \hat{\mathbf{w}}_j}{\left(\hat{\tilde{\mathbf{z}}}_j^3\right)^\top\mathbf{x}_j} = \dfrac{\left(\hat{\tilde{\mathbf{z}}}_j^3\right)^\top \left(\mathbf{y}-\mathbf{x}_{-j}\hat{\boldsymbol{\beta}}_{-j}\right)}{\left(\hat{\tilde{\mathbf{z}}}_j^3\right)^\top\mathbf{x}_j} \overset{H_0}{=} \beta_j + \dfrac{\left(\hat{\tilde{\mathbf{z}}}_j^3\right)^\top\mathbf{x}_{-j}\left(\boldsymbol{\beta}_{-j} -\hat{\boldsymbol{\beta}}_{-j}\right)/n}{\left(\hat{\tilde{\mathbf{z}}}_j^3\right)^\top\mathbf{x}_j /n} + \dfrac{\left(\hat{\tilde{\mathbf{z}}}_j^3\right)^\top \boldsymbol{\eps}}{\left(\hat{\tilde{\mathbf{z}}}_j^3\right)^\top\mathbf{x}_j}.
\end{equation*}

Finally, we are interested in the difference between $\hat{\beta}^{HOLS}_j$ and $\hat{\beta}^{OLS}_j$. Under suitable assumptions for the sparsity, the moment matrix, and the tail behaviour of $\mathbf{X}$ and $\mathcal{E}$, we can derive the limiting Gaussian distribution of this difference allowing for asymptotically valid tests.
We apply Algorithm \ref{alg:HOLS-HD-alt} where we make use of the (asymptotic) normality of the non-vanishing term in this difference. For non-Gaussian $\mathcal{E}$, a multiplicity correction method that does not rely on exact Gaussianity of this remainder might be preferred since the CLT does not apply for dimensions growing too fast.

We provide here the main result to justify Algorithm \ref{alg:HOLS-HD-alt} invoking additional assumptions on the dimensionality and sparsity of the problem. 
We use the definitions $s \coloneqq \left\Vert \boldsymbol{\beta} \right \Vert_0$, $s_j \coloneqq \left\Vert \boldsymbol{\gamma}_j \right \Vert_0$ and $\tilde{s}_j \coloneqq \left\Vert \tilde{\boldsymbol{\gamma}}_j \right \Vert_0$ to denote the different levels of sparsity.

\begin{assc}\label{ass:B-cov} 
The design matrix $\mathbf{x}$ has i.i.d.\ sub-Gaussian rows. The moment matrix $\boldsymbol{\Sigma}^{\mathbf{X}}$ has strictly positive smallest eigenvalue $\Lambda_\text{min}^2$ satisfying $1/\Lambda_\text{min}^2 = \mathcal{O}\left(1\right).$ Also, $\underset{j}{\text{max}} \ \boldsymbol{\Sigma}^{\mathbf{X}}_{j,j} = \mathcal{O}\left(1\right).$
\end{assc}
\begin{assc}\label{ass:B-s} 
$s = {\scriptstyle \mathcal{O}}\left(\dfrac{n^{1/2}}{\log\left(p\right)^3}\right)$.
\end{assc}
\begin{assc}\label{ass:B-ssj} 
$s s_j ^2 ={\scriptstyle \mathcal{O}}\left(\dfrac{n^{3/2}}{\log\left(p\right)^3}\right)$, $ss_j={\scriptstyle \mathcal{O}}\left(\dfrac{n}{\log\left(p\right)^{5/2}}\right)$ and $s s_j^{1/2}={\scriptstyle \mathcal{O}}\left(\dfrac{n^{1/2}}{\log\left(p\right)^{3/2}}\right)$.
\end{assc}
\begin{multicols}{3}
\begin{assc}\label{ass:B-sj} 
$s_j = {\scriptstyle \mathcal{O}}\left(\dfrac{n^{3/5}}{\log\left(p\right)}\right)$.
\end{assc}
\begin{assc}\label{ass:B-slam}
$\sqrt{n} s \lambda \tilde{\lambda}_j ={\scriptstyle\mathcal{O}} \left(1 \right)$.
\end{assc}
\begin{assc}\label{ass:B-sjlam}
$\tilde{s}_j \tilde{\lambda}_j^2 ={\scriptstyle\mathcal{O}} \left(1 \right)$.
\end{assc}
\end{multicols}

\newpage
\begin{theo}\label{theo:HD-local-gauss}
Assume that the data follows the model \eqref{eq:mvmodel} with sub-Gaussian $\mathcal{E}$ and that \ref{ass:B-cov} -  \ref{ass:B-sjlam} hold ($\forall j$). Let $\hat{\boldsymbol{\beta}}$ come from Lasso regression with $\lambda \asymp \sqrt{\log\left(p\right)/n}$, $\hat{\mathbf{z}}_j$ from nodewise Lasso regression using $\lambda_j \asymp \sqrt{\log\left(p\right)/n}$, and $\hat{\tilde{\mathbf{z}}}_j^3$ from nodewise Lasso regression of $\hat{\mathbf{z}}_j^3$ versus $\mathbf{x}_{-j}$ using \\$\tilde{\lambda}_j \asymp \text{max}\left\{\log\left(p\right)^{5/2}n^{-1/2},s_j^2 \log\left(p\right)^{5/2} n^{-3/2}, s_j \log\left(p\right)^{2} n^{-1}, \sqrt{s_j} \log\left(p\right) n^{-1/2} \right\}$. Let $\hat{\sigma}$ be any consistent estimator for $\sigma$. Then,
\begin{equation*}
\dfrac{\sqrt{n}\left(\hat{\beta}_j^{HOLS}-\hat{\beta}_j^{OLS}\right)}{\sqrt{\hat{\sigma}^2 \dfrac{1}{n}\hat{\mathbf{v}}_j^\top \hat{\mathbf{v}}_j}}\overset{\mathbb{D}}{\to}\mathcal{N}\left(0,1\right)\quad \text{where} \quad \hat{\mathbf{v}}_j =\dfrac{\left( \hat{\tilde{\mathbf{z}}}_j^3\right)}{\dfrac{1}{n}\left(\hat{\tilde{\mathbf{z}}}_j^3\right)^\top\mathbf{x}_j} - \dfrac{\hat{\mathbf{z}}_j}{\dfrac{1}{n}\hat{\mathbf{z}}_j^\top\mathbf{x}_j}.
\end{equation*}
\end{theo}

We defer the technical details to Section \ref{app:HD} of the supplemental material. 
Simulation results concerning high-dimensional data can be found in Section \ref{app:sim} of the supplemental material.

\section{The confounded case and local null hypotheses}\label{lin-conf}
In this section and the following, we mainly exploit confounding in linear models as the alternative hypothesis since these are the models where our tests for the local null hypotheses $H_{0,j}$ are most informative. For a discussion of which interpretations might carry over to more general data generating distributions, we refer to Section \ref{non-lin}.

Note that everything that is discussed in Sections \ref{lin-conf} and \ref{ex-models} implicitly applies to high-dimensional data as well under suitable assumptions. We refrain from going into detail for the sake of brevity. Thus, Theorems \ref{theo:conf} - \ref{theo:U-rec} which contain our main asymptotic results for the local intepretation are designed explicitly for the fixed $p$ case.

We look at the causal model
\begin{align}\label{eq:mvconfmodel}
\begin{split}
\mathbf{X} & \leftarrow \boldsymbol{\rho}\mathbf{H} + \mathcal{E}_{\mathbf{X}} \\
Y & \leftarrow \mathbf{X}^{\top}\boldsymbol{\beta} + \mathbf{H}^\top\boldsymbol{\alpha}+ \mathcal{E},
\end{split}
\end{align}
where $\mathbf{H} \in \mathbb{R}^d$, $\mathcal{E}_{\mathbf{X}} \in \mathbb{R}^p$ and $\mathcal{E} \in \mathbb{R}$ are independent and centered random variables, and $\boldsymbol{
\alpha} \in \mathbb{R}^d$ and $\boldsymbol{\rho}\in \mathbb{R}^{p \times d}$ are fixed model parameters. Thus, there exists some hidden confounder $\mathbf{H}$. For the inner product matrices, it holds that
\begin{equation*}
\boldsymbol{\Sigma}^{\mathbf{X}} = \boldsymbol{\Sigma}^{\mathcal{E}_{\mathbf{X}}} + \boldsymbol{\rho}\boldsymbol{\Sigma}^{\mathbf{H}}\boldsymbol{\rho}^\top.
\end{equation*}
Furthermore, we have
\begin{equation}\label{eq:bols}
\boldsymbol{\beta}^{OLS} =\left( \boldsymbol{\Sigma}^{\mathbf{X}}\right)^{-1}\mathbf{E}\left[\mathbf{X}Y\right] = \left( \boldsymbol{\Sigma}^{\mathbf{X}}\right)^{-1}\left(\boldsymbol{\Sigma}^{\mathbf{X}}\boldsymbol{\beta} + \boldsymbol{\rho}\boldsymbol{\Sigma}^{\mathbf{H}}\boldsymbol{\alpha} \right) = \boldsymbol{\beta} + \left( \boldsymbol{\Sigma}^{\mathbf{X}}\right)^{-1}\boldsymbol{\rho}\boldsymbol{\Sigma}^{\mathbf{H}}\boldsymbol{\alpha}
\end{equation}
We will generally refer to $\beta_j^{OLS} \neq \beta_j$, where $\beta_j$ is according to model \eqref{eq:mvconfmodel}, as confounding bias on $\beta_j^{OLS}$. Further, when writing directly confounded, we mean covariate indices $j$ for which $X_j \neq \mathcal{E}_{X_j}$.

Note that we can always decompose $Y$ both globally and locally as follows
\begin{alignat}{5}
Y & = \mathbf{X}^\top \boldsymbol{\beta}^{OLS} + \tilde{\mathcal{E}}, \quad && \text{with} \quad \EE\left[\mathbf{X}\tilde{\mathcal{E}}\right] = \mathbf{0},\quad && \EE\left[\tilde{\mathcal{E}}\right]=0 \quad \text{but (potentially)} \quad \mathbf{X}\not\perp\tilde{\mathcal{E}} \label{eq:refmodel} \\
W_j & = Z_j \beta_j^{OLS} + \tilde{\mathcal{E}}, \quad && \text{with} \quad \EE\left[Z_j\tilde{\mathcal{E}}\right] = \mathbf{0},\quad && \EE\left[\tilde{\mathcal{E}}\right]=0 \quad \text{but (potentially)} \quad Z_j\not\perp\tilde{\mathcal{E}} \label{eq:refmodel-1D}
\end{alignat}
using the definitions from \eqref{eq:z-w-def}.
We now want to see how $\boldsymbol{\beta}^{OLS}$ relates to $\boldsymbol{\beta}$ in certain models. Especially, we are interested in whether there is some potential local interpretation in the sense of distinguishing between ``confounded'' and ``unconfounded'' variables. From \eqref{eq:bols}, we see that this is linked to the structure of the covariance matrices as well as $\boldsymbol{\rho}$ and $\boldsymbol{\alpha}$. We define the sets
\begin{equation}\label{eq:UV-def}
V = \left\{j:\beta_j^{OLS}=\beta_j\right\} \quad \text{and} \quad   U=\left\{j:\beta_j^{OLS}=\beta_j^{HOLS}\right\}=\left\{j: H_{0,j} \ \text{is true}\right\}.
\end{equation}
Using the Woodbury matrix identity, we find a sufficient condition
\begin{align}\label{eq:no-neighbour}
j \in V \quad \text{if} \quad \boldsymbol{\rho}^\top \left(\boldsymbol{\Sigma}^{\mathcal{E}_{\mathbf{X}}}\right)^{-1}_j & = \mathbf{0} \quad \text{which is implied by} \nonumber \\
 \Big\{k \in \left\{1,\ldots,p\right\}: \ \left(\boldsymbol{\Sigma}^{\mathcal{E}_{\mathbf{X}}}\right)^{-1}_{jk} \neq 0 \Big\} \cap \Big\{l \in \left\{1,\ldots,p\right\}: \ \left\Vert \mathbf{e}_l^\top \boldsymbol{\rho}\right\Vert >0 \Big\}&=\emptyset.
\end{align}
Thus, if the intersection between covariates that have linear predictive power for $X_j$ and covariates that are directly confounded is empty, it must hold that $\beta_j^{OLS} = \beta_j$. Therefore, we can indeed say that for these variables we estimate the true causal effect using ordinary least squares.

To correctly detect $V$, we would like $\beta_j^{HOLS}=\beta_j^{OLS}=\beta_j$. As $\beta_j^{HOLS}$ involves higher-order moments, knowledge of the covariance structure is not sufficient to check this. From \eqref{eq:refmodel-1D}, we see that $\EE\left[Z_j^3 \tilde{\mathcal{E}}\right]=0$ is necessary and sufficient to ensure $j \in U$. In Section \ref{inter-U}, we discuss the two cases where detection fails, i.e., $U \setminus V \neq \emptyset$ and $V \setminus U \neq \emptyset$. We present models for which we can characterize a set of variables which are in $U \cap V =\left\{j:\beta_j^{HOLS}=\beta_j^{OLS}=\beta_j\right\}$ in Section \ref{ex-models}.

\subsection{Sample estimates}\label{conv-par}
For a confounded model, the hope is that the global test $\underset{j}{\text{min}}P_j \leq \alpha$, where $P_j$ is the adjusted p-value according to Step 10 in Algorithm \ref{alg:HOLS}, leads to a rejection of $H_0$, i.e., the modelling assumption \eqref{eq:mvmodel}. One could further examine the local structure and, based on the corrected p-values $P_j$, distinguish the predictors for which we have evidence that $\beta_j^{HOLS} \neq \beta_j^{OLS}$.
We consider in the following this local interpretation, showing that we asymptotically control the type-I error and receive power approaching $1$. Implicitly, we assume that $U$ is a useful proxy for $V$.

For all asymptotic results in this section, we assume $p$ to be fixed and $n \rightarrow \infty$ as in Theorem \ref{theo:non-gauss}.
\begin{theo}\label{theo:conf}
Assume that the data follows the model \eqref{eq:refmodel} and that \ref{ass:invertible}-\ref{ass:mom6} hold. Assume further $\sigma_{\tilde{\mathcal{E}}}^2=\EE\left[\tilde{\mathcal{E}}^2\right] < \infty$. Then,
\begin{equation*}
\hat{\beta}_j^{OLS} = \beta_j^{OLS} + { \scriptstyle \mathcal{O}}_p\left(1\right), \quad \hat{\beta}_j^{HOLS} = \beta_j^{HOLS} + { \scriptstyle \mathcal{O}}_p\left(1\right) \quad \text{and} \quad \widehat{\text{Var}}\left(\hat{\beta}_j^{HOLS} - \hat{\beta}_j^{OLS} \right)=\mathcal{O}_p \left(\dfrac{1}{n}\right),
\end{equation*}
where $\widehat{\text{Var}}\left(\hat{\beta}_j^{HOLS} - \hat{\beta}_j^{OLS}\right)$ is according to \eqref{eq:var-hat}.
\end{theo}
Thus, for some fixed alternative $\left\vert \beta_j^{HOLS} - \beta_j^{OLS} \right\vert > 0$, the absolute z-statistics increases as $\sqrt{n}$.

In order to get some local interpretation, the behaviour for variables $j \in U$ is of importance. If $ \left\vert \beta_j^{HOLS} - \beta_j^{OLS} \right\vert = 0$, Theorem \ref{theo:conf} is not sufficient to understand the asymptotic behaviour. We refine the results using additional assumptions.
\begin{multicols}{2}
\begin{assa}\label{ass:mixmom}
$\EE\left[\left(X_j\tilde{\mathcal{E}}\right)^2\right] < \infty \ \forall j$ \\
\end{assa}
\begin{assb}\label{ass:LLN-cond}
$\EE\left[Z_j^2 X_k \tilde{\mathcal{E}}\right]=0 \ \forall k \neq j$
\end{assb}
\begin{assb}\label{ass:z-ind}
$Z_j \perp \tilde{\mathcal{E}}$
\end{assb}
\begin{assb}\label{ass:z3-ind}
$\tilde{Z}_j^3 \perp \tilde{\mathcal{E}}$
\end{assb}
\end{multicols}
Note that we use different letters for the assumptions to distinguish between those that are essentially some (mild) moment conditions and those that truly make nodes unconfounded.
Obviously, \ref{ass:z-ind} is not necessary for $\beta_j^{OLS}=\beta_j^{HOLS}$, but we will focus on these variables as these are the ones that are truly unconfounded in the sense that the projected single variable model \eqref{eq:refmodel-1D} has an independent error term, while as for other variables it can be rather considered an unwanted artefact of our method. Furthermore, the derived asymptotic variance results only hold true when assuming \ref{ass:z-ind}  and \ref{ass:z3-ind} as well. Assumption \ref{ass:mixmom} implies a further moment condition. Especially, when considering nonlinearities, there exist cases for which \ref{ass:mixmom} is not implied by \ref{ass:mom6}. We discuss Assumptions \ref{ass:LLN-cond}, \ref{ass:z-ind} and \ref{ass:z3-ind} for certain models in Section \ref{ex-models}. Condition \ref{ass:LLN-cond} seems to be a bit artificial but is invoked in the proofs. We argue in Section \ref{ex-models} that it is naturally linked to the models in scope.

\begin{theo}\label{theo:local-standard-gauss}
Assume that the data follows the model \eqref{eq:refmodel} and that \ref{ass:invertible} - \ref{ass:mixmom} hold. Let $j$ be some covariate with $\beta_j^{OLS}=\beta_j^{HOLS}$ for which \ref{ass:LLN-cond} - \ref{ass:z3-ind} hold. Then,
\begin{equation*}
\dfrac{\sqrt{n}\left(\hat{\beta}_j^{HOLS}-\hat{\beta}_j^{OLS}\right)}{\sqrt{\widehat{\text{Var}}\left(\sqrt{n}\left(\hat{\beta}_j^{HOLS}-\hat{\beta}_j^{OLS}\right)\right)}}\overset{\mathbb{D}}{\to}\mathcal{N}\left(0,1\right).
\end{equation*}
\end{theo}
Thus, for these predictors we receive asymptotically valid tests.

\paragraph*{Multiplicity correction} In order not to falsely reject the local null hypothesis $H_{0,j}$ for any covariate with $j \in U$ (with probability at least $1-\alpha$), we need to invoke some multiplicity correction. Analogously to Section \ref{mv}, one can see that $\hat{\boldsymbol{\beta}}^{HOLS} - \hat{\boldsymbol{\beta}}^{OLS} = \hat{\mathbf{v}}^\top \tilde{\boldsymbol{\eps}} / n$, which enables the multiplicity correction as in Algorithm \ref{alg:HOLS}.
\begin{theo}\label{theo:V_U}
Assume that the data follows the model \eqref{eq:refmodel} and that \ref{ass:invertible} - \ref{ass:mixmom} hold. Let $U' $ be the set of variables $j$ for which $j \in U$ and \ref{ass:LLN-cond} - \ref{ass:z3-ind} hold. Then, 
\begin{align*}
\sqrt{n} \left(\hat{\boldsymbol{\beta}}_{U'}^{HOLS}-\hat{\boldsymbol{\beta}}_{U'}^{OLS}\right) & \overset{\mathbb{D}}{\to} \mathcal{N}\left(\mathbf{0}, \sigma_{\tilde{\mathcal{E}}}^2 \EE\left[\mathbf{V}_{U'}\mathbf{V}_{U'}^\top\right]\right)\\
\dfrac{1}{n}\hat{\mathbf{v}}_{U'}^\top\hat{\mathbf{v}}_{U'} & \overset{\mathbb{P}}{\to} \EE\left[\mathbf{V}_{U'}\mathbf{V}_{U'}^\top\right]
\end{align*}
\end{theo}
\begin{corr}\label{corr:alpha-H0j}
Assume the conditions of Theorem \ref{theo:V_U}. Consider the decision rule to reject $H_{0,j}$ iff $ P_j \le \alpha$, where $P_j$ is as in Step 10 of Algorithm \ref{alg:HOLS}. Then, the familywise error rate amongst the set $U'$ is asymptotically controlled at $\alpha$.
\end{corr}

\subsection{Inferring $V$ from $U$}\label{inter-U}
Recall the definitions in \eqref{eq:UV-def}. $U$ is the set that we try to infer with our HOLS check. Naturally, one would rather be interested in the set $V$, which consists of the variables for which we can consistently estimate the true linear causal effect according to \eqref{eq:mvconfmodel} through linear regression. We discuss here when using $U$ as proxy for $V$ might fail and especially analyse how variables could belong to the difference between the sets. For this, recall our formulation of the model when the global null hypothesis does not hold true in \eqref{eq:refmodel} and \eqref{eq:refmodel-1D}. Note that $j \in U$ is equivalent to $\EE\left[Z_j^3\tilde{\mathcal{E}}\right]=0$. 

For any variable $j \in U \setminus V$, certain modelling assumptions, that we discuss in the sequel, cannot be fulfilled but they are not necessary for $\EE\left[Z_j^3\tilde{\mathcal{E}}\right]=0$. Especially, the last equality always holds if both $\mathcal{E}_{\mathbf{X}}$ and $H$ jointly have Gaussian kurtosis. If they are even jointly Gaussian, then it is clear that $\mathbf{X} \perp \tilde{\mathcal{E}}$ such that the model $ \eqref{eq:refmodel}$ has independent Gaussian error. Thus, when using only observational data, it behaves exactly like a model under the global null hypothesis and, naturally, we cannot infer the confounding effect. Apart from Gaussian kurtosis, $j \in U \setminus V$ would be mainly due to special constellations implying cancellation of terms that one does not expect to encounter in normal circumstances.

For $j \in V \setminus U$, $Z_j$ and $\tilde{\mathcal{E}}$ must not be independent. As $\ Z_j \not\perp \tilde{\mathcal{E}}$, the single-covariate model \eqref{eq:refmodel-1D} is not a linear causal model with independent error term as given in \eqref{eq:1D-model}. Therefore, from a causal inference perspective, one can argue that rejecting the local null hypothesis $H_{0,j}$ is the right thing to do in this case. Furthermore, having variables $j \in V$ is usually related to certain model assumptions except for very specific data setups that lead to cancellation of terms. Under these assumptions, $Z_j \perp \tilde{\mathcal{E}}$ is then usually implied. An example where $\boldsymbol{\beta}^{OLS} = \boldsymbol{\beta}$, but (potentially) $\mathbf{X} \not \perp \tilde{\mathcal{E}}$ is data for which $\boldsymbol{\rho}\boldsymbol{\Sigma}^{\mathbf{H}}\boldsymbol{\alpha} = 0$ using the definitions from model \eqref{eq:mvconfmodel}.

\paragraph{Recovery of $U$} \label{recov-U}
Based on our asymptotic results when the global null does not hold true, we would like to construct a method that perfectly detects the unconfounded variables as $n \rightarrow \infty$. Define
\begin{equation}\label{eq:Uhat-def}
\hat{U} = \left\{j: H_{0,j} \ \text{not rejected}\right\}
\end{equation}
The question is how and when can we ensure that 
\begin{equation*}
\underset{n \rightarrow \infty}{\text{lim}} \mathbb{P}\left[\hat{U} = U\right]=1.
\end{equation*}
Suppose that we conduct our local $z$-tests at level $\alpha_n$, which varies with the sample size such that $\alpha_n \rightarrow 0 $ as $n \rightarrow \infty$. It will be more convenient to interpret this as a threshold on the (scaled) absolute $z$-statistics, say, $\tau_n$ that grows with $n$, where the $z$-statistics is defined as
\begin{equation*}
t_j=\dfrac{\sqrt{n}\left(\hat{\beta}_j^{HOLS}-\hat{\beta}_j^{OLS}\right)}{\sqrt{\widehat{\text{Var}}\left(\sqrt{n}\left(\hat{\beta}_j^{HOLS}-\hat{\beta}_j^{OLS}\right)\right)}}.
\end{equation*}
We refrain from calling it $z_j$ to avoid confusion. We use an additional assumption which is a relaxed version of \ref{ass:z3-ind}.
\begin{assa}\label{ass:6}
$\EE\left[\left(\tilde{Z}_j^3\tilde{\mathcal{E}}\right)^2\right] < \infty$
\end{assa}
\begin{theo}\label{theo:U-rec}
Assume that the data follows the model \eqref{eq:refmodel} and that \ref{ass:invertible} - \ref{ass:mixmom} hold. Assume that \ref{ass:LLN-cond} and \ref{ass:6} hold $\forall j \in U$. Let $\tau_n$ be the threshold on the absolute $z$-statistics to reject the according null hypothesis with $\tau_n = { \scriptstyle \mathcal{O}}\left(\sqrt{n} \right)$ and $1/\tau_n ={ \scriptstyle \mathcal{O}}\left(1\right)$. Then,
\begin{equation*}
\underset{n \rightarrow \infty}{\text{lim}} \mathbb{P}\left[\hat{U} = U\right]=1.
\end{equation*}
\end{theo}
In other words, we can choose $\tau_n$ to grow at any rate slower than $\sqrt{n}$.

\section{Specific models}\label{ex-models}
In this section, we discuss two types of models where the local interpretation applies. In these settings, there are variables for which $\beta_j = \beta_j^{OLS} = \beta_j^{HOLS}$ and assumptions \ref{ass:LLN-cond}-\ref{ass:z3-ind} hold even though the overall data follows the model \eqref{eq:mvconfmodel}. We note here first that the model of jointly Gaussian $\mathcal{E}_{\mathbf{X}}$, for which the method is suited, is a special case of the model in Section \ref{lin-SEM} below.

\subsection{Block independence of $\mathcal{E}_{\mathbf{X}}$}\label{block}
Assume that the errors $\mathcal{E}_{\mathbf{X}}$ can be grouped into two or more independent and disjoint blocks. Denote the block that includes $j$ by $B\left(j\right)$. Then, it is clear that $\left(\boldsymbol{\Sigma}^{\mathcal{E}_{\mathbf{X}}}\right)^{-1}_{jk} =0 $ if $B\left(j\right) \neq B\left(k\right)$. If $\mathbf{X}_{B\left(j\right)} = \mathcal{E}_{\mathbf{X}_{B\left(j\right)}}$, i.e., the confounder has no effect onto $\mathbf{X}_{B\left(j\right)}$, \eqref{eq:no-neighbour} holds for all covariates in $B\left(j\right)$. Then, no variable in $\mathbf{X}_{B\left(j\right)}$ contributes to the best linear predictor for $\mathbf{H}^\top \boldsymbol{\alpha}$. Due to the block independence, this yields $\mathbf{X}_{B\left(j\right)} \perp \tilde{\mathcal{E}}$ and $Z_j \perp \tilde{\mathcal{E}}$, i.e., \ref{ass:z-ind} is fulfilled. This also ensures $\EE\left[Z_j^3 \tilde{\mathcal{E}}\right]=0$. We consider the remaining assumptions: Naturally, the regression $Z_j^3$ versus $\mathbf{X}_{-j}$ only involves $\mathbf{X}_{B\left(j\right) \setminus j }$ and \ref{ass:z3-ind} holds as well. For \ref{ass:LLN-cond}, separately consider the case $k \in B\left(j\right)$ and $k \not \in B\left(j\right)$. In the first case, $\EE\left[Z_j^2 X_k \tilde{\mathcal{E}}\right]= \EE\left[Z_j^2 X_k \right]\EE\left[\tilde{\mathcal{E}}\right]=0$. In the second case, $\EE\left[Z_j^2 X_k \tilde{\mathcal{E}}\right]= \EE\left[Z_j^2\right]\EE\left[X_k \tilde{\mathcal{E}}\right]=0$.
\begin{theo}
Assume the data follows the model \eqref{eq:mvconfmodel} with errors $\mathcal{E}_{\mathbf{X}}$ that can be grouped into independent blocks. Then,
\begin{align*}
\beta_j^{HOLS} = \beta_j^{OLS} = \beta_j \ & \forall j \quad \text{for which} \quad \mathbf{X}_{B\left(j\right)} = \mathcal{E}_{\mathbf{X}_{B\left(j\right)}}. \text{ Further,}\\
\text{\ref{ass:LLN-cond} -\ref{ass:z3-ind} hold} \ &\forall j \quad \text{for which} \quad \mathbf{X}_{B\left(j\right)} = \mathcal{E}_{\mathbf{X}_{B\left(j\right)}}.
\end{align*}
\end{theo}
In some cases, block independence may be a restrictive assumption. Testing this assumption is not an easy problem, and will remain out of the scope of this paper. However, the HOLS check still provides an indirect check of such an assumption since HOLS would likely reject the local null-hypotheses for all covariates, at least for large data-sets, if there is no block that is unaffected by the confounding.
\subsection{Linear structural equation model}\label{lin-SEM}
From the previous sections, we know that locally unconfounded structures, in the sense that $\beta_j^{OLS} = \beta_j$, are strongly related to zeroes in the precision matrix. Thus, the question arises for what type of models having zeroes in the precision matrix is a usual thing. Besides block independence, which we have discussed in Section \ref{block}, this will mainly be the case if the data follows a linear structural equation model (SEM). Thus, we will focus on these linear SEMs for the interpretation of local, i.e., by parameter, null hypotheses.

To start, assume that there are no hidden variables. So, let $\mathbf{X}$ be given by the following linear SEM
\begin{equation}\label{eq:SEM-model}
X_j \leftarrow \Psi_j + \sum_{k \in \text{PA}\left(j\right)} \theta_{j,k} X_k \quad j=1,\ldots ,p,
\end{equation}
where the $\Psi_1,\ldots ,\Psi_p$ are independent and centered random variables. We use the notation $\text{PA}\left(j\right)$, $\text{CH}\left(j\right)$ and $\text{AN}\left(j\right)$ for $j$'s parents, children and ancestors. Further, assume that there exists a directed acyclic graph (DAG) representing this structure. For this type of model, we know that a variable's Markov boundary consists of its parents, its children, and its children's other parents. For every other variable $k$ outside of $j$'s Markov boundary, we have $\left(\boldsymbol{\Sigma}^{{\mathbf{X}}}\right)^{-1}_{jk}= 0$. Thus, these $0$ partial correlations are very usual. In the following, we will analyse how our local tests are especially applicable to this structure.

In the context of linear SEMs, hidden linear confounders can be thought of as unmeasured variables. Therefore, we split $\mathbf{X}$ which contains all possible predictors into two parts. Let $\mathbf{X}_M$ be the measured variables and $\mathbf{X}_N$ the hidden confounder variables. Let $\boldsymbol{\Psi} = \begin{pmatrix}
\Psi_1, \ldots, \Psi_p
\end{pmatrix}^\top $ with the according subsets $\boldsymbol{\Psi}_M$ and $\boldsymbol{\Psi}_N$. Then, we can write
\begin{equation*}
\mathbf{X} = \boldsymbol{\omega} \boldsymbol{\Psi} , \quad \mathbf{X}_M = \boldsymbol{\omega}_{M,M} \boldsymbol{\Psi}_M + \boldsymbol{\omega}_{M,N}\boldsymbol{\Psi}_N \quad \text{and} \quad \mathbf{X}_N = \boldsymbol{\omega}_{N,M} \boldsymbol{\Psi}_M + \boldsymbol{\omega}_{N,N}\boldsymbol{\Psi}_N
\end{equation*}
for some suitable $\boldsymbol{\omega} \in \mathbb{R}^{p\times p}$, where $\omega_{k,l} = 0$ for $k\neq l$ if $l \not \in \text{AN}\left(k\right)$ and $\omega_{k,k}=1$. Note that $\boldsymbol{\omega}_{M,M}$ is always invertible since it can be written as a triangular matrix with ones on the diagonal if permuted properly. Under model \eqref{eq:mvconfmodel}, $Y$ can be thought of as a sink node in \eqref{eq:SEM-model}.
To avoid confusion, we call the parameter if all predictors were observed $\boldsymbol{\beta}^*$. This leads to the definitions
\begin{align}\label{eq:lin-SEM-H-def}
\mathcal{E}_{\mathbf{X}} & \coloneqq \boldsymbol{\omega}_{M,M}  \boldsymbol{\Psi}_M, \quad \boldsymbol{\rho} \coloneqq \boldsymbol{\omega}_{M,N} \quad \text{and} \quad \mathbf{H} \coloneqq \boldsymbol{\Psi}_N\quad \text{such that} \\
\mathbf{X}_M & = \mathcal{E}_{\mathbf{X}} + \boldsymbol{\rho} \mathbf{H} \quad \text{with} \quad \mathcal{E}_{\mathbf{X}} \perp \mathbf{H} \nonumber \\
Y -\mathcal{E}& = \mathbf{X}^\top \boldsymbol{\beta}^* = \mathbf{X}_M^\top \boldsymbol{\beta}^*_M + \mathbf{X}_N^\top \boldsymbol{\beta}^*_N \nonumber \\
& = \mathbf{X}_M^\top\left(\boldsymbol{\beta}^*_M + \left( \boldsymbol{\omega}_{N,M} \boldsymbol{\omega}_{M,M}^{-1} \right)^\top \boldsymbol{\beta}^*_N\right) + \mathbf{H}^\top\left(\boldsymbol{\omega}_{N,N} - \boldsymbol{\omega}_{N,M} \boldsymbol{\omega}_{M,M}^{-1} \boldsymbol{\omega}_{M,N}\right)^\top \boldsymbol{\beta}^*_N \nonumber \\
&  \coloneqq \mathbf{X}_M^\top \boldsymbol{\beta} + \mathbf{H}^\top \boldsymbol{\alpha}. \nonumber 
\end{align}
When only the given subset is observed we are interested in the parameter $\boldsymbol{\beta}$ as before. We have $\beta_j=\beta^*_j$ iff $\big(\big(\boldsymbol{\omega}_{N,M} \boldsymbol{\omega}_{M,M}^{-1} \big)^\top \boldsymbol{\beta}^*_N \big)_j=0$.
\begin{theo}\label{theo:lin-SEM-betaM}
Assume that the data follows the model \eqref{eq:SEM-model} and \eqref{eq:lin-SEM-H-def}. Let $\mathbf{X}_M$ and $\mathbf{X}_N$ be the observed and hidden variables. Denote by $\text{PA}^M\left(k\right) $ the closest ancestors of $k$ that are in $M$. Consider some $j \in M$.
\begin{equation*}
\text{If} \quad \not \exists k \in N: \left(\ j \in \text{PA}^M\left(k\right) \ \text{and} \ \beta_k \neq 0\right), \quad \text{then} \quad \beta_j=\beta^*_j.
\end{equation*}
\end{theo}

In other words, the causal parameter can only change for variables that have at least one direct descendant in the hidden set which is a parent of $Y$ itself. By direct descendant, we mean that there is a path from $j$ to $k$ that does not pass any other observed variable. We analyse for which variables we can reconstruct this causal parameter using ordinary least squares regression. 

\begin{theo}\label{theo:lin-SEM-beta-eq}
Assume that the data follows the model \eqref{eq:SEM-model} and \eqref{eq:lin-SEM-H-def}. Let $\mathbf{X}_M$ and $\mathbf{X}_N$ be the observed and hidden variables. Then,
\begin{align*}
\beta_j^{HOLS} =\beta_j^{OLS} = \beta_j \ &\forall j\in M \quad \text{that are not in the Markov boundary of any hidden variable.}\\
\text{\ref{ass:LLN-cond} -\ref{ass:z3-ind} hold} \ &\forall j\in M \quad \text{that are not in the Markov boundary of any hidden variable.}
\end{align*}
\end{theo}
Thus, for those variables, we can a) correctly retrieve the causal parameter using ordinary least squares regression and b) detect that this is the true parameter by comparing it to $\beta_j^{HOLS}$. 

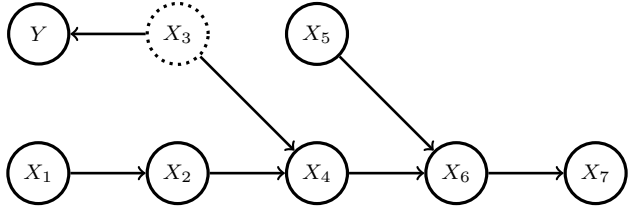
\begin{figure}[t!]
\centering
\begin{tikzpicture}[roundnode/.style={circle, very thick, minimum size=8mm}]
\node[draw, roundnode, text centered] (x1) {${\scriptstyle X_1}$};
\node[draw, roundnode, right =of x1, text centered] (x2) {${\scriptstyle X_2}$};
\node[draw, dotted, roundnode, above =of x2, text centered] (x3) {${\scriptstyle X_3}$};
\node[draw, roundnode, right =of x2, text centered] (x4) {${\scriptstyle X_4}$};
\node[draw, roundnode, above =of x4, text centered] (x5) {${\scriptstyle X_5}$};
\node[draw, roundnode, right=of x4, text centered] (x6) {${\scriptstyle X_6}$};
\node[draw, roundnode, right=of x6, text centered] (x7) {${\scriptstyle X_7}$};
\node[draw, roundnode, left=of x3, text centered] (y) {${\scriptstyle Y}$};

\draw[->, line width= 1] (x1) -- (x2);
\draw[->, line width= 1] (x2) -- (x4);
\draw[->, line width= 1] (x3) -- (x4);
\draw[->, line width= 1] (x4) -- (x6);
\draw[->, line width= 1] (x5) -- (x6);
\draw[->, line width= 1] (x6) -- (x7);
\draw[->, line width= 1] (x3) -- (y);
\end{tikzpicture} 
 \caption[DAG of the linear SEM]
 {DAG of the linear SEM. $X_3$ is assumed to be hidden which is depicted by the dashed circle. We use the following specifications: $\Psi_1 \overset{\mathbb{D}}{=}\Psi_3\overset{\mathbb{D}}{=}\Psi_5 \sim t_7 /\sqrt{7/5}$, $\Psi_2 \overset{\mathbb{D}}{=}\Psi_6\overset{\mathbb{D}}{=}\Psi_7  \sim \mathcal{N}\left(0, 1/2\right)$, $\Psi_4 \sim \text{Unif}\left[-\sqrt{3/2}, \sqrt{3/2}\right]$ and $\mathcal{E} \sim \mathcal{N}\left(0,1\right)$. $\theta_{2,1}=\theta_{7,6} = \sqrt{1/2}$, $\theta_{4,2} = \theta_{4,3} = \theta_{6,4} = \theta_{6,5} =0.5$ and $\beta^*_3 = \sqrt{5/2}$.}
 \label{fig:SEM-DAG}
\end{figure}

\paragraph{Simulation example}
We assess the performance of our HOLS method in a linear SEM using a simple example. In Figure \ref{fig:SEM-DAG}, we show the DAG that represents the setup.

For simplicity, the parameters are set such that $X_1$ to $X_7$ all have unit variance. $X_3$ is the only parent of $Y$ and we apply the HOLS method using all but $X_3$ as predictors, i.e., $X_3$ is treated as hidden variable. Following Theorem \ref{theo:lin-SEM-beta-eq}, we know that for variables $X_1$ and $X_5$ to $X_7$ the causal effect on $Y$ is consistently estimated with OLS, while we chose the detailed setup such that there is a detectable confounding bias on $\beta_2^{OLS}$ and $\beta_4^{OLS}$. Thus, ideally, our local tests reject the null hypothesis for those two covariates but not for the rest.

\begin{figure}[b!]
 \centering
 \includegraphics[width=0.8\textwidth]{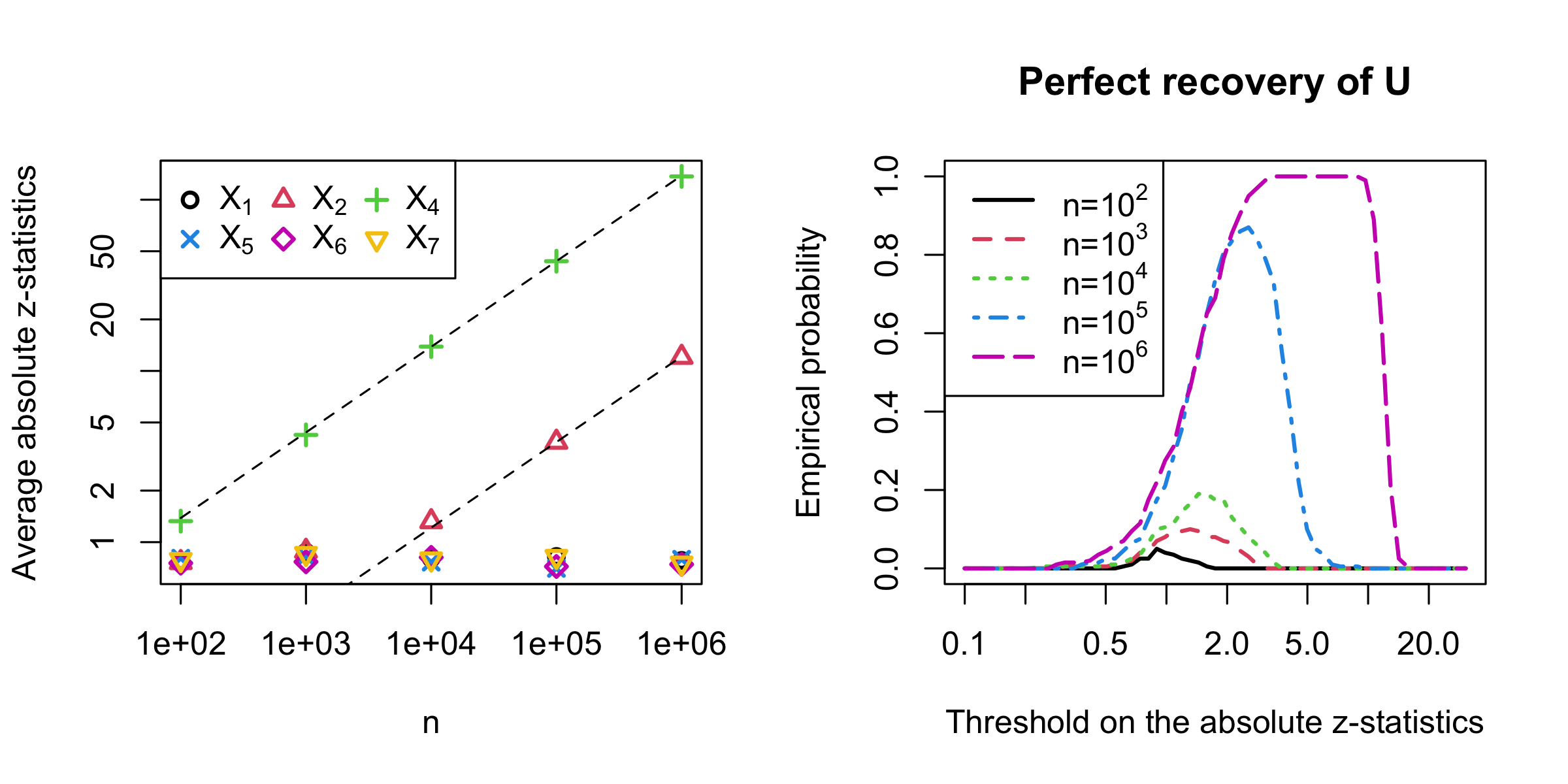}
 \caption[Simulation in a linear SEM]
 {Simulation in a linear SEM corresponding to Figure \ref{fig:SEM-DAG}. The results are based on $200$ simulation runs. On the left: Average absolute $z$-statistics per covariate for different sample sizes. The dotted lines grow as $\sqrt{n}$ and are fit to match perfectly at $n=10^5$. On the right: Empirical probability of perfectly recovering $U$ (cf.\ \eqref{eq:UV-def}) for different sample sizes.}
 \label{fig:SEM-missing-x3}
\end{figure}

For numerical results, we let the sample size grow from $10^2$ to $10^6$. For each sample size, we do $200$ simulation runs. On the left-hand side of Figure \ref{fig:SEM-missing-x3}, we show the average absolute $z$-statistics per predictor for the different sample sizes. For $X_2$ and $X_4$ we see the expected $\sqrt{n}$-growth. For the other variables, the empirical averages are close to the theoretical mean, which equals $\sqrt{2/\pi} \approx 0.8$, with a minimum of $0.70$ and a maximum of $0.88$. Further, we see that the confounding bias on the OLS parameter for $X_4$, which is a child of the hidden variable, is easier to detect than the bias onto the parameter for $X_2$, which is a child's other parent. The multiplicity corrected p-value for $X_4$ is rejected at level $\alpha = 0.05$ in $91.5\%$ of the cases for $n=10^3$, while as the null hypothesis for $X_2$ is only rejected with a empirical probability of $3\%$. For $X_2$, it takes $n=10^5$ samples to reject the local null hypothesis in $89\%$ of the simulation runs.

Following Section \ref{recov-U}, we should be able to perfectly recover the set $U$ (cf.\ \eqref{eq:UV-def}) as $n \rightarrow \infty$ if we let the threshold on the absolute $z$-statistics grow at the right rate. Therefore, we plot on the right-hand side of Figure \ref{fig:SEM-missing-x3} the empirical probability of perfectly recovering $U$ over a range of possible thresholds for the different sample sizes. For $n=10^6$, we could achieve an empirical probability of $1$. For $n=10^5$ the optimum probability is $87\%$, while as for $n=10^4$ it is only $19\%$.

\begin{figure}[b!]
 \centering
 \includegraphics[width=0.8\textwidth]{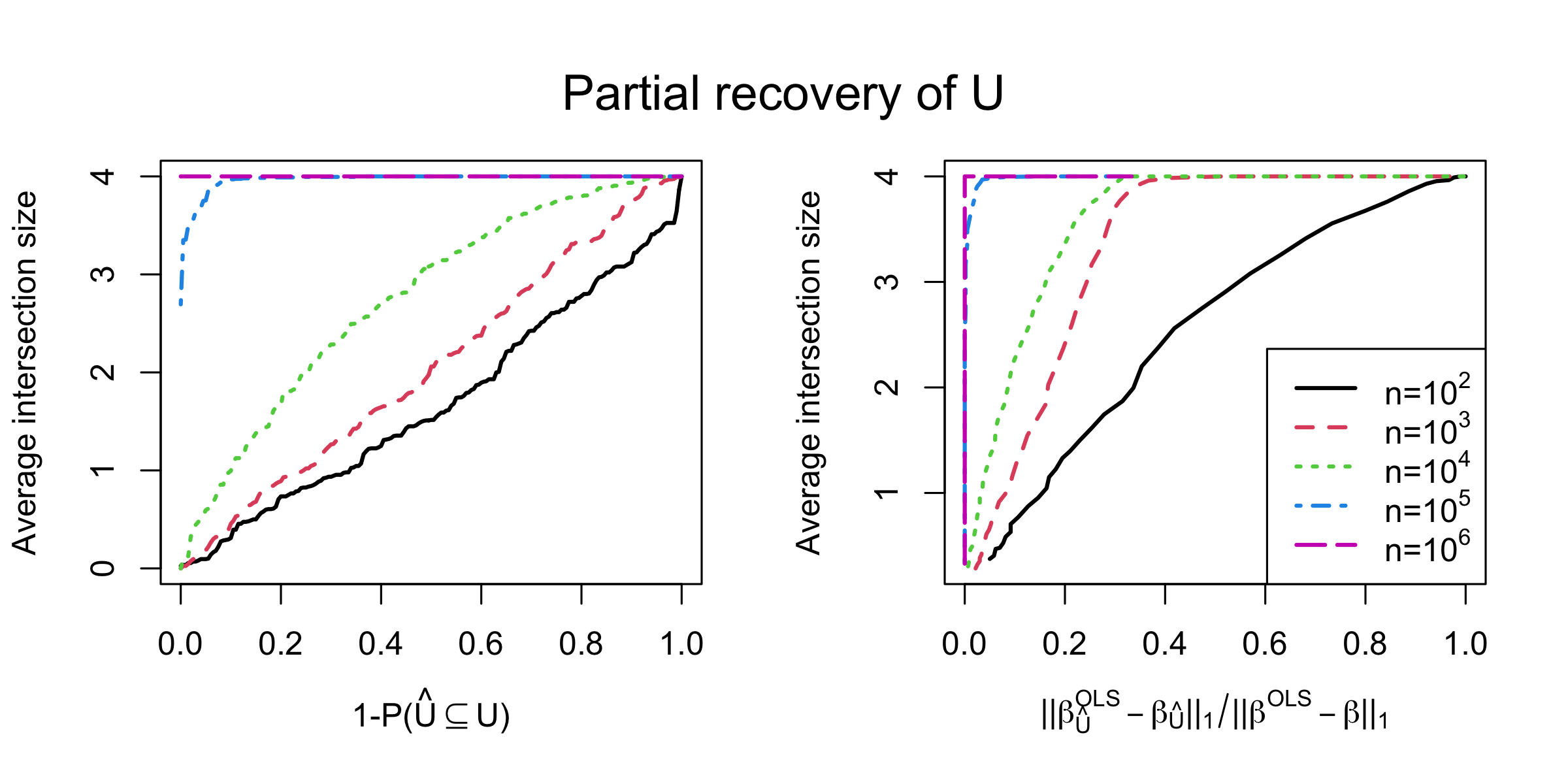}
 \caption[Simulation in a linear SEM]
 {Simulation in a linear SEM corresponding to Figure \ref{fig:SEM-DAG}. The results are based on $200$ simulation runs. On the left: Probability of not falsely including a variable in $\hat{U}$ versus average intersection size $\left\vert \hat{U} \cap U \right\vert$ (cf.\ \eqref{eq:Uhat-def}). On the right: average remaining fraction of confounding signal versus average intersection size $\left\vert \hat{U} \cap U \right\vert$. It holds that $\left\vert U \right\vert = 4$. Both curves use the threshold on the absolute $z$-statistics as implicit curve parameter. Note that the legend applies to either plot.}
 \label{fig:SEM-missing-x3-avg-size}
\end{figure}

Naturally, perfectly recovering $U$ is a very ambitious goal for smaller sample sizes, and one might want to consider different objectives. In Figure \ref{fig:SEM-missing-x3-avg-size}, we plot two different performance metrics. On the left-hand side, we plot the empirical probability of not falsely including any variable in $\hat{U}$ against the average intersection size $\left\vert\hat{U} \cap U\right\vert$. The curve is parametrized implicitly by the threshold on the absolute $z$-statistics in order to reject the local null hypothesis for some variable. Thus, the graphic considers the question of how many variables in $U$ can be recovered while keeping the probability of not falsely including any low. For a sample size of $10^5$, we have an average intersection size of $3.97$ allowing for a $10\%$ probability of false inclusions. For $10^4$, it is still $0.995$. Thus, we can find (almost) one of the $4$ variables in $U$ on average. As we see in Figure \ref{fig:SEM-missing-x3}, the bias on $\beta_4^{OLS}$ is much easier to detect than the bias on $\beta_2^{OLS}$. Thus, keeping the probability of including $X_2$ in $\hat{U}$ low is still an ambitious task. Therefore, we analyse on the right-hand side of Figure \ref{fig:SEM-missing-x3-avg-size} how many variables in $U$ we can find while removing a certain amount of confounding signal. We define the remaining fraction as
\begin{equation*}
\dfrac{\left\Vert \boldsymbol{\beta}_{\hat{U}}^{OLS}-\boldsymbol{\beta}_{\hat{U}}\right\Vert_1 }{\left\Vert \boldsymbol{\beta}^{OLS}-\boldsymbol{\beta}\right\Vert_1},
\end{equation*}
i.e., how much of the difference $\boldsymbol{\beta}^{OLS}-\boldsymbol{\beta}$ persists in terms of $\ell_1$ norm.

In this SEM, $\beta_4^{OLS}$ caries $2/3$ of the confounding signal, $\beta_2^{OLS}$ only $1/3$. Accepting $1/3$ of remaining confounding signal, we receive an average intersection size of $3.885$ for a sample size of $10^3$. For $10^4$, the average is $4$. Thus, if we allow for false inclusion of $X_2$ we can almost perfectly retrieve all of $U$ for sample size $10^3$ already.

\paragraph{What if $\mathbf{X}$ includes descendants of $Y$?}\label{anc-detect}
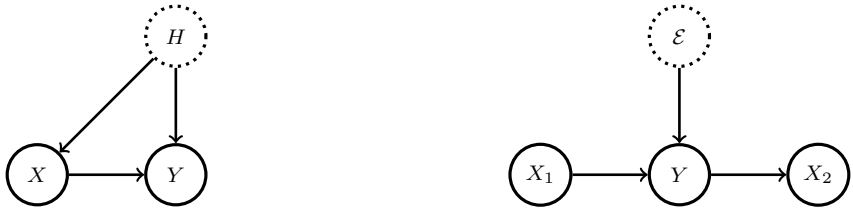
\begin{figure}[b!]
\centering

\begin{tikzpicture}[roundnode/.style={circle, very thick, minimum size=8mm}]
\node[draw, roundnode, text centered] (x) {${\scriptstyle X}$};
\node[draw, roundnode, right =of x, text centered] (y) {${\scriptstyle Y}$};
\node[draw, dotted, roundnode, above =of y, text centered] (h) {${\scriptstyle H}$};
\node[draw, roundnode, text centered, right=4cm of y] (x1) {${\scriptstyle X_1}$};
\node[draw, roundnode, right =of x1, text centered] (yn) {${\scriptstyle Y}$};
\node[draw, dotted, roundnode, above =of yn, text centered] (eps) {${\scriptstyle \mathcal{E}}$};
\node[draw, roundnode, right =of yn, text centered] (x2) {${\scriptstyle X_2}$};

\draw[->, line width= 1] (x) -- (y);
\draw[->, line width= 1] (h) -- (x);
\draw[->, line width= 1] (h) -- (y);
\draw[->, line width= 1] (x1) -- (yn);
\draw[->, line width= 1] (eps) -- (yn);
\draw[->, line width= 1] (yn) -- (x2);
\end{tikzpicture}
 \caption[DAG of the linear SEM with ANC]
 {Left: SEM with a hidden confounder. Right: SEM with a descendant of $Y$.}
 \label{fig:SEM-DE}
\end{figure}
So far, we have only considered the case where $\mathbf{X}$ causally affects $Y$, but potentially, some of $Y$'s parents are missing leading to a confounding effect. However, another possibility for $\boldsymbol{\beta}^{OLS}$ to not denote a causal effect is that there are descendants of $Y$ amongst the predictors. The two different situations are depicted in Figure \ref{fig:SEM-DE}. The case with descendants in the set of predictors fits our theory from before if interpreted properly. If the model \eqref{eq:mvmodel} for $Y$ holds true using only the parents as predictors, $Y$ can be naturally included in the assumed linear SEM for $\mathbf{X}$ in \eqref{eq:SEM-model}. Then, one can also think of $\mathcal{E}$ as an unobserved confounder. The Markov boundary of $\mathcal{E}$ with respect to the observed predictors is the same as the Markov boundary of $Y$. Of course, it holds $\boldsymbol{\beta} = \boldsymbol{\beta}^*$, i.e., $\beta_j = 0 \ \forall j \not \in \text{PA}\left(Y\right)$. Using Theorem \ref{theo:lin-SEM-beta-eq}, we find
\begin{equation*}
\beta_j^{HOLS} = \beta_j^{OLS} = \beta_j =0 \ \forall j\in M \quad \text{that are not in the Markov boundary of }Y.
\end{equation*}
Thus, for all variables outside $Y$'s Markov boundary, one can correctly detect that they have no causal effect onto $Y$ ceteris paribus. The variables in the boundary, which includes all parents, are up to term cancellations all confounded. This can be detected under some conditions, as discussed in Section \ref{inter-U}.

\subsection{Beyond linearity}\label{non-lin}
We have mainly focused on linear models, i.e., the data is either generated by model \eqref{eq:mvmodel} or model \eqref{eq:mvconfmodel}. Naturally, this assumption might be questionable in practice. Therefore, we provide some intuition about how HOLS might be applied in a more general setup. As we only detect misspecification of the OLS coefficient without identifying the type of misspecification, one should not try to over-interpret the effect of the regressors in $\hat{U}^c = \left\{1,\ldots,p\right\} \setminus \hat{U}$ (cf. Section \ref{recov-U}). However, the linear effect of the variables in $\hat{U}$ can always be interpreted to be well-specified, meaning that $\EE\left[W_j\vert Z_j\right]=Z_j \beta_j^{OLS}$ or at least ``sufficiently'' well-specified such that no misspecification is detected in the data. Generally, we can write
\begin{align*}
Y & = \mathbf{X}^\top \boldsymbol{\beta}^{OLS} + f_{nonlinear}\left(\mathbf{X}\right) + \mathcal{E}, \ \text{where }  f_{nonlinear}\left(\mathbf{X}\right) = \EE\left[Y\vert X\right] -\mathbf{X}^\top \boldsymbol{\beta}^{OLS}, \ \EE\left[\mathcal{E} \vert \mathbf{X}\right] = 0.\\
W_j & = Z_j \beta_j^{OLS} + f_{nonlinear}\left(\mathbf{X}\right) + \mathcal{E}.
\end{align*}
Thus, well-specification of $\beta_j^{OLS}$ implies $\EE\left[f_{nonlinear}\left(\mathbf{X}\right) \vert Z_j\right] =0$, i.e., after linearly adjusting for $\mathbf{X}_{-j}$, $X_j$ does not have any predictive power for $f_{nonlinear}\left(X\right)$. If it does not have any predictive power after linear adjustment, it would be a natural conclusion that it does not have predictive power after optimally adjusting for $\mathbf{X}_{-j}$ implying that $f_{nonlinear}\left(\mathbf{X}\right)$ can be written as function of $\mathbf{X}_{-j}$ only. This would then imply 
\begin{equation*}
\EE\left[Y \vert X_j = x_j + 1, \mathbf{X}_{-j} = \mathbf{x}_{-j}\right] -\EE\left[Y \vert X_j = x_j, \mathbf{X}_{-j} = \mathbf{x}_{-j}\right] = \beta_j^{OLS} \quad \forall x_j, \mathbf{x}_{-j}.
\end{equation*}
Except for Gaussian data, such a linear relationship must be either causal or due to very pathological data setups. Excluding such unusual cancellations, the conclusion is that for $j \in U$ there must be a true linear causal effect from $X_j$ to $Y$ keeping the other predictors fixed, which can be consistently estimated using OLS. Of course, if there are no locally linear structures, it might well be that $U = \emptyset$ such that the local tests are not more informative than the global test. However, there is also nothing to be lost by exploiting this local view.

Note that the asymptotic results presented in Sections \ref{conv-par} and \ref{recov-U} hold for nonlinear data as well since they only assume model \eqref{eq:refmodel} - \eqref{eq:refmodel-1D}, which is the most general formulation.
\section{Real data example}\label{sachs}
We analyse the flow cytometry dataset presented by \cite{sachs2005causal}. It contains cytometry measurements of 11 phosphorylated proteins and
phospholipids. There is a ``ground truth'' on how these quantities affect each other, the so-called consensus network \citep{sachs2005causal}. Data is available from various experimental conditions, some of which are interventional environments. The dataset has been further analysed in various projects, see, e.g., \cite{mooij2013cyclic}, \cite{meinshausen2016methods} and \cite{taeb2021perturbations}. Following these works, we consider data from 8 different environments, 7 of which are interventional. The sample size per environment ranges from $707$ to $913$.

In our analysis, we focus on the consensus network from \cite{sachs2005causal}. For each node, we go through all environments, fit a linear model using all its claimed parents as predictors, and assess the goodness of fit of the model using our HOLS check. In the consensus network, there is one bidirected edge between the variables PIP2 and PIP3. We include it as a parent for either direction. For each suggested edge, we also collect the p-values from the linear model fit in all environments, keeping only those where the edge passes the local HOLS check at level $\alpha=5\%$ without multiplicity correction. We omit the multiplicity correction here to lower the tendency to falsely claim causal detection. In Table \ref{tab:cons}, we report the minimum p-value from OLS, over the environments where the HOLS check is passed, sorted by increasing p-values.
Additionally, we show the number of environments in which the check is passed and out of these the number where the edge is significant at level $\alpha=5\%$ in the respective linear model fit (with Bonferroni correction over all $8$ environments and $17$ edges, i.e., we require a p-value of at most $0.05/136$). Note that there is one p-value $0$ reported corresponding to a t-value of $174$, which exceeds the precision that can be obtained with the standard \textsf{R}-function \texttt{lm}.

\begin{table}[t!]
\centering
\begin{tabular}{|P{0.17\textwidth}R{0.10\textwidth}R{0.17\textwidth}R{0.10\textwidth}|}
\hline
Edge & Passing HOLS & Significant in linear model & minimum p-value \\
\hline
RAF $\rightarrow$ MEK & 3  &  2  &  0\\
PKA $\rightarrow$ Akt & 3  &  3  &  1.5e-120\\
PKA $\rightarrow$ Erk & 5  &  5  &  3.8e-69\\
PKC $\rightarrow$ JNK & 3  &  3  &  5.9e-55\\
PIP2 $\rightarrow$ PIP3 & 1  &  1  &  6.5e-40\\
PIP3 $\rightarrow$ PLCg & 5  &  1  &  1.4e-36\\
PKC $\rightarrow$ p38 & 1  &  1  &  7.1e-34\\
PIP3 $\rightarrow$ PIP2 & 1  &  1  &  9.6e-08\\
PLCg $\rightarrow$ PKC & 6  &  0  &  0.016\\
PLCg $\rightarrow$ PIP2 & 1  &  0  &  0.027\\
PKC $\rightarrow$ RAF & 8  &  0  &  0.046\\
PKC $\rightarrow$ PIP2 & 8  &  0  &  0.057\\
PKA $\rightarrow$ RAF & 8  &  0  &  0.086\\
PKA $\rightarrow$ p38 & 8  &  0  &  0.12\\
PIP3 $\rightarrow$ Akt & 8  &  0  &  0.2\\
PKA $\rightarrow$ JNK & 8  &  0  &  0.21\\
MEK $\rightarrow$ Erk & 8  &  0  &  0.42\\
\hline
\end{tabular}
\caption{
The working model is taken from the consensus network. The second column reports the number of environments in which the edge passes the HOLS check (among 8 possible ones). The third column additionally shows, in how many of these it is also significant in the respective linear model fit. The p-value is the minimum of the p-values from linear regression in environments, where the edge passes the HOLS check.
}
\label{tab:cons}
\end{table}
We see that the edge RAF $\rightarrow$ MEK is the most significant.
Further, every edge of the consensus network passes the HOLS check in at least one environment. Frequently, we see that edges pass the HOLS check in certain environments without being significant in the linear model. Considering our discussion around linear SEMs, this could easily happen if the alleged predictor node is not actually in the Markov boundary of the response. In fact, there are seven edges that pass the HOLS check in every environment which are not significant based on the linear model fits. This is in agreement with \cite{taeb2021perturbations}, where none of them is reported.

As we cannot guarantee that the data follows a linear SEM as in \eqref{eq:SEM-model}, we shall not interpret the edges that do not pass the HOLS check to be subject to hidden confounding. However, the fact that we still find a decent number of suggested edges that pass the HOLS check, at least in some environments, leads to evidence that the assumption of some local unconfounded linear structures is not unrealistic, see also the discussion in Section \ref{non-lin}.

We can also analyse our results in the light of invariant causal prediction, see, e.g., \cite{peters2016causal}, where one typically assumes that interventions do not change the underlying graph except for edges that point towards the node that is intervened on. This assumption is highly questionable in practice, and our findings, which vary a lot over different environments, indicate that the assumption is likely not fulfilled in the given setup.

\section{Discussion}\label{diss}
We have introduced the so-called HOLS check to assess the goodness of fit of linear causal models. It is based on the dependence between residuals and predictors in misspecified models, leading to non-vanishing higher moments. Besides checking whether the overall model might hold true, the method allows to detect a set of variable for which linear regression consistently estimates a true (unconfounded) causal effect for certain model classes.

We extend the HOLS method to high-dimensional datasets based on the idea of the debiased Lasso \citep{zhang2014confidence,van2014asymptotically}. This extension comes very naturally as our HOLS check involves nodewise regression just as the debiased Lasso.

Of particular interest are linear structural equation models, for which our method allows for very precise characterizations regarding which least squares parameters are causal effects.
The result requires some non-Gaussianity. We complement our theory with a simulation study as well as a real data example.

A drawback of our method is that it does not distinguish whether a model is misspecified due to confounding or due to nonlinearities in the model. Therefore, an interesting follow-up direction would be to extend our methodology and theory from linear to nonlinear SEM using more flexible regression methods. This could allow to detect local causal structures in nonlinear settings as well.

Further simulation results as well as proofs and extended theory can be found in the supplemental material. Code scripts to reproduce the results presented in this paper are available
here \textbf{\href{https://github.com/cschultheiss/HOLS}{https://github.com/cschultheiss/HOLS}}.

\section*{Acknowledgment}
C. Schultheiss and P. B\"uhlmann have received funding from the European Research Council (ERC) under the European Union’s Horizon 2020 research and innovation programme (grant agreement No. 786461). M. Yuan was supported in part by NSF Grants DMS-2015285 and DMS-2052955. Part of the work was done while M. Yuan was visiting the Institute for Theoretical Studies at ETH Z\"urich, Switzerland, and he wishes to thank the institute for their support and hospitality.

\bibliographystyle{apalike} 
\bibliography{references}

\clearpage

\addtocontents{toc}{\vspace{.5\baselineskip}}
\appendix
\allowdisplaybreaks
\section{Simulation results}\label{app:sim}
\subsection{Global null}\label{sim:non-gauss}
We create data that follows the model \eqref{eq:mvmodel}. We chose the sample size and dimensionality to be $n=100$ and $p=30$ and sample $\mathbf{X}$ as follows. Let $\Psi_1,\ldots ,\Psi_p$ be i.i.d.\ random variables. Each of these follows a mixture distribution such that every copy comes from a $\mathcal{N}\left(0, 0.5\right)$ distribution with probability $2/3$ or from a $\mathcal{N}\left(0, 2\right)$ distribution with probability $1/3$. Thus, they are $0$ mean and unit variance random variables. Then, set $X_1 = \Psi_1$ and
\begin{equation*}
X_j = r X_{j-1} + \sqrt{1-r^2} \Psi_j \quad (\forall j>1)
\end{equation*}
This leads to a Toeplitz covariance structure $\boldsymbol{\Sigma}^{\mathbf{x}}$ with
${\Sigma}^{\mathbf{x}}_{ij} = r^{|i-j|}$, where we set $r=0.6$. The coefficient vector $\boldsymbol{\beta}$ is
5-sparse, and the active predictors are $\left\{1,5,10,15,20\right\}$, each
of which having a coefficient equal to $1$.
The random error $\mathcal{E}$ follows the same non-Gaussian distribution as the $\Psi_j$.

\begin{figure}[b!]
 \centering
 \includegraphics[width=0.8\textwidth]{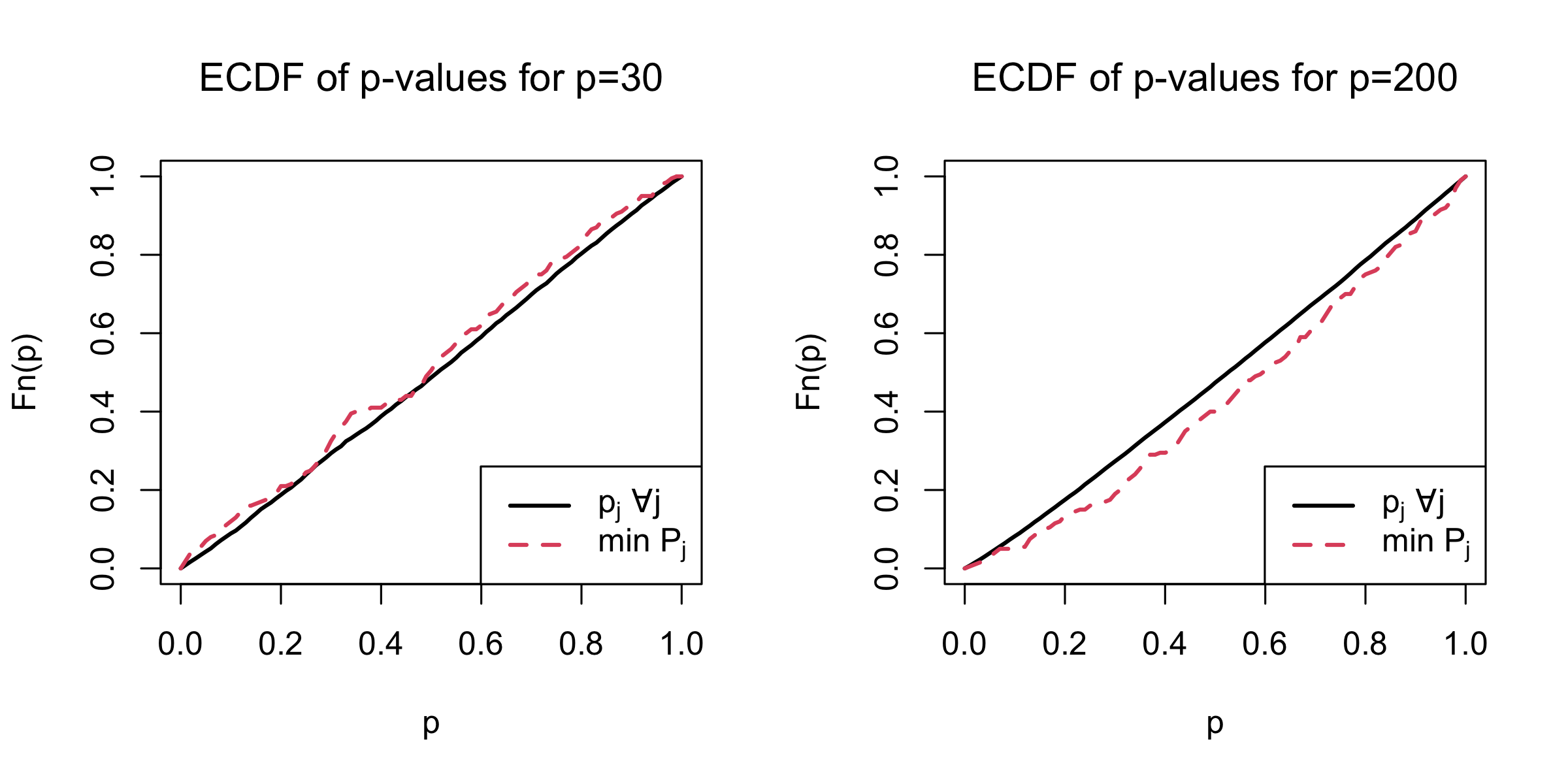}
 \caption[Results for non-Gaussian design under the global null]
 {Simulation under the global null. The results are based on $200$ simulation runs. On the left: ECDF of p-values for low-dimensional data $p_j$ (unadjusted as in Step 9 of Algorithm~\ref{alg:HOLS}) over all $p=30$ predictors and of $\underset{j}{\text{min}}P_j$ (multiplicity corrected as in Step 10 of Algorithm~\ref{alg:HOLS}). On the right: ECDF of p-values for high-dimensional data $p_j$ (unadjusted as in Step 10 of Algorithm~\ref{alg:HOLS-HD-alt}) over all $p=200$ predictors and of $\underset{j}{\text{min}}P_j$ (multiplicity corrected as in Step 11 of Algorithm~\ref{alg:HOLS-HD-alt}. 
 }
 \label{fig:non-gauss-null}
\end{figure}

We run $200$ simulations of this setup. For every simulation run, we calculate the p-value per predictor (without multiplicity adjustment) as well as the minimum of the multiplicity corrected p-values. Asymptotically,
these p-values would be uniformly distributed as the model assumptions hold true. On the left-hand side of Figure \ref{fig:non-gauss-null}, we analyse these p-values by looking at their empirical cumulative density function (ECDF). For $p_j$, the curve is combined over all the $p = 30$ covariates. Thus, it is based on $6000$ p-values. 

We see that even though the error does not have a Gaussian distribution, the p-values still are very close to being uniformly distributed. Furthermore, we see that the curve for the raw p-values is closer to the uniform distribution than the one for the minimum of the multiplicity corrected p-values. This is not surprising since the ECDF is based on more observations and as the CLT for multiple dimensions might take longer to kick in.

\paragraph{High-dimensional data } We extend the simulation to a high-dimensional case. We reuse the setup with the only exception that $p=200 > n$. Thus, we add an extra $170$ predictors that do not actually have any influence on $Y$.

To calculate $\hat{\mathbf{z}}_j$, $\hat{\mathbf{w}}_j$, and $\hat{\sigma}$ we use the default implementation of the debiased Lasso, available in the \textsf{R}-package \texttt{hdi} \citep{dezeure2015high}. To get the estimate $\hat{\tilde{\mathbf{z}}}_j^3$, we run a second level of nodewise regression with cross-validated $\tilde{\lambda}_j$. It shall be noted that all the estimated $\hat{\tilde{\boldsymbol{\gamma}}}_j = \mathbf{0}$. Thus, $\mathbf{x}_{-j}$ does not appear to contain strong enough information on $\mathbf{z}_j^3$, at least for the given sample size.
 
Again, we look at all the obtained raw p-values and plot the ECDF on the right-hand side of Figure \ref{fig:non-gauss-null}. We see some deviations from the uniform distribution. Especially, very low p-values become more unlikely. Thus, the procedure is a bit too conservative. This is emphasized by the ECDF for the minimum of the multiplicity corrected p-values as this minimum is affected by the distribution of very low p-values.
The issue might be related to $\sigma$ being overestimated: the empirical average of $\hat{\sigma}^2$ is $1.35$ and $\hat{\sigma} > \sigma = 1$ occurred in $88.5\%$ of the cases. However, when replacing the estimate with the true $\sigma$ the p-values become too liberal. 

In summary, after increasing the number of predictors but keeping the sample size the same, the results deviate a bit more from the optimal distribution. Though, the behaviour is still fairly close to what one would aim for, supporting the benefit of our method.

\subsection{Missing variable in a linear SEM}\label{SEM-sim}
\begin{figure}[b!]
 \centering
 \includegraphics[width=1\textwidth]{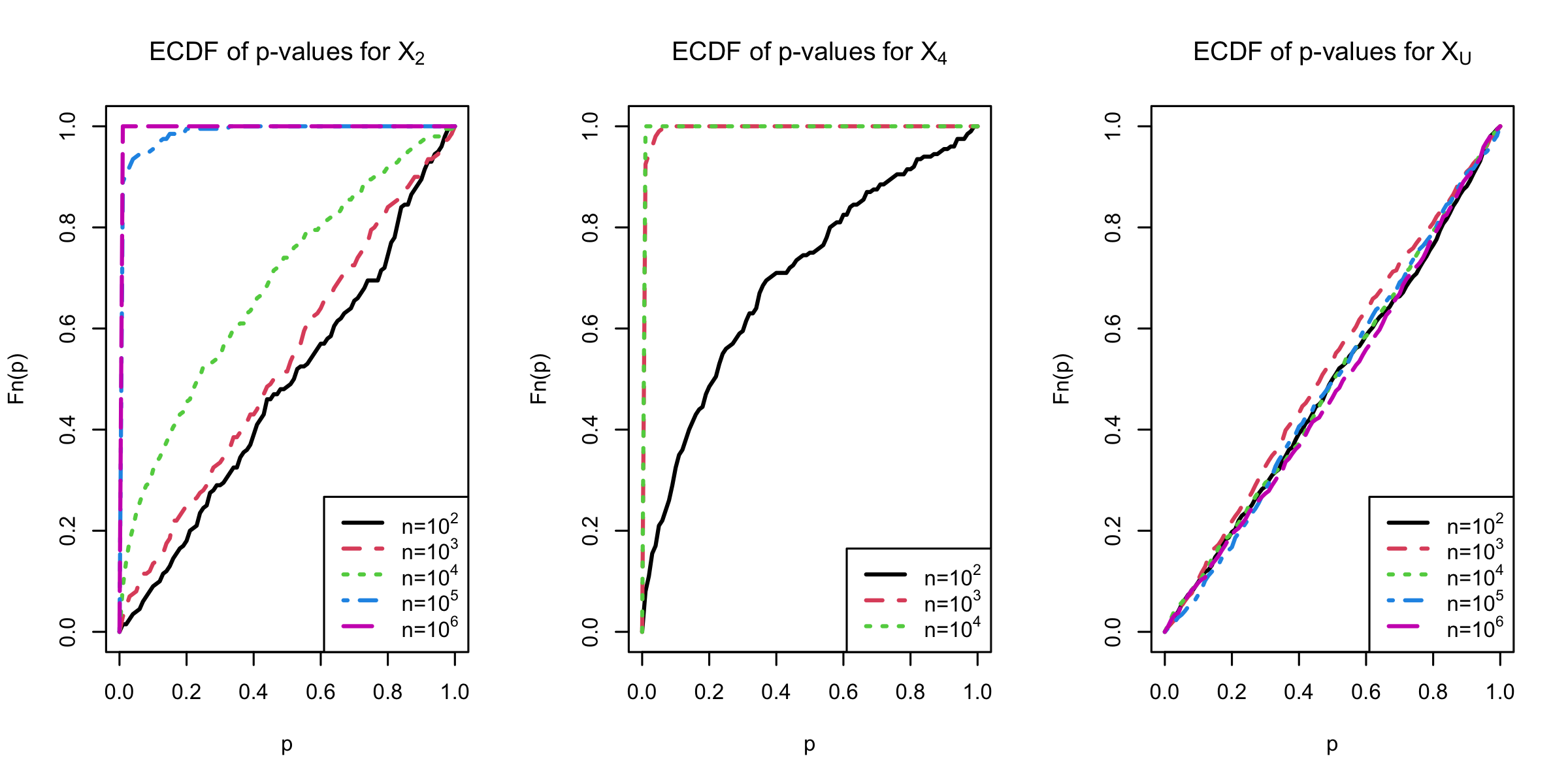}
 \caption[Simulation in a linear SEM corresponding to Figure \ref{fig:SEM-DAG}]
 {Simulation in a linear SEM corresponding to Figure \ref{fig:SEM-DAG}. The results are based on $200$ simulation runs. Depicted is the ECDF of the p-values for different predictors (unadjusted $p_j$ as in Step 9 of Algorithm~\ref{alg:HOLS}).}
 \label{fig:ecdf-x3}
\end{figure}
For the sake of comparison with the results under the global null, we provide here an additional analysis of the simulation example in Section \ref{lin-SEM}. Namely, we show in Figure \ref{fig:ecdf-x3} the empirical cumulative distribution function of the p-values obtained for different predictors. The predictors $X_2$ and $X_4$, for which the local null hypothesis should be rejected, are depicted separately, while as the rest is grouped together. Note that for $X_4$ sample sizes of more than $10^4$ are not in the plot anymore as they look just the same.

We see that the distribution of the p-values for the covariates outside the hidden variable's Markov boundary are close to the desired uniform distribution even for low sample sizes. Furthermore, in accordance with the z-statistics shown in Section \ref{lin-SEM}, the confounding bias on $\beta_4^{OLS}$ is much easier to detect than the bias on $\beta_2^{OLS}$.

\subsection{High-dimensional data: missing variable in a linear SEM}
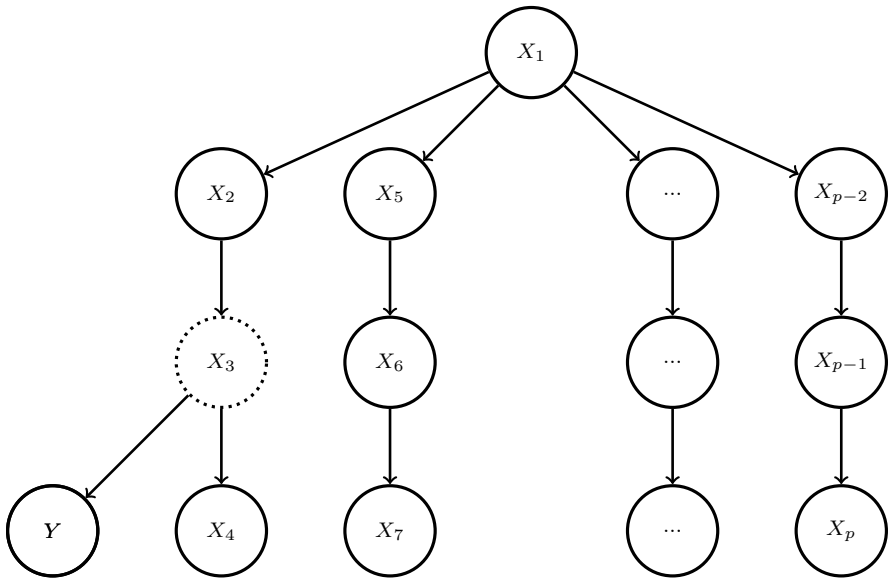
\begin{figure}[h!]
\centering
\begin{tikzpicture}[roundnode/.style={circle, very thick, minimum size=12mm}]
\node[draw, roundnode, text centered] (x1) {${\scriptstyle X_1}$};
\node[draw, roundnode, below left = of x1, text centered] (x5) {${\scriptstyle X_5}$};
\node[draw, roundnode, below = of x5, text centered] (x6) {${\scriptstyle X_6}$};
\node[draw, roundnode, below = of x6, text centered] (x7) {${\scriptstyle X_7}$};
\node[draw, roundnode, left = of x5, text centered] (x2) {${\scriptstyle X_2}$};
\node[draw, dotted, roundnode, below =of x2, text centered] (x3) {${\scriptstyle X_3}$};
\node[draw, roundnode, text centered, below=of x3] (x4) {${\scriptstyle X_4}$};
\node[draw, roundnode, text centered, left = 1cm of x4] (y) {${\scriptstyle Y}$};
\node[draw, roundnode, below right = of x1, text centered] (xd1) {${\scriptstyle \ldots}$};
\node[draw, roundnode, below =of xd1, text centered] (xd2) {${\scriptstyle \ldots}$};
\node[draw, roundnode, text centered, below=of xd2] (xd3) {${\scriptstyle \ldots}$};
\node[draw, roundnode, right = of xd1, text centered] (xp2) {${\scriptstyle X_{p-2}}$};
\node[draw, roundnode, below =of xp2, text centered] (xp1) {${\scriptstyle X_{p-1}}$};
\node[draw, roundnode, text centered, below=of xp1] (xp) {${\scriptstyle X_p}$};
\node[draw, roundnode, text centered, left = 1cm of x4] (y) {${\scriptstyle Y}$};

\draw[->, line width= 1] (x1) -- (x2);
\draw[->, line width= 1] (x2) -- (x3);
\draw[->, line width= 1] (x3) -- (x4);
\draw[->, line width= 1] (x3) -- (y);
\draw[->, line width= 1] (x1) -- (x5);
\draw[->, line width= 1] (x5) -- (x6);
\draw[->, line width= 1] (x6) -- (x7);
\draw[->, line width= 1] (x1) -- (xp2);
\draw[->, line width= 1] (xp2) -- (xp1);
\draw[->, line width= 1] (xp1) -- (xp);
\draw[->, line width= 1] (x1) -- (xd1);
\draw[->, line width= 1] (xd1) -- (xd2);
\draw[->, line width= 1] (xd2) -- (xd3);
\end{tikzpicture}
 \caption[DAG of the linear SEM with ANC]
 {Linear SEM used for the high-dimensional simulation.}
 \label{fig:SEM-HD}
\end{figure}
We want to assess how well our method for high-dimensional data can detect deviations from the null hypothesis. We create data from the linear SEM depicted in Figure \ref{fig:SEM-HD}, where all predictors but $X_3$ are observed. According to our theoretical results, $\beta_j^{OLS}=\beta_j^{HOLS}=\beta_j \ \forall j \not \in \left\{2,4\right\}$. We set the number of variables to $p = 1.5 n + 1$. Thus, there are $0.5 n$ of these blocks. We consider $n=10^2$ and $n=10^3$. For a comparison with low-dimensional HOLS, we also assess the performance using just $p=13$ predictors, i.e., four blocks. We let all $\Psi_j$ follow a centered uniform distribution and set $Y=X_3$. To execute the high-dimensional HOLS test, we proceed as in Section \ref{sim:non-gauss}.

\begin{figure}[t!]
 \centering
 \includegraphics[width=0.8\textwidth]{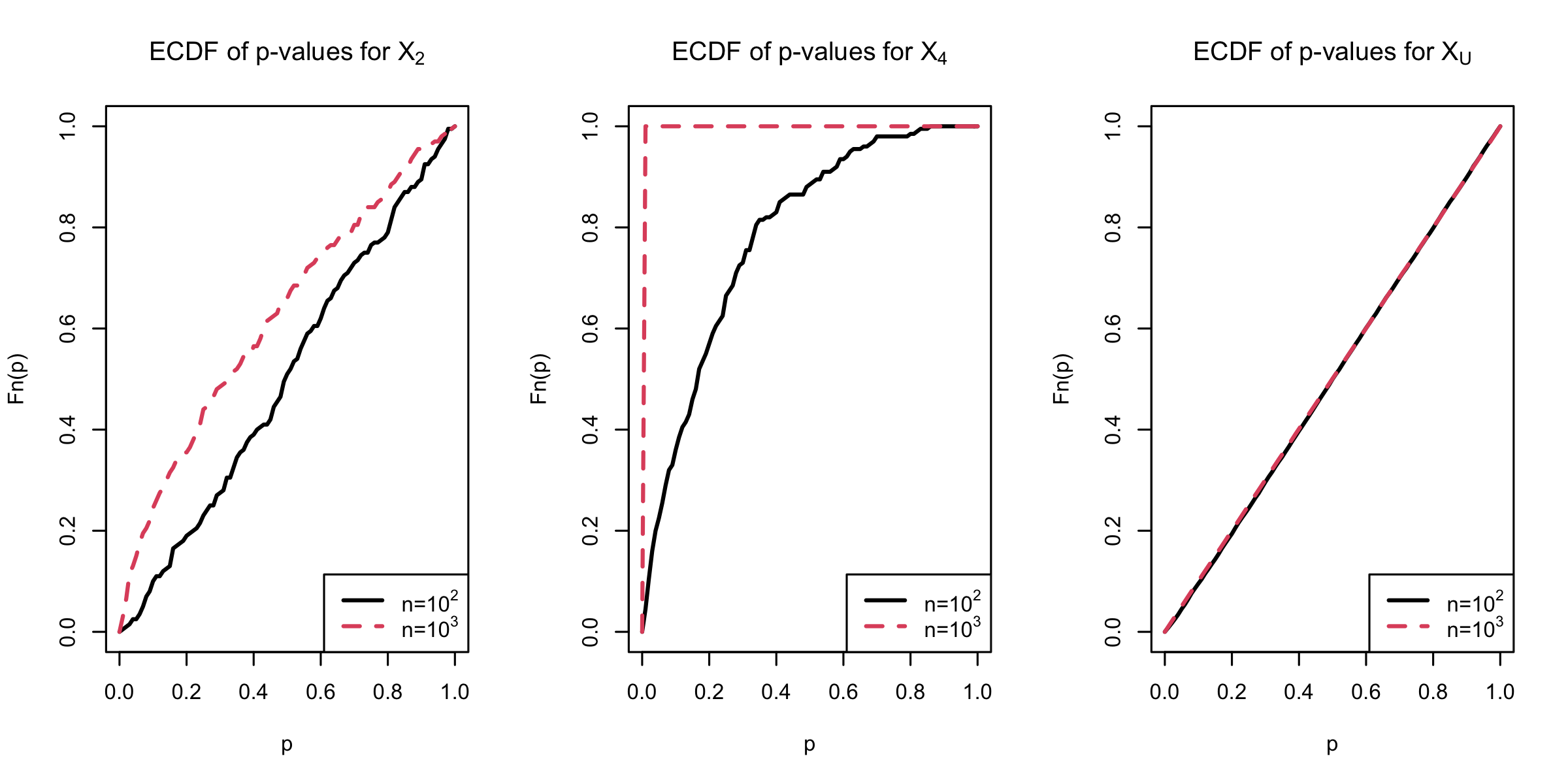}
 \caption[Simulation for a missing variable in a linear SEM corresponding to Figure \ref{fig:SEM-HD} for high-dimensional data]
 {Simulation for a missing variable in a linear SEM corresponding to Figure \ref{fig:SEM-HD} for high-dimensional data. The results are based on $200$ simulation runs. Depicted is the ECDF of the p-values for different predictors (unadjusted $p_j$ as in Step 10 of Algorithm~\ref{alg:HOLS-HD-alt}). }
 \label{fig:ecdf-hd}
\end{figure}

In Figure \ref{fig:ecdf-hd}, we show the empirical cumulative distribution function of the p-values obtained over $200$ simulation runs. We look at the p-values for $X_2$ and $X_4$ separately and for all other variables combined. The latter should roughly follow a uniform distribution. Similar to our results in Sections \ref{lin-SEM} and \ref{SEM-sim}, we see that it is much easier to reject the null hypothesis for the hidden variable's child $X_4$ than for the child's other parent (with respect to the observed covariates) $X_2$. With a sample size of $10^3$, no p-value $p_4$ larger than $1.4 * 10^{-4}$ was obtained.  Finally, we see that for the other covariates the obtained p-values are indeed close to being uniformly distributed.

\begin{figure}[b!]
 \centering
 \includegraphics[width=0.8\textwidth]{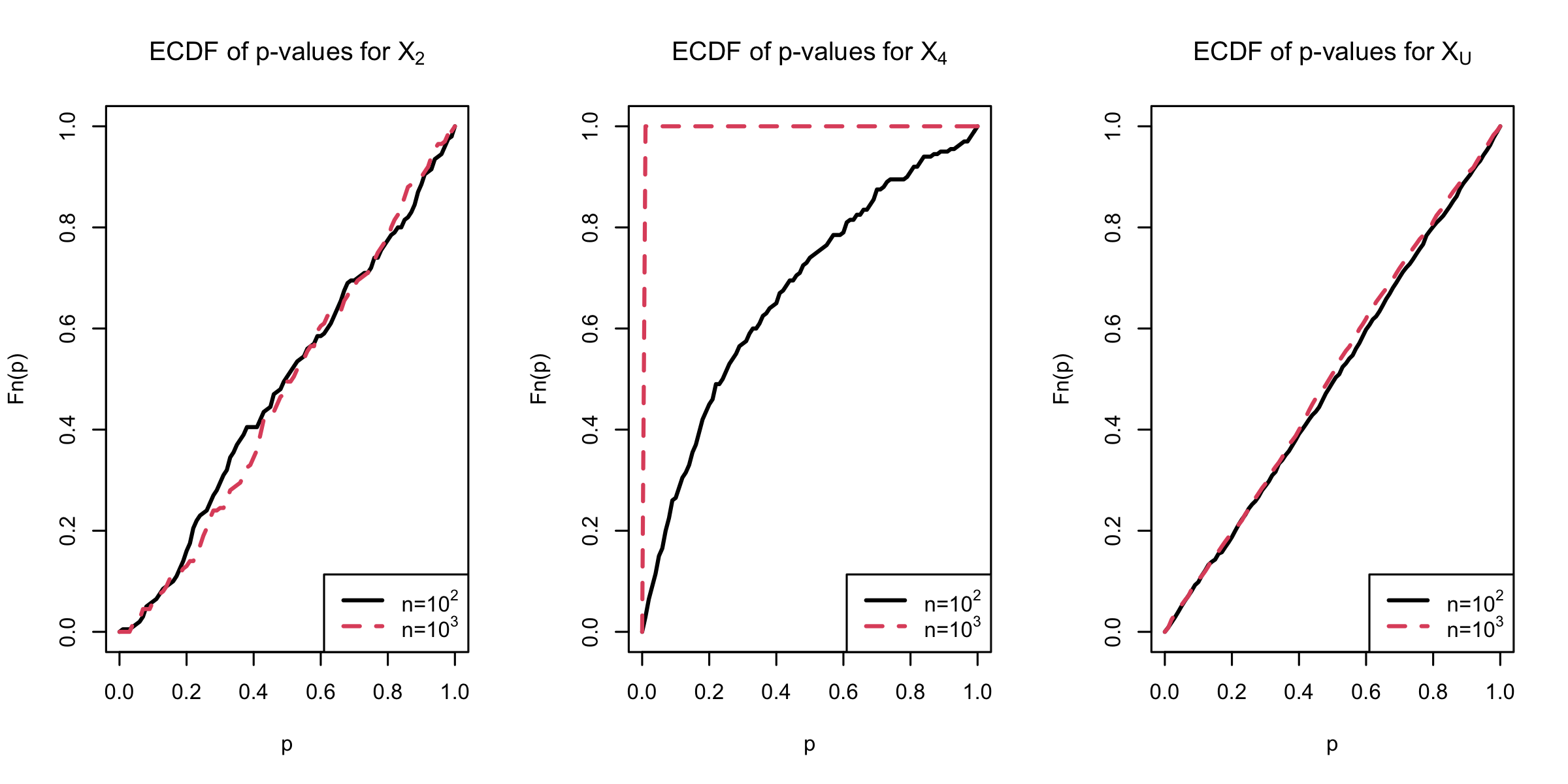}
 \caption[Results for a missing variable in a linear SEM for low-dimensional data]
 {Simulation for a missing variable in a linear SEM corresponding to Figure \ref{fig:SEM-HD} for low-dimensional data. The results are based on $200$ simulation runs. Depicted is the ECDF of the p-values for different predictors (unadjusted $p_j$ as in Step 9 of Algorithm~\ref{alg:HOLS}). }
 \label{fig:ecdf-ld}
\end{figure}

In Figure \ref{fig:ecdf-ld}, we consider the same statistics for the low-dimensional HOLS test applied to the same data but with just the first $12$ observed covariates. We note that for this data the distribution of the p-values for $X_2$ and $X_4$ is more distinct from the uniform distribution in the high-dimensional setup than in the low-dimensional setup. Other than that, the conclusion regarding the algorithm's performance remains the same.

\subsection{Confounding onto block-independent $\mathcal{E}_\mathbf{X}$}
To simulate block-independent data, we make use of the Boston housing data, available in the \textsf{R}-package \texttt{MASS} \citep{venables2002modern}. We use all variables but the variable \texttt{medv}, which is typically the response variable for regression. Then, we create two independent bootstrap samples and concatenate those such that we have two independent blocks forming the matrix $\boldsymbol{\eps}_\mathbf{x}$. There are $13$ covariates per block. Before this bootstrap sampling, we make all variables have $0$ mean and unit variance. We let $H$ be a standard normal random variable and set $X_1 = \mathcal{E}_{X_1} + H$ and $X_7 = \mathcal{E}_{X_7} - H$. For the remaining variables, we let $X_j = \mathcal{E}_{X_j}$. Finally, we set $Y=H$ for simplicity. Thus, the first block is confounded with $Y$, but the second is not. As the covariance $\boldsymbol{\Sigma}^{\mathcal{E}_{\mathbf{X}}}$ is defined by the empirical correlation of the Boston housing data, there are no vanishing entries in each of the blocks. Thus, none of the covariates $X_1$ to $X_{13}$ fulfils \eqref{eq:no-neighbour}, and there is a confounding bias on each of the OLS parameters.

\begin{figure}[b!]
 \centering
 \includegraphics[width=0.8\textwidth]{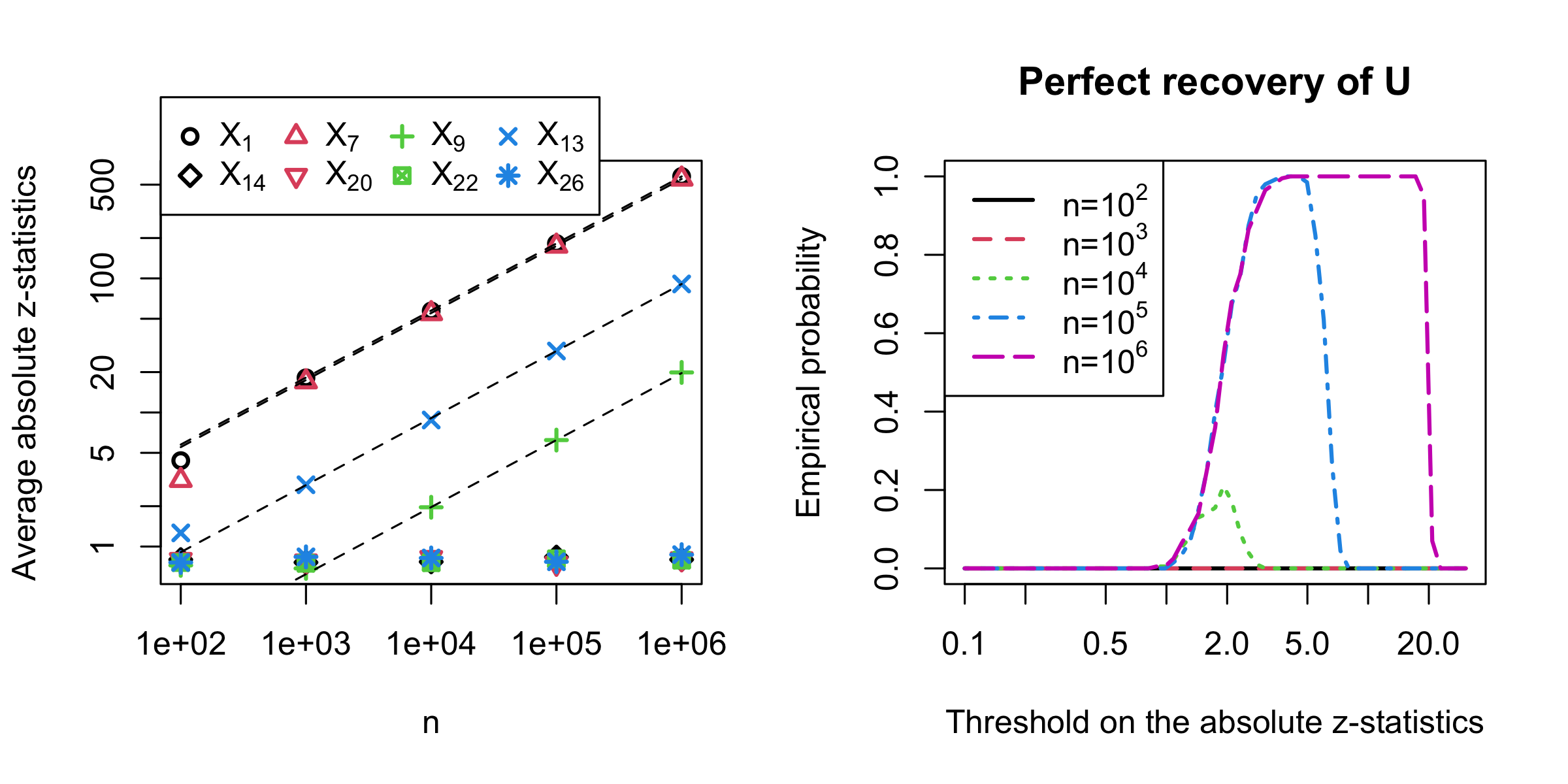}
 \caption[Confounding onto block independent $\mathcal{E}_\mathbf{X}$]
 {Simulation for confounding onto block independent $\mathcal{E}_\mathbf{X}$. The results are based on $200$ simulation runs. On the left: Average absolute $z$-statistics per covariate for different sample sizes. The dotted lines grow as $\sqrt{n}$ and are fit to match perfectly at $n=10^5$. On the right: Empirical probability (over $200$ simulation runs) of perfectly recovering $U$ (cf.\ \eqref{eq:UV-def}) for different sample sizes.}
 \label{fig:block-ind}
\end{figure}

We vary the sample size from $10^2$ to $10^6$, doing $200$ simulation runs for each sample size. Thus, it is a ``$m$ out of $n$'' bootstrap, where $m$ can be smaller than $n$ (for $m=10^2$) or larger (for the rest). In the remainder, we call the bootstrap sample size $n$ to keep the notation consistent and as the size of the real Boston housing data is not of primary interest. On the left-hand side of Figure \ref{fig:block-ind}, we plot the average absolute $z$-statistics for a representative subset of the predictors. Notably, it is the same four predictors once from the first block and once from the second. As expected, this average grows as $\sqrt{n}$ for variables in the confounded block, while it stays approximately constant for variables from the independent block. Further, we see that the two variables $X_1$ and $X_7$, which are directly confounded, are the easiest to detect as such. For only $10^2$ samples, the multiplicity corrected p-values for $X_1$ and $X_7$ lead to a rejection of the local null hypothesis in $78.5\%$ respectively $46\%$ of the simulation runs at level $\alpha = 0.05$. For some of the other variables in the confounded block, it takes many more samples to reliably reject the local null hypothesis. For $X_9$, the local null hypothesis is only rejected in $3\%$ of the cases for $n=10^4$ and only from $n=10^5$ it is always rejected.

On the right-hand side, we show the empirical probability of perfectly recovering $U$, i.e., rejecting the null hypothesis for all variables from the first block but not rejecting it for any variable from the second block. We see that for $n=10^5$ we are able to achieve this recovery with an empirical probability of $1$, and, for $n=10^6$, it is even possible for a larger range of thresholds. Comparing the two curves for $n=10^5$ and $n=10^6$, we see that they initially look very similar. This is as one would expect as the initial increase of the curve corresponds to reducing the type I error, which is independent of the sample size, assuming the CLT has kicked in sufficiently. The decrease of the curve depends on the $z$-statistics for the confounded variables, which we know to increase as $\sqrt{n}$. Thus, this decrease will appear later the larger $n$ gets.

\begin{figure}[b!]
 \centering
 \includegraphics[width=1\textwidth]{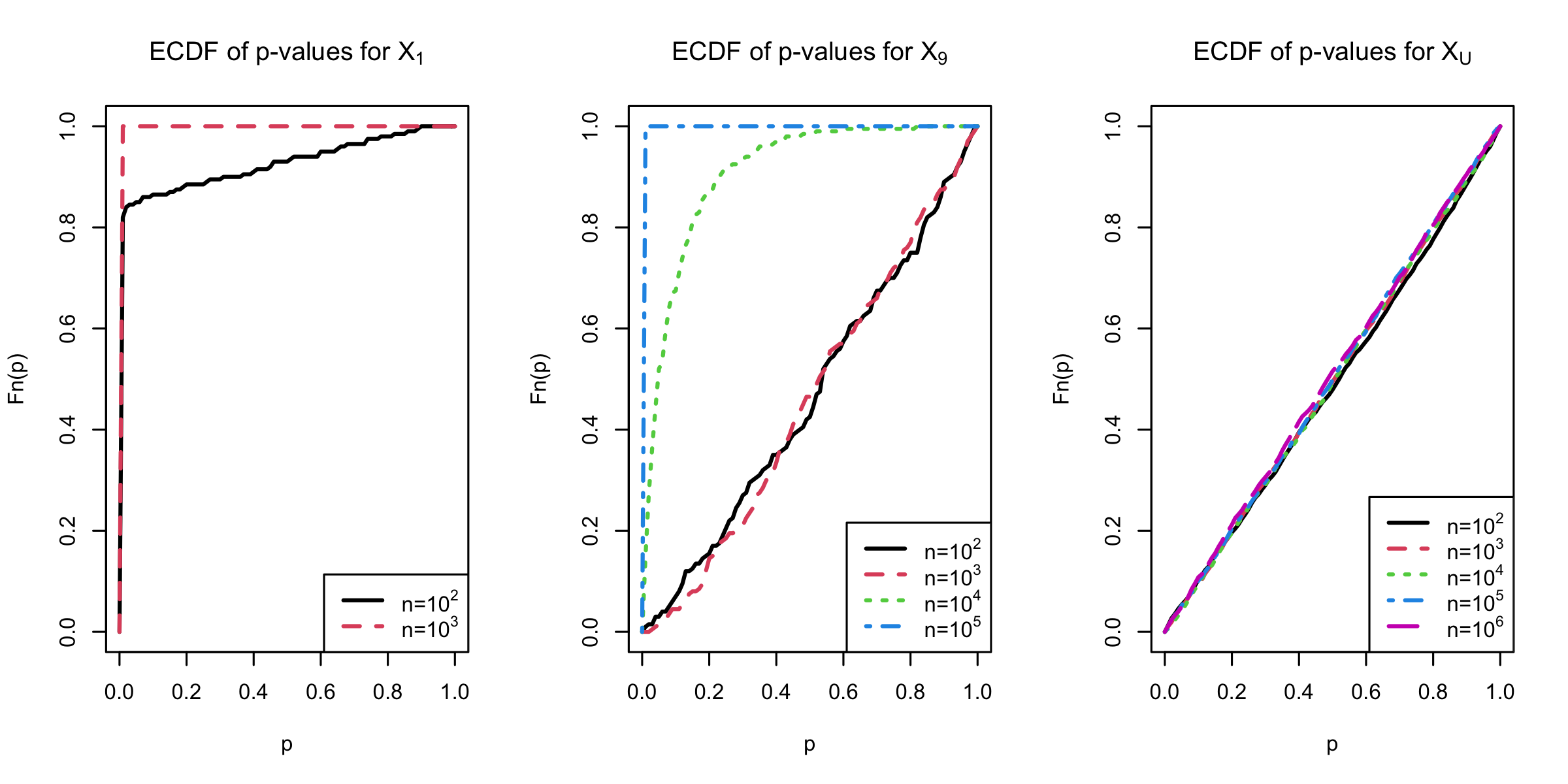}
 \caption
 {Simulation for confounding onto block independent $\mathcal{E}_\mathbf{X}$. The results are based on $200$ simulation runs. Depicted is the ECDF of the p-values for different predictors (unadjusted $p_j$ as in Step 9 of Algorithm~\ref{alg:HOLS}). }
 \label{fig:ecdf-block-ind}
\end{figure}

We show the empirical cumulative distribution function of the obtained p-values in Figure \ref{fig:ecdf-block-ind}. $X_1$ and $X_9$ are considered separately while as all p-values for the variables from the second block are grouped together. For the latter, the ECDF is close to the desired uniform distribution for every sample size. In accordance with the average $z$-statistics, the p-values for the directly confounded variable $X_1$ are more extreme than those for $X_9$.

\begin{figure}[t!]
 \centering
 \includegraphics[width=0.8\textwidth]{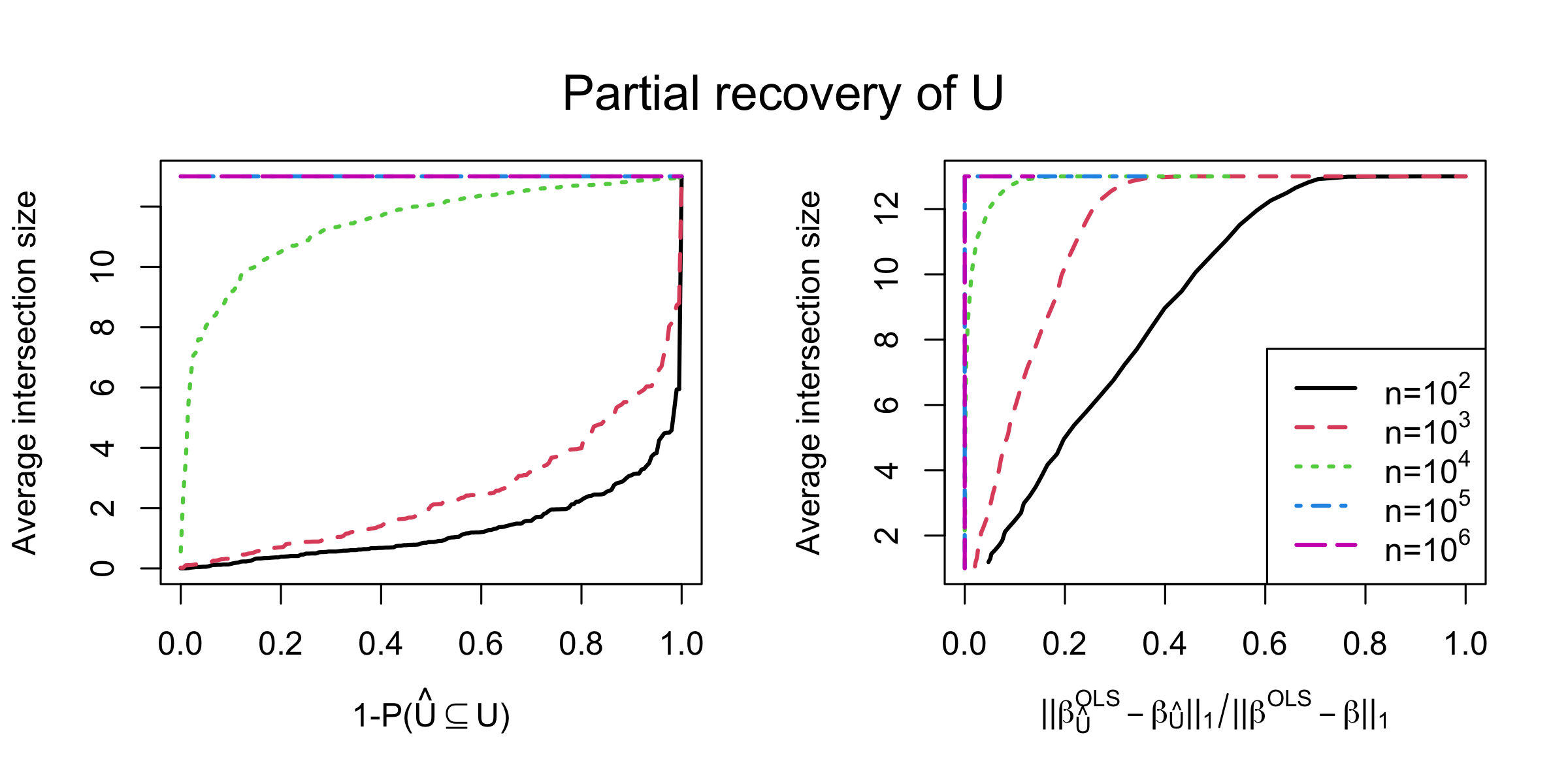}
 \caption[Confounding onto block independent $\mathcal{E}_\mathbf{X}$]
 {Simulation for confounding onto block independent $\mathcal{E}_\mathbf{X}$. The results are based on $200$ simulation runs. On the left: Probability of not falsely including a variable in $\hat{U}$ versus average intersection size $\left\vert \hat{U} \cap U \right\vert$ (cf.\ \eqref{eq:Uhat-def}). On the right: average remaining fraction of confounding signal versus average intersection size $\left\vert \hat{U} \cap U \right\vert$. It holds that $\left\vert U \right\vert = 13$. Both curves use the threshold on the absolute $z$-statistics as implicit curve parameter. Note that the legend applies to either plot.}
 \label{fig:block-ind-avg-size}
\end{figure}

In Figure \ref{fig:block-ind-avg-size}, we analyse the partial recovery of $U$ as in Section \ref{lin-SEM}. We see that for a sample size of $10^4$, for which perfect recovery is hardly achievable, we receive an average intersection size of $9.175$ (out of $13$) allowing for $10\%$ probability of false inclusion. For lower sample sizes, there is not much that can be found under this constraint. In this setup, there is a confounding bias onto $13$ OLS parameters with varying signal strength $\left\vert \beta_j^{OLS} -\beta_j\right\vert$. The two directly confounded variables amount to $45.254\%$ of the confounding signal. Thus, a remaining fraction of $54.746\%$ appears to be particularly achievable. We see that we can do even better than that. Namely, for a sample size of $10^3$, we can get an empirical probability of $1$ of including all variables of $U$ in $\hat{U}$ allowing for an average of $44.775\%$ of the confounding signal. For $10^4$, we can go down to a remaining fraction of $17.597\%$.

\section{Proofs} \label{app:proofs}
\subsection{Proof of Theorem \ref{theo:non-gauss}}
Note that model \eqref{eq:mvmodel} and \ref{ass:mom6} imply that \ref{ass:LLN-cond} - \ref{ass:z3-ind} hold for all $j$, i.e., $U' =\left\{1,\ldots,p\right\}$. Thus, we receive Theorem \ref{theo:non-gauss} for free by proving Theorem \ref{theo:V_U} and we receive Corollary \ref{corr:alpha-H0} for free by proving Corollary \ref{corr:alpha-H0j}.
\subsection{Proof of Theorem \ref{theo:conf}}\label{app:theo:conf}
Note first that Assumptions \ref{ass:invertible} and \ref{ass:mom6} imply
\begin{align}\label{eq:conv-inv}
\begin{split}
\dfrac{1}{n}\mathbf{x}_{-j}^\top \mathbf{x}_{-j}\overset{\mathbb{P}}{\to} \boldsymbol{\Sigma}^\mathbf{X}_{-j,-j} & \implies n \left(\mathbf{x}_{-j}^\top \mathbf{x}_{-j}\right)^{-1} \overset{\mathbb{P}}{\to} \left(\boldsymbol{\Sigma}^\mathbf{X}_{-j,-j}\right)^{-1} \\
 & \implies \left\Vert n \left(\mathbf{x}_{-j}^\top \mathbf{x}_{-j}\right)^{-1} \right\Vert \overset{\mathbb{P}}{\to} \left\Vert \left(\boldsymbol{\Sigma}^\mathbf{X}_{-j,-j}\right)^{-1} \right\Vert = \mathcal{O}\left(1\right),
\end{split}
\end{align}
where we use invertibility and the continuous mapping theorem.
In several occasions, we use bounds on multiplication with the projection matrix $\boldsymbol{P}_{-j}$, e.g.,
\begin{align}\label{eq:P_j-conf}
\begin{split}
\left\vert \mathbf{z}_j ^\top \boldsymbol{P}_{-j} \mathbf{w}_j \right \vert & = \left\vert \mathbf{z}_j^\top \mathbf{x}_{-j} \left( \mathbf{x}_{-j}^\top \mathbf{x}_{-j}\right)^{-1}\mathbf{x}_{-j}^\top \mathbf{w}_j \right \vert \leq \left\Vert \mathbf{z}_j^\top \mathbf{x}_{-j} \right\Vert_2 \left\Vert \left( \mathbf{x}_{-j}^\top \mathbf{x}_{-j}\right)^{-1} \right\Vert_2 \left\Vert \mathbf{x}_{-j}^\top \mathbf{w}_j \right\Vert_2 \\ 
& \leq \left\Vert \mathbf{z}_j^\top \mathbf{x}_{-j} \right\Vert_1 \left\Vert \left( \mathbf{x}_{-j}^\top \mathbf{x}_{-j}\right)^{-1} \right\Vert_2 \left\Vert \mathbf{x}_{-j}^\top \mathbf{w}_j \right\Vert_1 = \sum_{k \neq j } \left\vert \mathbf{z}_j^\top \mathbf{x}_k \right\vert \left\Vert \left( \mathbf{x}_{-j}^\top \mathbf{x}_{-j}\right)^{-1} \right\Vert_2 \sum_{k \neq j } \left\vert \mathbf{x}_k^\top \mathbf{w}_j \right\vert \\
& = \mathcal{O}_p\left(\sqrt{n}\right)\mathcal{O}_p\left(\dfrac{1}{n}\right){ \scriptstyle \mathcal{O}}_p\left(n\right).
\end{split}
\end{align}
For the last equality, we used Chebyshev's inequality, \eqref{eq:conv-inv}, and the LLN together with $\EE\left[Z_jX_k\right] =0$, $\EE\left[\left(Z_jX_k\right)^2\right] < \infty$ and $\EE\left[X_k W_j\right] =0$.

Theorem \ref{theo:conf} consists of three parts. Consider $\hat{\beta}_j^{OLS}$.
\begin{align*}
\dfrac{1}{n}\left\vert\hat{\mathbf{z}}_j^\top\hat{\mathbf{w}}_j - \mathbf{z}_j^\top \mathbf{w}_j \right\vert & = \dfrac{1}{n}\left\vert\mathbf{z}_j^\top \boldsymbol{P}^{\perp}_{-j} \boldsymbol{P}^{\perp}_{-j} \mathbf{w}_j - \mathbf{z}_j^\top \mathbf{w}_j \right\vert= \dfrac{1}{n} \left\vert \mathbf{z}_j^\top \boldsymbol{P}_{-j} \mathbf{w}_j\right\vert\\ &
 = \dfrac{1}{n}\mathcal{O}_p\left(\sqrt{n}\right)\mathcal{O}_p\left(\dfrac{1}{n}\right){\scriptstyle \mathcal{O}}_p\left(n\right) = {\scriptstyle \mathcal{O}}_p\left(\dfrac{1}{\sqrt{n}}\right). \text{ Thus,}\\
\dfrac{1}{n}\hat{\mathbf{z}}_j^\top\hat{\mathbf{w}}_j & = \dfrac{1}{n}\mathbf{z}_j^\top \mathbf{w}_j + {\scriptstyle \mathcal{O}}_p\left(\dfrac{1}{\sqrt{n}}\right) = \EE\left[Z_jW_j\right] + {\scriptstyle \mathcal{O}}_p\left(1\right) + {\scriptstyle \mathcal{O}}_p\left(\dfrac{1}{\sqrt{n}}\right) = \EE\left[Z_jW_j\right] + {\scriptstyle \mathcal{O}}_p\left(1\right). \\
 \dfrac{1}{n}\hat{\mathbf{z}}_j^\top\hat{\mathbf{z}}_j & = \EE\left[Z_j^2\right] + \mathcal{O}_p\left(\dfrac{1}{\sqrt{n}}\right)\text{ follows analogously such that} \\
 \hat{\beta}_j^{OLS} & = \dfrac{\EE\left[Z_j W_j\right]}{\EE\left[Z_j^2\right]} + { \scriptstyle \mathcal{O}}_p\left(1\right).
\end{align*}
For $\hat{\beta}_j^{HOLS}$, we first consider some intermediate results.
\begin{align*}
\left\Vert \hat{\boldsymbol{\gamma}}_j - \boldsymbol{\gamma}_j\right\Vert_2 & = \left\Vert \left(\mathbf{x}_{-j}^\top \mathbf{x}_{-j}\right)^{-1}\mathbf{x}_{-j}^\top \mathbf{z}_j\right\Vert_2 \leq \left\Vert \left(\dfrac{1}{n}\mathbf{x}_{-j}^\top \mathbf{x}_{-j}\right)^{-1}\right\Vert_2\left\Vert\dfrac{1}{n}\mathbf{x}_{-j}^\top \mathbf{z}_j\right\Vert_2 \\
& = \mathcal{O}_p\left(1\right)\mathcal{O}_p\left(\dfrac{1}{\sqrt{n}}\right)=\mathcal{O}_p\left(\dfrac{1}{\sqrt{n}}\right) \text{ such that} \\
\left\Vert \hat{\mathbf{z}}_j - \mathbf{z}_j\right\Vert_\infty & = \left\Vert \mathbf{x}_{-j}\left(\boldsymbol{\gamma}_j - \hat{\boldsymbol{\gamma}}_j\right)\right\Vert_\infty \leq \left\Vert \mathbf{x}_{-j}\right\Vert_\infty \left\Vert \hat{\boldsymbol{\gamma}}_j - \boldsymbol{\gamma}_j\right\Vert_1\leq \left\Vert \mathbf{x}_{-j}\right\Vert_\infty \sqrt{p}\left\Vert \hat{\boldsymbol{\gamma}}_j - \boldsymbol{\gamma}_j\right\Vert_2 \\
 & =\mathcal{O}_p\left(K\right)\mathcal{O}_p\left(\dfrac{1}{\sqrt{n}}\right) = \mathcal{O}_p\left(\dfrac{K}{\sqrt{n}}\right),
\end{align*}
using fixed $p$. Note that we denote the bound on $\left\Vert \mathbf{x}_{-j}\right\Vert_\infty $ by $K$. \ref{ass:mom6} induces a worst-case bound of $K=n^{1/6}$. This could be heavily improved for certain assumptions on the distribution of $\mathbf{X}$, e.g., $K=\sqrt{\log \left(n\right)}$ for Gaussian data. To keep things more general, we will use generic $K$ in the following. Further,
\begin{align*}
\left\Vert \hat{\mathbf{z}}_j - \mathbf{z}_j \right\Vert_2^2 =\left\Vert \boldsymbol{P}^{\perp}_{-j} \mathbf{z}_j -\mathbf{z}_j \right\Vert_2^2 =\left\Vert \boldsymbol{P}_{-j} \mathbf{z}_j\right\Vert_2^2 = \mathcal{O}_p\left(1\right) \quad \text{and analogously} \quad \left\Vert \hat{\mathbf{w}}_j - \mathbf{w}_j \right\Vert_2^2 ={\scriptstyle\mathcal{O}}_p\left(n\right).
\end{align*}
We invoke the following identity 
\begin{equation*}
\left(a^3 - b^3\right) =\left(a-b\right)^3 - 3a\left(a-b\right)^2 + 3a^2 \left(a-b\right)
\end{equation*}
to find
\begin{align}\label{eq:z3-conv}
\begin{split}
\left\Vert \mathbf{z}_j ^3 -\hat{\mathbf{z}}_j ^3 \right\Vert_2 & \leq \left\Vert\left(\mathbf{z}_j -\hat{\mathbf{z}}_j\right)^3\right\Vert_2 + 3 \left\Vert \mathbf{z}_j \odot \left(\mathbf{z}_j -\hat{\mathbf{z}}_j\right)^2\right\Vert_2 + 3 \left\Vert \mathbf{z}_j^2 \odot\left(\mathbf{z}_j -\hat{\mathbf{z}}_j\right)\right\Vert_2 \\
& \leq \left\Vert\left(\mathbf{z}_j -\hat{\mathbf{z}}_j\right)^2\right\Vert_\infty \left\Vert\mathbf{z}_j -\hat{\mathbf{z}}_j\right\Vert_2 + 3 \left\Vert \mathbf{z}_j\right\Vert_\infty \left\Vert \mathbf{z}_j -\hat{\mathbf{z}}_j\right\Vert_\infty \left\Vert \mathbf{z}_j -\hat{\mathbf{z}}_j\right\Vert_2 + 3 \left\Vert \mathbf{z}_j^2\right\Vert_\infty\left\Vert \left(\mathbf{z}_j -\hat{\mathbf{z}}_j\right)\right\Vert_2 \\
& = \mathcal{O}_p\left(\dfrac{K^2}{n}\right)+ \mathcal{O}_p\left(\dfrac{K^2}{\sqrt{n}}\right)+ \mathcal{O}_p\left(K^2\right)=\mathcal{O}_p\left(K^2\right).
\end{split}
\end{align}
With this at hand, we find
\begin{align*}
& \dfrac{1}{n}\left\vert \left(\mathbf{z}_j ^3\right)^\top \mathbf{w}_j - \left(\hat{\mathbf{z}}_j ^3\right)^\top \hat{\mathbf{w}}_j \right\vert\\
 = & \dfrac{1}{n}\left\vert \left(\mathbf{z}_j ^3-\hat{\mathbf{z}}^3_j\right)^\top \mathbf{w}_j + \left(\hat{\mathbf{z}}^3_j-\mathbf{z}_j ^3\right)^\top\left(\mathbf{w}_j - \hat{\mathbf{w}}_j\right)+\left(\mathbf{z}_j ^3\right)^\top\left(\mathbf{w}_j-\hat{\mathbf{w}}_j\right)\right\vert \\
 \leq & \dfrac{1}{n}\left(\left\vert \left(\mathbf{z}_j ^3-\hat{\mathbf{z}}^3_j\right)^\top \mathbf{w}_j \right\vert + \left\vert \left(\hat{\mathbf{z}}^3_j-\mathbf{z}_j ^3\right)^\top\left(\mathbf{w}_j - \hat{\mathbf{w}}_j\right)\right\vert + \left\vert\left(\mathbf{z}_j ^3\right)^\top \left(\mathbf{w}_j-\hat{\mathbf{w}}_j\right)\right\vert\right)
\\
 \leq & \dfrac{1}{n}\left(\left\Vert \left(\mathbf{z}_j ^3-\hat{\mathbf{z}}^3_j\right)\right\Vert_2 \left\Vert \mathbf{w}_j \right\Vert_2 + \left\Vert \left(\hat{\mathbf{z}}^3_j-\mathbf{z}_j ^3\right)\right\Vert_2 \left\Vert\left(\mathbf{w}_j - \hat{\mathbf{w}}_j\right)\right\Vert_2 + \left\Vert\left(\mathbf{z}_j ^3\right)\right\Vert_2 \left\Vert \left(\mathbf{w}_j-\hat{\mathbf{w}}_j\right)\right\Vert_2\right)\\
 = & \dfrac{1}{n}\left(\mathcal{O}_p\left(K^2\right)\mathcal{O}_p\left(\sqrt{n}\right) + \mathcal{O}_p\left(K^2\right){\scriptstyle\mathcal{O}}_p\left(\sqrt{n}\right) + \mathcal{O}_p\left(\sqrt{n}\right){\scriptstyle\mathcal{O}}_p\left(\sqrt{n}\right)\right) = {\scriptstyle\mathcal{O}}_p\left(1\right). \text{ Thus,} \\
 \dfrac{1}{n}\left(\hat{\mathbf{z}}_j ^3\right)^\top \hat{\mathbf{w}}_j = &\dfrac{1}{n}\left(\mathbf{z}_j ^3\right)^\top \mathbf{w}_j + {\scriptstyle\mathcal{O}}_p\left(1\right) = \EE\left[Z_j^3 W_j\right]+ {\scriptstyle\mathcal{O}}_p\left(1\right). \\
\dfrac{1}{n}\left(\hat{\mathbf{z}}_j ^3\right)^\top \hat{\mathbf{z}}_j=& \EE\left[Z_j^4\right]+ \mathcal{O}_p\left(\dfrac{K^2}{\sqrt{n}}\right) \text{ follows analogously such that} \\
\hat{\beta}_j^{HOLS} = & \dfrac{\EE\left[Z_j^3 W_j\right]}{\EE\left[Z_j^4\right]} = \dfrac{\EE\left[Z_j^3 W_j\right]}{\EE\left[Z_j^4\right]} + { \scriptstyle \mathcal{O}}_p\left(1\right).
\end{align*}
The last part of Theorem \ref{theo:conf} considers the variance estimate (cf.\ \eqref{eq:var-hat}) and is implied by the following Lemma which is a more precise statement.
\begin{lemm}\label{lemm:var-est-conv}
Assume that the data follows the model \eqref{eq:refmodel} and that \ref{ass:invertible} - \ref{ass:mom6} hold. Then,
\begin{equation*}
\widehat{\text{Var}}\left(\sqrt{n}\left(\hat{\beta}_j^{HOLS}-\hat{\beta}_j^{OLS}\right)\right) \overset{\mathbb{P}}{\to} \sigma_{\tilde{\mathcal{E}}}^2\left(\dfrac{\EE\left[\left(\tilde{Z}_j^3\right)^2 \right]}{\EE\left[Z_j^4 \right]^2}-\dfrac{1}{\EE\left[Z_j^2 \right]}\right) \ \forall j.
\end{equation*}
Note that we defined
\begin{equation*}
\widehat{\text{Var}}\left(\sqrt{n}\left(\hat{\beta}_j^{HOLS}-\hat{\beta}_j^{OLS}\right)\right) \coloneqq n\widehat{\text{Var}}\left(\left(\hat{\beta}_j^{HOLS}-\hat{\beta}_j^{OLS}\right)\right).
\end{equation*}
\end{lemm}
\subsubsection{Proof of Lemma \ref{lemm:var-est-conv}}
For $\hat{\mathbf{z}}_j^\top\hat{\mathbf{z}}_j$ and $\left(\hat{\mathbf{z}}_j ^2\right)^\top \left(\hat{\mathbf{z}}_j\right)^2=\left(\hat{\mathbf{z}}_j ^3\right)^\top \hat{\mathbf{z}}_j$, we have established convergence already. It remains to look at the other terms in \eqref{eq:var-hat}, i.e., $\left(\hat{\mathbf{z}}_j ^3\right)^\top \boldsymbol{P}^{\perp}_{-j}\left(\hat{\mathbf{z}}_j ^3\right)$ and $\hat{\sigma}^2$.
We find
\begin{align*}
 \dfrac{1}{n} \left\vert\left(\hat{\mathbf{z}}_j^3\right)^{\top}\boldsymbol{P}^\perp_{-j}\left(\hat{\mathbf{z}}_j^3\right)- \left(\mathbf{z}_j^3\right)^\top\boldsymbol{P}^\perp_{-j}\left(\mathbf{z}_j^3\right)\right\vert &  = \dfrac{1}{n} \left\vert\left(\hat{\mathbf{z}}_j^3-\mathbf{z}_j^3\right)^{\top}\boldsymbol{P}^\perp_{-j}\left(\hat{\mathbf{z}}_j^3-\mathbf{z}_j^3\right) + 2 \left(\mathbf{z}_j^3\right)^\top\boldsymbol{P}^\perp_{-j}\left(\hat{\mathbf{z}}_j^3-\mathbf{z}_j^3\right)\right\vert \\ 
 \leq \dfrac{1}{n} \left\Vert\hat{\mathbf{z}}_j^3-\mathbf{z}_j^3\right\Vert_2^2 + \dfrac{2}{n} \left\Vert\mathbf{z}_j^3\right\Vert_2 \left\Vert\hat{\mathbf{z}}_j^3-\mathbf{z}_j^3\right\Vert_2 & = \mathcal{O}_p\left(\dfrac{K^4}{n}\right) + \mathcal{O}_p\left(\dfrac{K^2}{\sqrt{n}}\right)=\mathcal{O}_p\left(\dfrac{K^2}{\sqrt{n}}\right), \text{ and} \\
 \dfrac{1}{n} \left\vert \left(\mathbf{z}_j^3\right)^\top\boldsymbol{P}^\perp_{-j}\left(\mathbf{z}_j^3\right)-\left(\tilde{\mathbf{z}}_j^3\right)^\top \left(\tilde{\mathbf{z}}_j^3\right)\right\vert & =\dfrac{1}{n} \left\vert \left(\tilde{\mathbf{z}}_j^3\right)^\top\boldsymbol{P}^\perp_{-j}\left(\tilde{\mathbf{z}}_j^3\right)-\left(\tilde{\mathbf{z}}_j^3\right)^\top \left(\tilde{\mathbf{z}}_j^3\right)\right\vert = \dfrac{1}{n} \left\vert \left(\tilde{\mathbf{z}}_j^3\right)^\top\boldsymbol{P}_{-j}\left(\tilde{\mathbf{z}}_j^3\right)\right\vert \\
 & = { \scriptstyle \mathcal{O}}_p\left(1\right) \text{ such that} \\
\dfrac{1}{n} \left(\hat{\mathbf{z}}_j^3\right)^{\top}\boldsymbol{P}^\perp_{-j}\left(\hat{\mathbf{z}}_j^3\right) & = \dfrac{1}{n} \left(\tilde{\mathbf{z}}_j^3\right)^\top \left(\tilde{\mathbf{z}}_j^3\right) + { \scriptstyle \mathcal{O}}_p\left(1\right) = \EE\left[\left(\tilde{Z}_j^3\right)^2\right] + { \scriptstyle \mathcal{O}}_p\left(1\right).
\end{align*}
This ensures convergence of the per variable error scaling. It remains to estimate the variance of $\tilde{\mathcal{E}}$. Although the error is now only uncorrelated but not independent from $\mathbf{X}$ (cf. \eqref{eq:refmodel}), the variance can still be estimated consistently using the standard formula. Let 
\begin{equation*}
\hat{\boldsymbol{\eps}} = \mathbf{y}-\mathbf{x}\hat{\boldsymbol{\beta}}^{OLS}=\boldsymbol{P}^{\perp}_{-j} \tilde{\boldsymbol{\eps}},
\end{equation*}
which is used for variance estimation. We find
\begin{align*}
\dfrac{1}{n-p}\left\vert \tilde{\boldsymbol{\eps}}^\top \tilde{\boldsymbol{\eps}}- \hat{\boldsymbol{\eps}}^\top \hat{\boldsymbol{\eps}} \right\vert & = \dfrac{1}{n-p}\left\vert \tilde{\boldsymbol{\eps}}^\top \boldsymbol{P}_{-j} \tilde{\boldsymbol{\eps}} \right\vert = \mathcal{O}_p\left(\dfrac{1}{n}\right){\scriptstyle \mathcal{O}}_p\left(n\right)\mathcal{O}_p\left(\dfrac{1}{n}\right){\scriptstyle \mathcal{O}}_p\left(n\right)={\scriptstyle \mathcal{O}}_p\left(1\right) \text{ such that} \\
\hat{\sigma}^2& =\dfrac{\left\Vert \hat{\boldsymbol{\eps}} \right\Vert_2^2}{n-p} = \dfrac{\left\Vert \tilde{\boldsymbol{\eps}} \right\Vert_2^2}{n-p} + {\scriptstyle \mathcal{O}}_p\left(1\right)=\EE\left[\tilde{\mathcal{E}}^2\right] + {\scriptstyle \mathcal{O}}_p\left(1\right).
\end{align*}
\subsection{Proof of Theorem \ref{theo:local-standard-gauss}}
We provide a supporting Lemma.
\begin{lemm}\label{lemm:par-est-conv}
Assume that the data follows the model \eqref{eq:refmodel} and that \ref{ass:invertible} - \ref{ass:mixmom} hold . Let $j$ be some covariate with $\beta_j^{OLS}=\beta_j^{HOLS}$ for which \ref{ass:LLN-cond} and \ref{ass:6} hold. Then,
\begin{equation*}
\sqrt{n}\left(\hat{\beta}_j^{HOLS}-\hat{\beta}_j^{OLS}\right)\overset{\mathbb{D}}{\to}\mathcal{N}\left(0,\text{Var}\left(\dfrac{\tilde{Z}_j^3 \tilde{\mathcal{E}}}{\EE\left[Z_j^4 \right]}-\dfrac{Z_j \tilde{\mathcal{E}}}{\EE\left[Z_j^2 \right]}\right)\right).
\end{equation*}
If \ref{ass:z-ind} and \ref{ass:z3-ind} hold as well for $j$, this can be refined as 
\begin{equation*}
\sqrt{n}\left(\hat{\beta}_j^{HOLS}-\hat{\beta}_j^{OLS}\right)\overset{\mathbb{D}}{\to}\mathcal{N}\left(0,\sigma_{\tilde{\mathcal{E}}}^2\left(\dfrac{\EE\left[\left(\tilde{Z}_j^3\right)^2 \right]}{\EE\left[Z_j^4 \right]^2}-\dfrac{1}{\EE\left[Z_j^2 \right]}\right)\right).
\end{equation*}
\end{lemm}
Note that \ref{ass:6} is implied by \ref{ass:z3-ind}. Theorem \ref{theo:local-standard-gauss} follows from Lemmata \ref{lemm:var-est-conv} and \ref{lemm:par-est-conv}, applying Slutsky's theorem. Thus, it remains to prove Lemma \ref{lemm:par-est-conv}.
\subsubsection{Proof of Lemma \ref{lemm:par-est-conv}}
We look at the scaled estimates $\sqrt{n}\hat{\beta}_j^{OLS}$ and $\sqrt{n}\hat{\beta}_j^{HOLS}$ for some variable with $\beta_j^{OLS} = \beta_j^{HOLS}$ fulfilling \ref{ass:LLN-cond} and \ref{ass:6}. Note that since we assume \ref{ass:mixmom}, we can sharpen $\left\vert \mathbf{x}_k^\top \mathbf{w}_j \right\vert = \mathcal{O}_p\left(\sqrt{n}\right)$ instead of just ${\scriptstyle \mathcal{O}}_p\left(n\right)$.
\begin{align*}
\sqrt{n}\hat{\beta}_j^{OLS} & = \dfrac{\sqrt{n}\dfrac{1}{n}\hat{\mathbf{z}}_j^\top\hat{\mathbf{w}}_j}{\dfrac{1}{n}\hat{\mathbf{z}}_j^\top\hat{\mathbf{z}}_j} = \dfrac{\sqrt{n}\dfrac{1}{n}\mathbf{z}_j^\top \mathbf{w}_j + \mathcal{O}_p\left(1/\sqrt{n}\right)}{\dfrac{1}{n}\mathbf{z}_j^\top \mathbf{z}_j + \mathcal{O}_p\left(1/n\right)} = \dfrac{\sqrt{n}\dfrac{1}{n}\mathbf{z}_j^\top \mathbf{w}_j }{\dfrac{1}{n}\mathbf{z}_j^\top \mathbf{z}_j } + \mathcal{O}_p\left(1/\sqrt{n}\right) \\
&=\sqrt{n}\beta_j^{OLS} + \dfrac{\sqrt{n}\dfrac{1}{n}\mathbf{z}_j^\top \tilde{\boldsymbol{\eps}}}{\dfrac{1}{n}\mathbf{z}_j^\top \mathbf{z}_j } + \mathcal{O}_p\left(1/\sqrt{n}\right)=\sqrt{n}\beta_j^{OLS} + \dfrac{\sqrt{n}\dfrac{1}{n}\mathbf{z}_j^\top \tilde{\boldsymbol{\eps}}}{\EE\left[Z_j^2\right] + \mathcal{O}_p\left(1/\sqrt{n}\right)} + \mathcal{O}_p\left(1/\sqrt{n}\right) \\
& =\sqrt{n}\beta_j^{OLS} + \dfrac{\sqrt{n}\dfrac{1}{n}\mathbf{z}_j^\top \tilde{\boldsymbol{\eps}}}{\EE\left[Z_j^2\right]} + \mathcal{O}_p\left(1/\sqrt{n}\right),
\end{align*}
where we used results from the previous section (together with the sharpening) and the fact that $\mathbf{w}_j = \mathbf{z}_j \beta_j^{OLS} + \tilde{\boldsymbol{\eps}}$.
For $\hat{\beta}_j^{HOLS}$, we analyse the numerator.
\begin{equation}\label{eq:z3-eps-conv-SEM}
\sqrt{n}\dfrac{1}{n}\left\vert \left(\mathbf{z}_j ^3\right)^\top\boldsymbol{P}^{\perp}_{-j}\tilde{\boldsymbol{\eps}} - \left(\hat{\mathbf{z}}_j ^3\right)^\top \boldsymbol{P}^{\perp}_{-j}\tilde{\boldsymbol{\eps}} \right\vert \leq \sqrt{n}\dfrac{1}{n}\left\vert \left(\mathbf{z}_j ^3-\hat{\mathbf{z}}^3_j\right)^\top \tilde{\boldsymbol{\eps}} \right\vert +\sqrt{n}\dfrac{1}{n}\left\vert\left(\mathbf{z}_j ^3-\hat{\mathbf{z}}^3_j\right)^\top\boldsymbol{P}_{-j}\tilde{\boldsymbol{\eps}} \right\vert.
\end{equation}
Using a derivation as in \eqref{eq:P_j-conf} and \ref{ass:mixmom}, we know $\left\Vert \boldsymbol{P}_{-j}\tilde{\boldsymbol{\eps}}\right\Vert_2 =\mathcal{O}_p\left(1\right)$. Thus, using \eqref{eq:z3-conv}, the second term is controlled. In \eqref{eq:z3-conv}, we have split $\left(\hat{\mathbf{z}}^3_j-\mathbf{z}_j ^3\right)$ in to three parts. Only the third part is critical concerning the convergence of the first term in \eqref{eq:z3-eps-conv-SEM} as the others lead to ${ \scriptstyle \mathcal{O}}_p\left(1\right)$ terms when applying Cauchy-Schwarz to the inner product. Therefore, we take a closer look at $\left(\mathbf{z}_j^2 \odot\left(\mathbf{z}_j -\hat{\mathbf{z}}_j\right)\right)^\top \tilde{\boldsymbol{\eps}}$.
\begin{align*}
& \sqrt{n}\dfrac{1}{n} \left\vert\left(\mathbf{z}_j^2 \odot\left(\mathbf{z}_j -\hat{\mathbf{z}}_j\right)\right)^\top \tilde{\boldsymbol{\eps}}\right\vert  =\sqrt{n}\dfrac{1}{n} \left\vert \sum_{i=1}^{n} z_{ij}^2\left(z_{ij}-\hat{z}_{ij}\right)\tilde{\eps}_i\right\vert =\sqrt{n}\dfrac{1}{n} \left\vert \sum_{i=1}^{n} z_{ij}^2\mathbf{x}_{i,-j}\left(\hat{\boldsymbol{\gamma}}_j-\boldsymbol{\gamma}_j\right)\tilde{\eps}_i \right\vert \\ 
=&\sqrt{n}\dfrac{1}{n} \left\vert \sum_{i=1}^{n} z_{ij}^2 \tilde{\eps}_i \sum_{k\neq j} x_{ik}\left(\hat{\gamma}_{jk}-\gamma_{jk}\right)\right\vert =\left\vert\sum_{k\neq j} \left(\hat{\gamma}_{jk}-\gamma_{jk}\right)\sqrt{n}\dfrac{1}{n} \sum_{i=1}^{n} z_{ij}^2 x_{ik}\tilde{\eps}_i \right\vert \\
 \leq & \sum_{k\neq j} \left\vert\hat{\gamma}_{jk}-\gamma_{jk}\right\vert\left\vert\sqrt{n}\dfrac{1}{n} \sum_{i=1}^{n} z_{ij}^2 x_{ik}\tilde{\eps}_i \right\vert = \sum_{k\neq j} \mathcal{O}_p \left(\dfrac{1}{\sqrt{n}}\right)\mathcal{O}_p \left(\sqrt{n}\right) { \scriptstyle \mathcal{O}}_p\left(1\right) = { \scriptstyle \mathcal{O}}_p\left(1\right).
\end{align*}
In the second to last inequality, we use \ref{ass:LLN-cond}. In short,
\begin{align*}
& \sqrt{n}\dfrac{1}{n}\left\vert \left(\mathbf{z}_j ^3\right)^\top\boldsymbol{P}^{\perp}_{-j}\tilde{\boldsymbol{\eps}} - \left(\hat{\mathbf{z}}_j ^3\right)^\top \boldsymbol{P}^{\perp}_{-j}\tilde{\boldsymbol{\eps}} \right\vert  ={ \scriptstyle \mathcal{O}}_p\left(1\right) \text{ such that}\\
& \sqrt{n}\dfrac{1}{n}\left\vert \left(\tilde{\mathbf{z}}_j ^3\right)^\top\tilde{\boldsymbol{\eps}} - \left(\hat{\mathbf{z}}_j ^3\right)^\top \boldsymbol{P}^{\perp}_{-j}\tilde{\boldsymbol{\eps}} \right\vert \\
 \leq & \sqrt{n}\dfrac{1}{n}\left\vert \left(\mathbf{z}_j ^3\right)^\top\boldsymbol{P}^{\perp}_{-j}\tilde{\boldsymbol{\eps}} - \left(\hat{\mathbf{z}}_j ^3\right)^\top \boldsymbol{P}^{\perp}_{-j}\tilde{\boldsymbol{\eps}} \right\vert + \sqrt{n}\dfrac{1}{n}\left\vert \left(\tilde{\mathbf{z}}_j ^3\right)^\top\tilde{\boldsymbol{\eps}} - \left({\mathbf{z}}_j ^3\right)^\top \boldsymbol{P}^{\perp}_{-j}\tilde{\boldsymbol{\eps}} \right\vert \\
 = & \sqrt{n}\dfrac{1}{n}\left\vert \left(\mathbf{z}_j ^3\right)^\top\boldsymbol{P}^{\perp}_{-j}\tilde{\boldsymbol{\eps}} - \left(\hat{\mathbf{z}}_j ^3\right)^\top \boldsymbol{P}^{\perp}_{-j}\tilde{\boldsymbol{\eps}} \right\vert + \sqrt{n}\dfrac{1}{n}\left\vert \left(\tilde{\mathbf{z}}_j ^3\right)^\top\boldsymbol{P}_{-j}\tilde{\boldsymbol{\eps}} \right\vert ={ \scriptstyle \mathcal{O}}_p\left(1\right).
\end{align*}
This leads to
\begin{align*}
\sqrt{n}\hat{\beta}_j^{HOLS} &= \dfrac{\sqrt{n}\dfrac{1}{n}\left(\hat{\mathbf{z}}_j^3\right)^\top \hat{\mathbf{w}_j}}{\dfrac{1}{n}\left(\hat{\mathbf{z}}_j^3\right)^\top \hat{\mathbf{z}_j}}=\sqrt{n}\beta_j^{OLS} + \dfrac{\sqrt{n}\dfrac{1}{n}\left(\hat{\mathbf{z}}_j ^3\right)^\top \boldsymbol{P}^{\perp}_{-j}\tilde{\boldsymbol{\eps}}}{\dfrac{1}{n}\left(\hat{\mathbf{z}}_j^3\right)^\top \hat{\mathbf{z}_j}} \\
&=\sqrt{n}\beta_j^{OLS} + \dfrac{\sqrt{n}\dfrac{1}{n} \left(\tilde{\mathbf{z}}_j ^3\right)^\top\tilde{\boldsymbol{\eps}}+{ \scriptstyle \mathcal{O}}_p\left(1\right)}{\EE\left[Z_j^4\right]+{ \scriptstyle \mathcal{O}}_p\left(1\right)} =\sqrt{n}\beta_j^{OLS} + \dfrac{\sqrt{n}\dfrac{1}{n} \left(\tilde{\mathbf{z}}_j ^3\right)^\top\tilde{\boldsymbol{\eps}}}{\EE\left[Z_j^4\right]}+{ \scriptstyle \mathcal{O}}_p\left(1\right).
\end{align*}
Combining the results for $\sqrt{n}\hat{\beta}_j^{OLS}$ and $\sqrt{n}\hat{\beta}_j^{HOLS}$, we find
\begin{equation}\label{eq:convinp}
\sqrt{n}\left(\hat{\beta}_j^{HOLS}-\hat{\beta}_j^{OLS}\right)= \sqrt{n}\dfrac{1}{n} \left(\dfrac{\left(\tilde{\mathbf{z}}_j ^3\right)^\top}{\EE\left[Z_j^4\right]}-\dfrac{\mathbf{z}_j^\top }{\EE\left[Z_j^2\right]}\right)\tilde{\boldsymbol{\eps}}+{ \scriptstyle \mathcal{O}}_p\left(1\right).
\end{equation}
Since the first term is a scaled sum of i.i.d.\ random variables, we can apply the CLT to it
\begin{equation}\label{eq:convind}
\sqrt{n}\dfrac{1}{n}\left(\dfrac{\tilde{\mathbf{z}}_j^3}{\EE\left[Z_j^4 \right]}-\dfrac{\mathbf{z}_j}{\EE\left[Z_j ^2\right]}\right)^\top \tilde{\boldsymbol{\eps}} \overset{\mathbb{D}}{\to} \mathcal{N}\left(0,\text{Var}\left(\dfrac{\tilde{Z}_j^3 \tilde{\mathcal{E}}}{\EE\left[Z_j^4 \right]}-\dfrac{Z_j \tilde{\mathcal{E}}}{\EE\left[Z_j^2 \right]}\right)\right).
\end{equation}
Note that $\EE\left[\tilde{Z}_j^3 \tilde{\mathcal{E}}\right]=0$ as $\beta_j^{OLS}=\beta_j^{HOLS}$. Combining \eqref{eq:convinp} and \eqref{eq:convind} leads to the first statement in Lemma \ref{lemm:par-est-conv}. Applying the independence relationship induced by \ref{ass:z-ind} and \ref{ass:z3-ind}, the second statement follows trivially.

\subsection{Proof of Theorem \ref{theo:V_U}}
From \eqref{eq:convinp}, we know $\sqrt{n}\dfrac{1}{n}\left\vert \hat{\mathbf{v}}_j ^\top \tilde{\boldsymbol{\eps}} -\mathbf{v}_j^\top \tilde{\boldsymbol{\eps}} \right\vert \overset{\mathbb{P}}{\to} 0 \ \forall j \in U' $ under the given assumptions. For fixed dimensions, we can easily make this statement multivariate, i.e., $\sqrt{n}\dfrac{1}{n}\left\Vert \hat{\mathbf{v}}_U ^\top \tilde{\boldsymbol{\eps}} -\mathbf{v}_U^\top \tilde{\boldsymbol{\eps}} \right\Vert \overset{\mathbb{P}}{\to} 0 $. Therefore, we inspect $\mathbf{v}_U^\top \tilde{\boldsymbol{\eps}}$ in the following. Note that this is a (scaled) sum of mean $0$ i.i.d random vectors. Obviously, this enables the multivariate CLT such that
\begin{equation*}
\sqrt{n}\dfrac{1}{n}\mathbf{v}_U^\top \tilde{\boldsymbol{\eps}} \overset{\mathbb{D}}{\to} \mathcal{N}\left(\boldsymbol{0}, \EE\left[\tilde{\mathcal{E}}\mathbf{V}_U\mathbf{V}_U^\top \tilde{\mathcal{E}}\right]\right) = \mathcal{N}\left(\boldsymbol{0}, \sigma_{\tilde{\mathcal{E}}}^2 \EE\left[\mathbf{V}_U\mathbf{V}_U^\top \right]\right),
\end{equation*}
which implies the first part of Theorem \ref{theo:V_U}. 
For the second part, note
\begin{align*}
\dfrac{1}{n}\hat{\mathbf{v}}_j^\top \hat{\mathbf{v}}_k & = \dfrac{1}{n} \left(\dfrac{\boldsymbol{P}_{-j}^\perp \left( \hat{\mathbf{z}}_j^3\right)}{\dfrac{1}{n}\left(\hat{\mathbf{z}}_j^2\right)^\top\left(\hat{\mathbf{z}}_j^2\right)} - \dfrac{\hat{\mathbf{z}}_j}{\dfrac{1}{n}\hat{\mathbf{z}}_j^\top\hat{\mathbf{z}}_j}\right)^\top \left(\dfrac{\boldsymbol{P}_{-k}^\perp \left( \hat{\mathbf{z}}_k^3\right)}{\dfrac{1}{n}\left(\hat{\mathbf{z}}_k^2\right)^\top\left(\hat{\mathbf{z}}_k^2\right)} - \dfrac{\hat{\mathbf{z}}_k}{\dfrac{1}{n}\hat{\mathbf{z}}_k^\top\hat{\mathbf{z}}_k}\right) \\
& = \dfrac{1}{n}\left(\dfrac{\left( \hat{\mathbf{z}}_j^3\right)^\top\boldsymbol{P}_{-j}^\perp \boldsymbol{P}_{-k}^\perp \left( \hat{\mathbf{z}}_k^3\right)}{\dfrac{1}{n}\left(\hat{\mathbf{z}}_j^2\right)^\top\left(\hat{\mathbf{z}}_j^2\right)\dfrac{1}{n}\left(\hat{\mathbf{z}}_k^2\right)^\top\left(\hat{\mathbf{z}}_k^2\right)}-\dfrac{\left( \hat{\mathbf{z}}_j^3\right)^\top\boldsymbol{P}_{-j}^\perp \hat{\mathbf{z}}_k}{\dfrac{1}{n}\left(\hat{\mathbf{z}}_j^2\right)^\top\left(\hat{\mathbf{z}}_j^2\right)\dfrac{1}{n}\hat{\mathbf{z}}_k^\top\hat{\mathbf{z}}_k}\right.\\
&\left. -\dfrac{\hat{\mathbf{z}}_j^\top \boldsymbol{P}_{-k}^\perp \left( \hat{\mathbf{z}}_k^3\right)}{\dfrac{1}{n}\hat{\mathbf{z}}_j^\top\hat{\mathbf{z}}_j\dfrac{1}{n}\left(\hat{\mathbf{z}}_k^2\right)^\top\left(\hat{\mathbf{z}}_k^2\right)} + \dfrac{\hat{\mathbf{z}}_j^\top \hat{\mathbf{z}}_k}{\dfrac{1}{n}\hat{\mathbf{z}}_j^\top\hat{\mathbf{z}}_j\dfrac{1}{n}\hat{\mathbf{z}}_k^\top\hat{\mathbf{z}}_k}\right)
\end{align*}
For each of the denominator terms, convergence has been established already. For the numerator terms, we can apply \eqref{eq:P_j-conf}, \eqref{eq:z3-conv}, and $\left\Vert \hat{\mathbf{z}}_j - \mathbf{z}_j \right\Vert_ 2 = \mathcal{O}_p \left(1\right)$.
\begin{align*}
&\dfrac{1}{n}\left\vert \left(\hat{\mathbf{z}}_j^3\right)^\top\boldsymbol{P}_{-j}^\perp \boldsymbol{P}_{-k}^\perp \left( \hat{\mathbf{z}}_k^3\right) - \left(\mathbf{z}_j^3\right)^\top\boldsymbol{P}_{-j}^\perp \boldsymbol{P}_{-k}^\perp \left( \mathbf{z}_k^3\right) \right\vert \\
=&\dfrac{1}{n}\left\vert\left(\hat{\mathbf{z}}_j^3-\mathbf{z}_j^3 \right)^\top \boldsymbol{P}_{-j}^\perp \boldsymbol{P}_{-k}^\perp \left(\hat{\mathbf{z}}_k^3-\mathbf{z}_k^3\right) + \left(\mathbf{z}_j^3\right)^\top \boldsymbol{P}_{-j}^\perp \boldsymbol{P}_{-k}^\perp \left(\hat{\mathbf{z}}_k^3-\mathbf{z}_k^3 \right) + \left(\hat{\mathbf{z}}_j^3-\mathbf{z}_j^3 \right)^\top \boldsymbol{P}_{-j}^\perp \boldsymbol{P}_{-k}^\perp \left(\hat{\mathbf{z}}_k^3\right) \right\vert \\
\leq & \dfrac{1}{n}\left\vert\left(\hat{\mathbf{z}}_j^3-\mathbf{z}_j^3 \right)^\top \boldsymbol{P}_{-j}^\perp \boldsymbol{P}_{-k}^\perp \left(\hat{\mathbf{z}}_k^3-\mathbf{z}_k^3\right)\right\vert + \dfrac{1}{n}\left\vert\left(\mathbf{z}_j^3\right)^\top \boldsymbol{P}_{-j}^\perp \boldsymbol{P}_{-k}^\perp \left(\hat{\mathbf{z}}_k^3-\mathbf{z}_k^3 \right)\right\vert + \dfrac{1}{n}\left\vert \left(\hat{\mathbf{z}}_j^3-\mathbf{z}_j^3 \right)^\top \boldsymbol{P}_{-j}^\perp \boldsymbol{P}_{-k}^\perp \left(\hat{\mathbf{z}}_k^3\right) \right\vert \\
\leq & \dfrac{1}{n} \left\Vert \boldsymbol{P}_{-j}^\perp \left(\hat{\mathbf{z}}_j^3-\mathbf{z}_j^3 \right) \right\Vert_2 \left\Vert \boldsymbol{P}_{-k}^\perp \left(\hat{\mathbf{z}}_k^3-\mathbf{z}_k^3 \right) \right\Vert_2 + \dfrac{1}{n} \left\Vert \boldsymbol{P}_{-j}^\perp \left(\hat{\mathbf{z}}_j^3\right) \right\Vert_2 \left\Vert \boldsymbol{P}_{-k}^\perp \left(\hat{\mathbf{z}}_k^3-\mathbf{z}_k^3 \right) \right\Vert_2 + \\
& \dfrac{1}{n} \left\Vert \boldsymbol{P}_{-j}^\perp \left(\hat{\mathbf{z}}_j^3-\mathbf{z}_j^3 \right) \right\Vert_2 \left\Vert \boldsymbol{P}_{-k}^\perp \left(\hat{\mathbf{z}}_k^3\right) \right\Vert_2 \\
\leq & \dfrac{1}{n} \left\Vert \left(\hat{\mathbf{z}}_j^3-\mathbf{z}_j^3 \right) \right\Vert_2 \left\Vert \left(\hat{\mathbf{z}}_k^3-\mathbf{z}_k^3 \right) \right\Vert_2 + \dfrac{1}{n} \left\Vert \hat{\mathbf{z}}_j^3 \right\Vert_2 \left\Vert \left(\hat{\mathbf{z}}_k^3-\mathbf{z}_k^3 \right) \right\Vert_2 + \dfrac{1}{n} \left\Vert \left(\hat{\mathbf{z}}_j^3-\mathbf{z}_j^3 \right) \right\Vert_2 \left\Vert \hat{\mathbf{z}}_k^3\right\Vert_2 \\
= & \mathcal{O}_p\left(\dfrac{K^2}{\sqrt{n}}\right) = {\scriptstyle \mathcal{O}}_p\left(1\right)\\
&\dfrac{1}{n}\left\vert \left(\mathbf{z}_j^3\right)^\top\boldsymbol{P}_{-j}^\perp \boldsymbol{P}_{-k}^\perp \left( \mathbf{z}_k^3\right) -\left(\tilde{\mathbf{z}}_j^3\right)^\top \left(\tilde{\mathbf{z}}_k^3\right) \right\vert = \dfrac{1}{n}\left\vert \left(\tilde{\mathbf{z}}_j^3\right)^\top\boldsymbol{P}_{-j}^\perp \boldsymbol{P}_{-k}^\perp \left( \tilde{\mathbf{z}}_k^3\right) -\left(\tilde{\mathbf{z}}_j^3\right)^\top \left(\tilde{\mathbf{z}}_k^3\right) \right \vert \\
 = & \dfrac{1}{n}\left\vert \left(\tilde{\mathbf{z}}_j^3\right)^\top\boldsymbol{P}_{-j} \boldsymbol{P}_{-k} \left( \tilde{\mathbf{z}}_k^3\right) + \left(\tilde{\mathbf{z}}_j^3\right)^\top \boldsymbol{P}_{-k} \left( \tilde{\mathbf{z}}_k^3\right) + \left(\tilde{\mathbf{z}}_j^3\right)^\top\boldsymbol{P}_{-j} \left( \tilde{\mathbf{z}}_k^3\right)  \right \vert \\
 \leq & \dfrac{1}{n}\left\vert \left(\tilde{\mathbf{z}}_j^3\right)^\top\boldsymbol{P}_{-j} \boldsymbol{P}_{-k} \left( \tilde{\mathbf{z}}_k^3\right)\right \vert + \dfrac{1}{n}\left\vert \left(\tilde{\mathbf{z}}_j^3\right)^\top \boldsymbol{P}_{-k} \left( \tilde{\mathbf{z}}_k^3\right)\right \vert + \dfrac{1}{n}\left\vert \left(\tilde{\mathbf{z}}_j^3\right)^\top\boldsymbol{P}_{-j} \left( \tilde{\mathbf{z}}_k^3\right)  \right \vert \\
 \leq & \dfrac{1}{n} \left\Vert \boldsymbol{P}_{-j}\left(\tilde{\mathbf{z}}_j^3\right)\right\Vert_2 \left\Vert \boldsymbol{P}_{-k}\left(\tilde{\mathbf{z}}_k^3\right)\right\Vert_2 + \dfrac{1}{n} \left\Vert\tilde{\mathbf{z}}_j^3\right\Vert_2 \left\Vert \boldsymbol{P}_{-k}\left(\tilde{\mathbf{z}}_k^3\right)\right\Vert_2 + \dfrac{1}{n} \left\Vert \boldsymbol{P}_{-j}\left(\tilde{\mathbf{z}}_j^3\right)\right\Vert_2 \left\Vert \tilde{\mathbf{z}}_k^3\right\Vert_2 = {\scriptstyle \mathcal{O}}_p\left(1\right) \text{ so}\\
 & \left(\hat{\mathbf{z}}_j^3\right)^\top\boldsymbol{P}_{-j}^\perp \boldsymbol{P}_{-k}^\perp \left( \hat{\mathbf{z}}_k^3\right) = \left(\tilde{\mathbf{z}}_j^3\right)^\top \left(\tilde{\mathbf{z}}_k^3\right) + {\scriptstyle \mathcal{O}}_p\left(1\right) = \EE\left[\tilde{Z}_j^3 \tilde{Z}_k^3\right] + {\scriptstyle \mathcal{O}}_p\left(1\right)
\end{align*}
The other terms follow in a very similar fashion such that Slutsky's theorem leads to
\begin{equation*}
\dfrac{1}{n}\hat{\mathbf{v}}_j^\top \hat{\mathbf{v}}_k = \EE\left[V_j V_k\right] + {\scriptstyle \mathcal{O}}_p\left(1\right)
\end{equation*}
For fixed $p$, this can be directly made multidimensional which proves the theorem's statement.

\subsubsection{Proof of Corollary \ref{corr:alpha-H0j}}
Consider $\mathbf{S}$ as given in Step 7 of Algorithm \ref{alg:HOLS}. Using the second part of Theorem \ref{theo:V_U} and a consistent estimate of $\hat{\sigma}$, we have
\begin{equation*}
\sqrt{n}\mathbf{S}_{U'} \overset{\mathbb{D}}{\to} \mathcal{N}\left(\mathbf{0}, \sigma_{\tilde{\mathcal{E}}}^2 \EE\left[\mathbf{V}_{U'}\mathbf{V}_{U'}^\top\right]\right).
\end{equation*}
Let $\mathbf{S}^* \sim \mathcal{N}\left(\mathbf{0}, \sigma_{\tilde{\mathcal{E}}}^2 \EE\left[\mathbf{V}_{U'}\mathbf{V}_{U'}^\top\right]\right)$ and denote the cumulative density function (CDF) of its maximum absolute value by $F^*$. Denote the CDF of $\sqrt{n}\left\Vert\hat{\boldsymbol{\beta}}_{U'}^{HOLS} -\hat{\boldsymbol{\beta}}_{U'}^{OLS} \right\Vert_\infty$ by $F_n$. Let $q$ be the quantile function and $\hat{q}$ the estimated quantile function using $\mathbf{s}^1, \ldots, \mathbf{s}^{n_{sim}}$. Then,
\begin{align*}
& \underset{n\rightarrow\infty}{\text{lim}} \ \underset{n_{sim}\rightarrow\infty}{\text{lim}}\PP \left(\exists j \in U' \text{ such that } H_{0,j} \text{ is rejected}\right) \\
=& \underset{n\rightarrow\infty}{\text{lim}} \ \underset{n_{sim}\rightarrow\infty}{\text{lim}} \PP\left(\left\Vert \hat{\mathbf{\beta}}_{U'}^{HOLS} - \hat{\mathbf{\beta}}_{U'}^{OLS}\right\Vert_\infty > \hat{q}_{1-\alpha}\left(\left\Vert \mathbf{S} \right\Vert_\infty \right)\right)\\
 \leq & \underset{n\rightarrow\infty}{\text{lim}} \ \underset{n_{sim}\rightarrow\infty}{\text{lim}} \PP\left(\left\Vert \hat{\mathbf{\beta}}_{U'}^{HOLS} - \hat{\mathbf{\beta}}_{U'}^{OLS}\right\Vert_\infty > \hat{q}_{1-\alpha}\left(\left\Vert \mathbf{S}_{U'} \right\Vert_\infty \right)\right) = 1 - \underset{n\rightarrow\infty}{\text{lim}} \ \underset{n_{sim}\rightarrow\infty}{\text{lim}} F_n\left(\hat{q}_{1-\alpha}\left(\left\Vert \mathbf{S}_{U'} \right\Vert_\infty \right)\right) \\
 = & 1 - \underset{n\rightarrow\infty}{\text{lim}} \ \underset{n_{sim}\rightarrow\infty}{\text{lim}} F^*\left(\hat{q}_{1-\alpha}\left(\left\Vert \mathbf{S}_{U'} \right\Vert_\infty \right)\right) = 1 - \underset{n\rightarrow\infty}{\text{lim}} F^*\left(q_{1-\alpha}\left(\left\Vert \mathbf{S}_{U'} \right\Vert_\infty \right)\right)  = 1 - F^*\left(q_{1-\alpha}\left(\left\Vert \mathbf{S}^* \right\Vert_\infty \right)\right)\\
 = & \alpha
\end{align*}
For the equality between the second and third line, note that $F_n \rightarrow F$ using Theorem \ref{theo:V_U} and the continuous mapping theorem. As the maximum of several Gaussian random variables has a continuous CDF, this convergence is uniform such the convergence also holds at $\hat{q}_{1-\alpha}\left(\left\Vert \mathbf{S}_{U'} \right\Vert_\infty \right)$ which is not constant in $n$ and $n_{sim}$.
For the convergence of empirical quantiles, see, e.g., the discussion in \cite[Chapter~21]{van2000asymptotic}.

\subsection{Proof of Theorem \ref{theo:U-rec}}
We split the goal into two problems, namely,
\begin{equation}\label{eq:u-conv2}
\underset{n \rightarrow \infty}{\text{lim}} \mathbb{P}\left[\hat{U} \subseteq U\right]=1 \ \text{and} \ \underset{n \rightarrow \infty}{\text{lim}} \mathbb{P}\left[\hat{U} \supseteq U\right]=1.
\end{equation}
The first one corresponds to rejecting the required $\ H_{0,j}$ and the second one to no wrong rejection.
We provide the supporting lemmata. Theorem \ref{theo:U-rec} follows directly by combining these.
\begin{lemm}\label{lemm:U-sub}
Assume that the data follows the model \eqref{eq:refmodel} and that \ref{ass:invertible} - \ref{ass:mom6} hold. Let $j \not \in U$. Then,
\begin{equation*}
\mathbb{P}\left[\left\vert t_j \right\vert \geq \tau_n \right] \geq \mathbb{P}\left[\left\vert \beta_j^{HOLS}-\beta_j^{OLS} \right\vert \geq \tau_n  \left\vert \mathcal{O}_p\left(1/ \sqrt{n} \right)\right\vert + \left\vert { \scriptstyle \mathcal{O}}_p\left(1\right) \right\vert \right].
\end{equation*}
\end{lemm}

This probability can be ensured to approach $1$ if we chose $\tau_n = { \scriptstyle \mathcal{O}}\left(\sqrt{n} \right)$. Under this condition, we find $\underset{n \rightarrow \infty}{\text{lim}} \mathbb{P}\left[\hat{U} \subseteq U\right]=1$. Notably, we assume $\left\vert \beta_j^{HOLS}-\beta_j^{OLS} \right\vert$ to be constant, i.e., we deal with a fixed alternative. We further remark that we could use a constant significance level $\alpha_n = \alpha$ to receive just the first convergence in \eqref{eq:u-conv2}.

Let us now turn to variables for which $H_{0,j}$ holds true. In order to reuse our convergence results from Section \ref{conv-par}, we have to additionally invoke \ref{ass:mixmom}, \ref{ass:LLN-cond} and \ref{ass:6}.
\begin{lemm}\label{lemm:U-sup}
Assume that the data follows the model \eqref{eq:refmodel} and that \ref{ass:invertible} - \ref{ass:mixmom} hold. Let $j$ be some covariate in $U$ for which \ref{ass:LLN-cond} and \ref{ass:6} hold. Then,
\begin{equation*}
\mathbb{P}\left[\left\vert t_j \right\vert \geq \tau_n \right] \leq \EE\left[\left(\left(\dfrac{\tilde{Z}_j^3}{\EE\left[Z_j^4\right]}-\dfrac{Z_j}{\EE\left[Z_j^2\right]}\right)\tilde{\mathcal{E}}\right)^2\right] / \left(\tau_n \left\vert \mathcal{O}_p\left(1\right)\right\vert / 2\right)^2 + \mathbb{P}\left[\left\vert { \scriptstyle \mathcal{O}}_p\left(1\right) \right\vert \geq \tau_n/2 \right] .
\end{equation*}
\end{lemm}
Either term vanishes if we choose $1/\tau_n ={ \scriptstyle \mathcal{O}}\left(1\right) $. Thus, as long as $\tau_n $ grows at any rate, we receive $\underset{n \rightarrow \infty}{\text{lim}} \mathbb{P}\left[\hat{U} \supseteq U\right]=1$.
\subsubsection{Proof of Lemma \ref{lemm:U-sub}}
From Theorem \ref{theo:conf},
 we know
\begin{align*}
\sqrt{n}\left(\hat{\beta}_j^{HOLS}-\hat{\beta}_j^{OLS}\right) & = \sqrt{n}\left(\beta_j^{HOLS}-\beta_j^{OLS} \right) + { \scriptstyle \mathcal{O}}_p\left(\sqrt{n}\right) \\
\widehat{\text{Var}}\left(\sqrt{n}\left(\hat{\beta}_j^{HOLS}-\hat{\beta}_j^{OLS}\right)\right) & = \mathcal{O}_p\left(1\right).
\end{align*}
Thus, we have
\begin{align*}
\left\vert t_j \right\vert & = \left\vert \dfrac{ \sqrt{n}\left(\beta_j^{HOLS}-\beta_j^{OLS} \right)}{\mathcal{O}_p\left(1\right)} + { \scriptstyle \mathcal{O}}_p\left(\sqrt{n}\right) \right\vert \geq \left\vert \dfrac{ \sqrt{n}\left(\beta_j^{HOLS}-\beta_j^{OLS}\right) }{\mathcal{O}_p\left(1\right)} \right\vert - \left\vert { \scriptstyle \mathcal{O}}_p\left(\sqrt{n}\right) \right\vert \\
\mathbb{P}\left[\left\vert t_j \right\vert \geq \tau_n \right] & \geq \mathbb{P}\left[\left\vert \dfrac{ \sqrt{n}\left(\beta_j^{HOLS}-\beta_j^{OLS}\right) }{\mathcal{O}_p\left(1\right)} \right\vert \geq \tau_n + \left\vert { \scriptstyle \mathcal{O}}_p\left(\sqrt{n}\right) \right\vert \right] \\ 
& = \mathbb{P}\left[\left\vert \beta_j^{HOLS}-\beta_j^{OLS} \right\vert  \geq \tau_n \left\vert \mathcal{O}_p\left(1/ \sqrt{n} \right)\right\vert + \left\vert { \scriptstyle \mathcal{O}}_p\left(1\right) \right\vert \right].
\end{align*}
\subsubsection{Proof of Lemma \ref{lemm:U-sup}}
For variables fulfilling \ref{ass:LLN-cond}, we know from \eqref{eq:convinp} and Theorem \ref{theo:conf}
\begin{align*}
\sqrt{n}\left(\hat{\beta}_j^{HOLS}-\hat{\beta}_j^{OLS}\right) &= \sqrt{n}\dfrac{1}{n}\left(\dfrac{\tilde{\mathbf{z}}_j^3}{\EE\left[Z_j^4 \right]}-\dfrac{\mathbf{z}_j}{\EE\left[Z_j ^2\right]}\right)^\top \tilde{\boldsymbol{\eps}}+{ \scriptstyle \mathcal{O}}_p\left(1\right). \\
\widehat{\text{Var}}\left(\sqrt{n}\left(\hat{\beta}_j^{HOLS}-\hat{\beta}_j^{OLS}\right)\right) & = \mathcal{O}_p\left(1\right).
\end{align*}
This yields
\begin{align*}
\left\vert t_j \right\vert & =\left\vert \sqrt{n}\dfrac{1}{n \mathcal{O}_p\left(1\right)}\left(\dfrac{\tilde{\mathbf{z}}_j^3}{\EE\left[Z_j^4 \right]}-\dfrac{\mathbf{z}_j}{\EE\left[Z_j ^2\right]}\right)^\top \tilde{\boldsymbol{\eps}}+{ \scriptstyle \mathcal{O}}_p\left(1\right) \right\vert \\
 & \leq  \left\vert \sqrt{n}\dfrac{1}{n \mathcal{O}_p\left(1\right)}\left(\dfrac{\tilde{\mathbf{z}}_j^3}{\EE\left[Z_j^4 \right]}-\dfrac{\mathbf{z}_j}{\EE\left[Z_j ^2\right]}\right)^\top \tilde{\boldsymbol{\eps}} \right\vert + \left\vert { \scriptstyle \mathcal{O}}_p\left(1\right) \right\vert \\
\mathbb{P}\left[\left\vert t_j \right\vert \geq \tau_n \right] & \leq \mathbb{P}\left[\left\vert \sqrt{n}\dfrac{1}{n \mathcal{O}_p\left(1\right)}\left(\dfrac{\tilde{\mathbf{z}}_j^3}{\EE\left[Z_j^4 \right]}-\dfrac{\mathbf{z}_j}{\EE\left[Z_j ^2\right]}\right)^\top \tilde{\boldsymbol{\eps}} \right\vert + \left\vert { \scriptstyle \mathcal{O}}_p\left(1\right) \right\vert \geq \tau_n \right] \\
& \leq \mathbb{P}\left[\left\vert \sqrt{n}\dfrac{1}{n \mathcal{O}_p\left(1\right)}\left(\dfrac{\tilde{\mathbf{z}}_j^3}{\EE\left[Z_j^4 \right]}-\dfrac{\mathbf{z}_j}{\EE\left[Z_j ^2\right]}\right)^\top \tilde{\boldsymbol{\eps}} \right\vert \geq \tau_n/2 \right] + \mathbb{P}\left[\left\vert { \scriptstyle \mathcal{O}}_p\left(1\right) \right\vert \geq \tau_n/2 \right] \\
& \leq \mathbb{P}\left[\left\vert \sqrt{n}\dfrac{1}{n}\left(\dfrac{\tilde{\mathbf{z}}_j^3}{\EE\left[Z_j^4 \right]}-\dfrac{\mathbf{z}_j}{\EE\left[Z_j ^2\right]}\right)^\top \tilde{\boldsymbol{\eps}} \right\vert \geq \tau_n \left\vert \mathcal{O}_p\left(1\right)\right\vert / 2\right] + \mathbb{P}\left[\left\vert { \scriptstyle \mathcal{O}}_p\left(1\right) \right\vert \geq \tau_n/2 \right] \\
& \leq \EE\left[\left(\left(\dfrac{\tilde{Z}_j^3}{\EE\left[Z_j^4\right]}-\dfrac{Z_j}{\EE\left[Z_j^2\right]}\right)\mathcal{E}\right)^2\right] / \left(\tau_n \left\vert \mathcal{O}_p\left(1\right)\right\vert / 2\right)^2 + \mathbb{P}\left[\left\vert { \scriptstyle \mathcal{O}}_p\left(1\right) \right\vert \geq \tau_n/2 \right],
\end{align*}
where the last step follows from Chebyshev's inequality, assuming the second moment exists (cf.\ \ref{ass:6}).
\subsection{Proof of Theorem \ref{theo:lin-SEM-betaM}}
From the definitions in \eqref{eq:lin-SEM-H-def}, we see that $\beta_j=\beta^*_j$ iff $\big(\big(\boldsymbol{\omega}_{N,M} \boldsymbol{\omega}_{M,M}^{-1} \big)^\top \boldsymbol{\beta}^*_N \big)_j=0$. We can inspect this further
\begin{equation*}
\left(\left(\boldsymbol{\omega}_{N,M} \boldsymbol{\omega}_{M,M}^{-1} \right)^\top \boldsymbol{\beta}^*_N \right)_j = \left(\left( \boldsymbol{\omega}_{M,M}^{-1} \right)^\top \boldsymbol{\omega}_{N,M}^\top \boldsymbol{\beta}^*_N \right)_j =\left( \boldsymbol{\omega}_{M,M}^{-1} \right)_j^\top \sum_{k \in N} \boldsymbol{\omega}_{k, M}^\top \beta^*_k = \sum_{k \in N} \left( \boldsymbol{\omega}_{M,M}^{-1} \right)_j^\top \boldsymbol{\omega}_{k, M}^\top \beta^*_k.
\end{equation*}
For some variable $k \in N$, we have
\begin{align*}
\EE\left[\boldsymbol{\omega}_{M,M} \boldsymbol{\Psi}_M \left(\boldsymbol{\omega}_{M,M} \boldsymbol{\Psi}_M\right)^\top\right] & = \boldsymbol{\omega}_{M,M} \boldsymbol{\Sigma}^{\boldsymbol{\Psi}_M} \boldsymbol{\omega}_{M,M}^\top \quad \text{and} \\
 \EE\left[\boldsymbol{\omega}_{M,M} \boldsymbol{\Psi}_M \left(\boldsymbol{\omega}_{k,M} \boldsymbol{\Psi}_M\right)^\top\right] & = \boldsymbol{\omega}_{M,M} \boldsymbol{\Sigma}^{\boldsymbol{\Psi}_M} \boldsymbol{\omega}_{k,M}^\top. \quad \text{Thus,} \\ 
\EE\left[\boldsymbol{\omega}_{M,M} \boldsymbol{\Psi}_M \left(\boldsymbol{\omega}_{M,M} \boldsymbol{\Psi}_M\right)^\top\right]^{-1} \EE\left[\boldsymbol{\omega}_{M,M} \boldsymbol{\Psi}_M \left(\boldsymbol{\omega}_{k,M} \boldsymbol{\Psi}_M\right)^\top\right] & =\left(\boldsymbol{\omega}_{M,M}^{-1} \right)^\top \boldsymbol{\omega}_{k, M}^\top
\end{align*}
is the regression parameter of the regression $\boldsymbol{\omega}_{k,M} \boldsymbol{\Psi}_M$ versus $\boldsymbol{\omega}_{M,M} \boldsymbol{\Psi}_M$. Naturally, $\boldsymbol{\omega}_{k,M} \boldsymbol{\Psi}_M$ can be perfectly recovered by a linear combination of $\boldsymbol{\omega}_{M,M} \boldsymbol{\Psi}_M$ using only $k$'s nearest ancestors in $M$, say, $\text{PA}^M\left(k\right)$. Thus, $\left( \boldsymbol{\omega}_{M,M}^{-1} \right)_j^\top \boldsymbol{\omega}_{k, M} = 0$ if $j \not \in \text{PA}^M\left(k\right)$. Extending this argument to all $k \in N$ the theorem's statement follows.

\subsection{Proof of Theorem \ref{theo:lin-SEM-beta-eq}}
We provide some supporting lemmata.
\begin{lemm}\label{lemm:zrep}
Assume that the data follows the model \eqref{eq:SEM-model} without hidden variables. Then,
\begin{equation*}
Z_j = \delta_{j,j} \Psi_j + \sum_{k \in \text{CH}\left(j\right)} \delta_{j,k} \Psi_k \quad j=1,\ldots ,p
\end{equation*}
for an appropriate set of parameters. Further, the support of $\boldsymbol{\gamma}_j$ (cf.\ \eqref{eq:z-w-def}) is restricted to $j$'s Markov boundary.
\end{lemm}
Thus, only the ``noise'' of $j$ itself or its children remains in $Z_j$. 

Now, consider the best regression of $Z_j^3$ versus $\mathbf{X}_{-j}$ as defined in \eqref{eq:ztilde-def}. We get an analogous result for the residuum $\tilde{Z}_j^3$.
\begin{lemm}\label{lemm:ztilde-rep}
Assume that the data follows the model \eqref{eq:SEM-model} without hidden variables. Then,
\begin{equation*}
\tilde{Z}_j^3 =Z_j^3 + \tilde{\delta}_{j,j} \Psi_j + \sum_{k \in \text{CH}\left(j\right)} \tilde{\delta}_{j,k} \Psi_k \quad j=1,\ldots ,p
\end{equation*}
for an appropriate set of parameters. Further, the support of $\tilde{\boldsymbol{\gamma}}_j$ (cf.\ \eqref{eq:ztilde-def}) is restricted to $j$'s Markov boundary.
\end{lemm}
Finally, we inspect the regression of $\Psi_k$ versus $\mathbf{X}_M$ for some $k \in N$. With a slight abuse of notation, define
\begin{equation}\label{eq:zk-def}
Z_k \coloneqq \Psi_k- \mathbf{X}_{M}^{\top}\boldsymbol{\beta}^k, \quad \text{where}\quad \boldsymbol{\beta}^k \coloneqq \underset{\mathbf{b} \in \mathbb{R}^{\left\vert M \right\vert}}{\text{argmin}}\EE\left[\left(\Psi_k-\mathbf{X}_{M}^\top \mathbf{b}\right)^2\right] = \EE\left[\mathbf{X}_{M}\mathbf{X}_{M}^\top\right]^{-1}\EE\left[\mathbf{X}_{M} \Psi_k\right].
\end{equation}
\begin{lemm}\label{lemm:zk-rep}
Assume that the data follows the model \eqref{eq:SEM-model}. Let $\mathbf{X}_M$ and $\mathbf{X}_N$ be the observed and the hidden variables, where $k \in N$. Then,
\begin{equation*}
Z_k = \sum_{l \in N}{\delta_{k,l} \Psi_l} + \sum_{m \in \text{CH}_N}{ \delta_{k,m} \Psi_m}, \quad \text{where} \quad \text{CH}_N = \left( \underset{l \in N}{\cup }\text{CH}\left(l\right)\right) \setminus N ,
\end{equation*}
for an appropriate set of parameters. Further, the support of $\boldsymbol{\beta}^k$ is restricted to the union of the hidden variables' Markov boundaries.
\end{lemm}
As $\tilde{\mathcal{E}}$ is a linear combination of these $Z_k$ and the independent $\mathcal{E}$, we can combine Lemmata \ref{lemm:zrep} and \ref{lemm:ztilde-rep} and \ref{lemm:zk-rep} to find $Z_j \perp \tilde{\mathcal{E}}$ and $\tilde{Z}_j^3 \perp \tilde{\mathcal{E}}$ for some variable $j$ outside the hidden variables' Markov boundaries. Furthermore, $\boldsymbol{\beta}^{OLS} - \boldsymbol{\beta}$ is a linear combination of the $\boldsymbol{\beta}^k$ for $k \in N$ such that outside the hidden variables' Markov boundaries the two parameters are equal as claimed.

To check \ref{ass:LLN-cond}, split $X_k$ into a part consisting of $\Psi_j$ and $\Psi_l \ \forall l \in \text{CH}\left(j\right)$, say $X_{k,1}$, independent from $\tilde{\mathcal{E}}$, and the remainder, say $X_{k,2}$, independent from $Z_j$. Then, we find
\begin{align*}
\EE\left[Z_j ^2 X_k \tilde{\mathcal{E}}\right] & =\EE\left[Z_j ^2 X_{k,1}\tilde{\mathcal{E}}\right] + \EE\left[Z_j ^2 X_{k,2}\tilde{\mathcal{E}}\right] = \EE\left[Z_j ^2 X_{k,1}\right] \EE\left[\tilde{\mathcal{E}}\right] + \EE\left[Z_j ^2 \right] \EE\left[ X_{k,2}\tilde{\mathcal{E}}\right] \\
& = \EE\left[Z_j ^2 \right] \EE\left[ X_{k,2}\tilde{\mathcal{E}}\right] = \EE\left[Z_j ^2 \right] \left(\EE\left[ X_k\tilde{\mathcal{E}}\right]-\EE\left[ X_{k,1}\tilde{\mathcal{E}}\right]\right) =0.
\end{align*}
\subsubsection{Proof of Lemma \ref{lemm:zrep}}\label{app:lemm:zrep}
Recall the representation
\begin{equation}\label{eq:z-rep}
Z_j = \delta_{j,j} \Psi_j + \sum_{k \in \text{CH}\left(j\right)} \delta_{j,k} \Psi_k.
\end{equation}
Assume first only the noise terms $\Psi_j$ and $\Psi_k \ \forall k \in \text{CH}_j$ exist, while all the other terms are set to $0$. Call the variables in this construction $X'_k$ and the residuum $Z'_j$. Obviously, $Z'_j$ has a representation as in \eqref{eq:z-rep}. Now by the definition of least squares, $Z'_j$ and $Z_j$ always have the smallest possible variance in their given model. If we add more independent noise terms to the model, the variance cannot decrease. Therefore, it holds $\text{Var}\left(Z_j\right) \geq \text{Var}\left(Z'_j\right)$. Thus, if there exists a parameter such that $X_j - \boldsymbol{\gamma}^\top \mathbf{X}_{-k} = Z'_j$, it must be optimal such that $Z_j=Z'_j$. Let now $\boldsymbol{\gamma} = -\boldsymbol{\delta}_j$. Then, we have 
\begin{align*}
X_j-\sum_{k \neq j} \gamma_k X_k & =X_j +\sum_{k \in \text{CH}_j} \delta_{j,k} X_k =X_j+\sum_{k \in \text{CH}\left(j\right)} \delta_{j,k} \left( \Psi_k + \theta_{k,j} X_j + \sum_{l \in \text{PA}\left(k\right) \setminus j} \theta_{k,l} X_l \right)\\
& = X_j \left(1 + \sum_{k \in \text{CH}\left(j\right)} \delta_{j,k} \theta_{k,j} \right) +\sum_{k \in \text{CH}\left(j\right)} \delta_{j,k} \Psi_k +\sum_{k \in \text{CH}\left(j\right)} \delta_{j,k} \sum_{l \in \text{PA}\left(k\right) \setminus j} \theta_{k,l}X_l.
\end{align*}
Now adjust $\boldsymbol{\gamma}$ by
\begin{itemize}
\item $\forall l \in \text{PA}\left(j\right)$ adding $\left(1 + \sum_{k \in \text{CH}\left(j\right)} \delta_{j,k} \theta_{k,j} \right)\theta_{j,l}$ to $\gamma_l$\\
\item $\forall k \in \text{CH}\left(j\right), \quad \forall l \in \text{PA}\left(k\right) \setminus j$ adding $\delta_{j,k} \theta_{k,l}$ to $\gamma_l$
\end{itemize}
This leads to
\begin{equation*}
X_j-\sum_{k \neq j} \gamma_k X_k = \Psi_j \left(1 + \sum_{k \in \text{CH}\left(j\right)} \delta_{j,k} \theta_{k,j} \right) +\sum_{k \in \text{CH}\left(j\right)} \delta_{j,k} \Psi_k.
\end{equation*}
This is almost the optimal $Z'_j$ as in \eqref{eq:z-rep}. It remains to argue that the term in the bracket equals $\delta_{j,j}$. For this, note that the weighted sum of terms that include $\Psi_k$ in the construction of $Z'_j$ must be exactly $\delta_{j,k}$. These terms can occur from adding a multiple of either $k$ itself or of descendants thereof (that are children of $j$ as well). These descendants have ``inherited'' the same multiple of $\Psi_k$ as of $\theta_{k,j}\Psi_j$. Therefore, there is a net contribution of $\delta_{j,k} \theta_{k,j}\Psi_j$ originating from variable $k$. Applying this argument to each child, we receive the desired sum. Naturally, the $1$ is the contribution from $X'_j = \Psi_j$ itself.

Thus, we receive the desired construction of $Z_j$. Further, we see that in this construction the support of $\boldsymbol{\gamma}_j$ is restricted to $j$'s parents, its children, and its children's other parents, which is exactly the second part of the lemma.
\subsubsection{Proof of Lemma \ref{lemm:ztilde-rep}}
This follows using very similar arguments as in Section \ref{app:lemm:zrep} and is omitted here for simplicity.
\subsubsection{Proof of Lemma \ref{lemm:zk-rep}}
Recall the construction
\begin{equation}\label{eq:zk-rep}
Z_k = \sum_{l \in N}{\delta_{k,l} \Psi_l} + \sum_{m \in \text{CH}_N}{ \delta_{k,m} \Psi_m}, \quad \text{where} \quad \text{CH}_N = \left( \underset{l \in N}{\cup }\text{CH}\left(l\right) \right) \setminus N.
\end{equation}
We argue as before: Assume first only these variables are nonzero leading to an optimal residuum $Z'_k$ which has a representation as in \eqref{eq:zk-rep}. Naturally, $Z_k = \Psi_k - \left(\boldsymbol{\beta}^k\right)^\top \mathbf{X}_M$ for some $\boldsymbol{\beta}^k$. If we find a parameter such that $\Psi_k -\left(\boldsymbol{\beta}^k\right)^\top \mathbf{X}_M = Z'_k$, it must be optimal as $\text{Var}\left(Z_k\right) \geq \text{Var}\left(Z'_k\right)$. We now construct such a parameter. Start by $\boldsymbol{\beta}^k = -\boldsymbol{\delta}_k$.
\begin{align*}
- \left(\boldsymbol{\beta}^k\right)^\top\mathbf{X}_M & = \sum_{m \in \text{CH}_N}{ \delta_{k,m} X_m} = \sum_{m \in \text{CH}_N}{ \delta_{k,m}\left( \mathcal{E}_{X_m} +\boldsymbol{\omega}_{m,N}\boldsymbol{\Psi}_N \right)} \\
& = \sum_{l \in N}{\Psi_l}\sum_{m \in \text{CH}_N}{\delta_{k,m}\omega_{m,l}}+ \sum_{m \in \text{CH}_N}{ \delta_{k,m}\mathcal{E}_{X_m}} \\
& = \sum_{l \in N}{\Psi_l\sum_{m \in \text{CH}_N}{\delta_{k,m}\omega_{m,l}}}+ \sum_{m \in \text{CH}_N}{ \delta_{k,m}\left(\Psi_m + \sum_{r \in \text{PA}^M\left(m\right)}{\theta'_{mr}\mathcal{E}_{X_r}}\right)} \\
& = \sum_{l \in N}{\Psi_l\sum_{m \in \text{CH}_N}{\delta_{k,m}\omega_{m,l}}}+ \sum_{m \in \text{CH}_N}{ \delta_{k,m}\left(\Psi_m + \sum_{r \in \text{PA}^M\left(m\right)}{\theta'_{mr}\left(X_r - \boldsymbol{\omega}_{r,H}\boldsymbol{\Psi}_H\right)}\right)} \\
& = \sum_{l \in N}{\Psi_l\sum_{m \in \text{CH}_N}{\delta_{k,m}}\left(\omega_{m,l}- \sum_{r \in \text{PA}^M\left(m\right)}{\theta'_{mr} \omega}_{r,l} \right)}\\
& + \sum_{m \in \text{CH}_N}{ \delta_{k,m}\left(\Psi_m + \sum_{r \in \text{PA}^M\left(m\right)}{\theta'_{mr}X_r}\right)},
\end{align*}
where we have used the fact that $\mathcal{E}_{\boldsymbol{X}}$ follows a linear SEM as well for some suitable set of parameters $\theta'_{m,r}$.
Now adjust $\boldsymbol{\beta}^k$ by
\begin{itemize}
\item $\forall m \in \text{CH}_N, \quad \forall r \in \text{PA}^M\left(m\right) $ adding $\delta_{k,m} \theta'_{m,r}$ to $\beta^k_r$.
\end{itemize}
This leads to
\begin{equation*}
\Psi_k - \left(\boldsymbol{\beta}^k\right)^\top \mathbf{X}_M = \Psi_k + \sum_{l \in N}{\Psi_l\sum_{m \in \text{CH}_N}{\delta_{k,m}}\left(\omega_{m,l}- \sum_{r \in \text{PA}^M\left(m\right)}{\theta'_{mr} \omega_{r,l}} \right)}+ \sum_{m \in \text{CH}_N}{ \delta_{k,m}\Psi_m},
\end{equation*}
which is as in \eqref{eq:zk-rep}. It remains to argue that the coefficient for $\Psi_l$ equals $\delta_{k,l}$. Note that $ \forall m \in \text{CH}_N$ there is a net contribution of $\delta_{k,m}\Psi_m$ coming from a weighted sum of $m$ and its descendants. There must be an according net contribution of all other parts that $X'_m$ does not inherit from its parents (in $M$). These must be multiples of $\Psi_l$ for $l \in N$. If $\omega_{r,l} \neq 0$, there is already a multiple of $\Psi_l$ in $X'_r$. Thus, $X'_m$ inherits $\theta'_{mr}\omega_{r,l}$ from $X'_r$. Extending this argument to all parents, a total of $ \sum_{r \in \text{PA}^M\left(m\right)}{\theta'_{mr} \omega_{r,l}}$ is inherited. The remainder, i.e., $\omega_{m,l} - \sum_{r \in \text{PA}^M\left(m\right)}{\theta'_{mr} \omega_{r,l}}$ must then originate from $X'_k$ itself and there is a contribution of $\delta_{k,m}$ times this remainder times $\Psi_l$. As this holds $ \forall m \in \text{CH}_N$, one can add all contributions leading to the desired sum.

Thus, we have established that $Z_k = Z'_k$ such that $Z_k$ and $\boldsymbol{\beta}^k$ must be optimal. We also see that the support of $\boldsymbol{\beta}^k$ is restricted to $\text{CH}_N$ and $ \underset{m \in \text{CH}_N}{\cup }\text{PA}^M\left(m\right)$, which is exactly the hidden variables' Markov boundary.

\section{Theory for the high-dimensional extension}\label{app:HD}
We provide additional details on the high-dimensional extension including proofs of the main results.

To understand Theorem \ref{theo:HD-local-gauss}, take a closer look at the difference between $\hat{\beta}_j^{OLS}$ and $\hat{\beta}_j^{HOLS}$, which can be written as
\begin{align} 
\begin{split}\label{eq:HOdiff-alt}
\sqrt{n}\left(\hat{\beta}^{HOLS}_j-\hat{\beta}^{OLS}_j\right) = & \sqrt{n} \dfrac{\left(\hat{\tilde{\mathbf{z}}}_j^3\right)^\top \mathbf{x}_{-j}/n}{\left(\hat{\tilde{\mathbf{z}}}_j^3\right)^\top \mathbf{x}_j/n}\left(\boldsymbol{\beta}_{-j}-\hat{\boldsymbol{\beta}}_{-j}\right) -\sqrt{n}\dfrac{\hat{\mathbf{z}}_j^\top \mathbf{x}_{-j}/n}{\hat{\mathbf{z}}_j^\top \mathbf{x}_j/n}\left(\boldsymbol{\beta}_{-j}-\hat{\boldsymbol{\beta}}_{-j}\right) + \\ & \left( \dfrac{\left(\hat{\tilde{\mathbf{z}}}_j^3\right)^\top /\sqrt{n}}{\left(\hat{\tilde{\mathbf{z}}}_j^3\right)^\top \mathbf{x}_j/n} - \dfrac{\hat{\mathbf{z}}_j^\top/\sqrt{n}}{\hat{\mathbf{z}}_j^\top\mathbf{x}_j /n}\right)\boldsymbol{\eps} \coloneqq \Delta_j^{HOLS} - \Delta_j^{OLS} + \sqrt{n}\dfrac{1}{n}\hat{\mathbf{v}}_j^\top \boldsymbol{\eps}
\end{split}\\
\sqrt{n} \left(\hat{\boldsymbol{\beta}}^{HOLS} - \hat{\boldsymbol{\beta}}^{HOLS}\right) = & \boldsymbol{\Delta}^{HOLS} - \boldsymbol{\Delta}^{OLS} + \sqrt{n}\dfrac{1}{n}\hat{\mathbf{v}}^\top \boldsymbol{\eps}. \label{eq:HOdiff-alt-pD}
\end{align}
Thus, it consists of two bias terms and an error term, whose variance we can estimate. In the following, we inspect the terms in \eqref{eq:HOdiff-alt} assuming model \eqref{eq:mvmodel} to justify Algorithm \ref{alg:HOLS-HD-alt} and Theorem \ref{theo:HD-local-gauss}. $\Delta_j^{OLS}$ is under control under certain conditions as discussed in \citep{van2014asymptotically}. For $\Delta_j^{HOLS} $ as well as the error scaling, we invoke our extra assumptions. 

We consider the bias term.
\begin{lemm}\label{lemm:HD-par-est-conv-alt}
Assume that the data follows the model \eqref{eq:mvmodel} with sub-Gaussian $\mathcal{E}$ and that \ref{ass:B-cov} - \ref{ass:B-ssj} and \ref{ass:B-slam} - \ref{ass:B-sjlam} hold ($\forall j$). Let $\hat{\boldsymbol{\beta}}$ come from Lasso regression with $\lambda \asymp \sqrt{\log\left(p\right)/n}$, $\hat{\mathbf{z}}_j$ from nodewise Lasso regression using $\lambda_j \asymp \sqrt{\log\left(p\right)/n}$, and $\hat{\tilde{\mathbf{z}}}_j^3$ from nodewise Lasso regression of $\hat{\mathbf{z}}_j^3$ versus $\mathbf{x}_{-j}$ using \\$\tilde{\lambda}_j \asymp \text{max}\left\{\log\left(p\right)^{5/2}n^{-1/2},s_j^2 \log\left(p\right)^{5/2} n^{-3/2}, s_j \log\left(p\right)^{2} n^{-1}, \sqrt{s_j} \log\left(p\right) n^{-1/2} \right\}$. Use the definitions in \eqref{eq:HOdiff-alt}. Then,
\begin{equation*}
\left\Vert \boldsymbol{\Delta}^{HOLS} \right\Vert_\infty = {\scriptstyle \mathcal{O}}_p\left(1\right) \quad \text{and} \quad \left\Vert \boldsymbol{\Delta}^{OLS} \right\Vert_\infty = {\scriptstyle \mathcal{O}}_p\left(1\right).
\end{equation*}
\end{lemm}
Thus, under suitable assumptions, the bias vanishes. To get powerful tests, we want the variance of the error term to stay bounded. For asymptotically valid tests, we must ensure that the estimated standard deviation is of higher order of magnitude than the bias term.
\begin{lemm}\label{lemm:HD-var-est-conv-alt}
Assume that the data follows the model \eqref{eq:mvmodel} with sub-Gaussian $\mathcal{E}$ and that \ref{ass:B-cov}, \ref{ass:B-s}, \ref{ass:B-sj} and \ref{ass:B-sjlam} hold ($\forall j$). Let $\hat{\boldsymbol{\beta}}$ come from Lasso regression with $\lambda \asymp \sqrt{\log\left(p\right)/n}$, $\hat{\mathbf{z}}_j$ from nodewise Lasso regression using $\lambda_j \asymp \sqrt{\log\left(p\right)/n}$, and $\hat{\tilde{\mathbf{z}}}_j^3$ from nodewise Lasso regression of $\hat{\mathbf{z}}_j^3$ versus $\mathbf{x}_{-j}$ using \\$\tilde{\lambda}_j \asymp \text{max}\left\{\log\left(p\right)^{5/2}n^{-1/2},s_j^2 \log\left(p\right)^{5/2} n^{-3/2}, s_j \log\left(p\right)^{2} n^{-1}, \sqrt{s_j} \log\left(p\right) n^{-1/2} \right\}$. Use the definitions in \eqref{eq:HOdiff-alt}. Then,
\begin{equation*}
\dfrac{1}{n} \hat{\mathbf{v}}_j ^\top \hat{\mathbf{v}}_j \overset{\PP}{\to} \EE\left[V_j^2\right] \quad \text{uniformly in } j
\end{equation*}
where $V_j = \dfrac{\tilde{Z}_j^3}{\EE\left[Z_j^4\right]}- \dfrac{Z_j}{\EE\left[Z_j^2\right]}$.
\end{lemm}
Note that the sub-Gaussian assumption ensures $\EE\left[V_j^2\right] < \infty$, while as $\EE\left[V_j^2\right] > 0$ if \\$\EE\left[\left(\tilde{Z}_j^3\right)^2\right]\EE\left[Z_j^2\right] > \EE\left[Z_j^4\right]^2$, which always holds if $\tilde{Z}_j^3$ is not a linear function of $Z_j$. Thus, the estimate of the standard error approaches a bounded positive constant, enabling asymptotically valid $z$-tests.

In \ref{ass:B-ssj}, we have joint conditions on the different sparsity levels $s$ and $s_j$. Thus, the larger the one is, the more restrictive the assumption on the other is. Let us consider some specific cases, namely, $s\approx s_j$, $s_j$ maximal according to \ref{ass:B-sj}, and $s$ maximal according to \ref{ass:B-s}.
\begin{align*}
\text{For } & s\approx s_j, \text{ we need } s\approx s_j={\scriptstyle \mathcal{O}}\left(\dfrac{n^{1/3}}{\log\left(p\right)}\right). \\
\text{For } & s_j={\scriptstyle \mathcal{O}}\left(\dfrac{n^{3/5}}{\log\left(p\right)}\right), \text{ we need } s=\mathcal{O}\left(\dfrac{n^{1/5}}{\log\left(p\right)}\right). \\
\text{For } & s={\scriptstyle \mathcal{O}}\left(\dfrac{n^{1/2}}{\log\left(p\right)^{3}}\right), \text{ we need } s_j=\mathcal{O}\left(\log \left(p\right)^3\right). \\
\end{align*}
Note that if $\tilde{\lambda}_j$ is chosen optimally with respect to $s_j$, \ref{ass:B-slam} is actually the same as \ref{ass:B-s} and \ref{ass:B-ssj} such that there is no extra assumption on $s$ that one has to invoke. In \ref{ass:B-sjlam}, we have joint conditions on the different sparsity levels $s_j$ and $\tilde{s}_j$. Thus, the larger the one is, the more restrictive the assumption on the other is. Let us consider some specific cases, namely, $ s_j\approx \tilde{s}_j$, $s_j$ maximal according to \ref{ass:B-sj}, and $\tilde{s}_j$ maximal according to \ref{ass:B-sjlam}.
\begin{align*}
\text{For } & s_j\approx \tilde{s}_j, \text{ we need } s_j\approx \tilde{s}_j={\scriptstyle \mathcal{O}}\left(\dfrac{n^{1/2}}{\log\left(p\right)}\right). \\
\text{For } & s_j={\scriptstyle \mathcal{O}}\left(\dfrac{n^{3/5}}{\log\left(p\right)}\right), \text{ we need } \tilde{s}_j=\mathcal{O}\left(\dfrac{n^{2/5}}{\log\left(p\right)}\right). \\
\text{For } & \tilde{s}_j={\scriptstyle \mathcal{O}}\left(\dfrac{n}{\log\left(p\right)^5}\right), \text{ we need } s_j=\mathcal{O}\left(\log \left(p\right)^3\right). 
\end{align*}
Naturally, $s_j$ and $\tilde{s}_j$ are to some extent related. In a linear SEM, the support of $\boldsymbol{\gamma}_j$ and the support of $\tilde{\boldsymbol{\gamma}}_j$ always lie within $j$'s Markov boundary as we argue in Section \ref{lin-SEM}. For completely arbitrary setups, it is typically even all of the boundary. Thus, $s_j=\tilde{s}_j$ would then be usual, except for ``sink'' nodes, such that the first case appears to be most interesting. Furthermore, if $\EE\left[Z_j^3 \mathbf{X}_{-j}\right]= \mathbf{0}\ \forall j$, it holds $\tilde{s}_j=0$ and \ref{ass:B-sjlam} is automatically fulfilled.
\subsection{Proof of Lemma \ref{lemm:HD-par-est-conv-alt}}
Following the proofs in \cite{van2014asymptotically}, the assumptions are sufficient to claim
\begin{align*}
\left\Vert \boldsymbol{\Delta}^{OLS}\right\Vert_\infty & ={\scriptstyle \mathcal{O}}_p\left(1\right), \quad \left\Vert \hat{\boldsymbol{\beta}} - \boldsymbol{\beta} \right \Vert_1 = \mathcal{O}_p\left(s \lambda\right), \quad \dfrac{1}{n} \left\Vert \mathbf{x}\left(\hat{\boldsymbol{\beta}} - \boldsymbol{\beta}\right) \right \Vert_2^2 = \mathcal{O}_p\left(s \lambda^2\right) , \\
 \quad \left\Vert \hat{\boldsymbol{\gamma}}_j - \boldsymbol{\gamma}_j \right \Vert_1 & = \mathcal{O}_p\left(s_j \lambda_j\right) \ \forall j \quad \text{and} \quad \dfrac{1}{n} \left\Vert \mathbf{x}_{-j}\left(\hat{\boldsymbol{\gamma}}_j - \boldsymbol{\gamma}_j\right) \right \Vert_2^2 = \mathcal{O}_p\left(s_j \lambda_j^2\right) \ \forall j.
\end{align*}
We now turn to $\left\Vert \boldsymbol{\Delta}^{HOLS}\right\Vert_\infty$. To control this, we want to ensure that
\begin{equation} \label{eq:goal-alt}
\left\vert\left(\hat{\tilde{\mathbf{z}}}_j^3\right)^\top\mathbf{x}_{-j}\left(\boldsymbol{\beta}_{-j} -\hat{\boldsymbol{\beta}}_{-j}\right)\right\vert/n = { \scriptstyle \mathcal{O}}_p\left(1/\sqrt{n}\right) \quad \text{and} \quad \left(\hat{\tilde{\mathbf{z}}}_j^3\right)^\top\mathbf{x}_j/n = \EE\left[Z_j^4\right] + {\scriptstyle \mathcal{O}}_p\left(1\right).
\end{equation}
Note that we always have $\big\Vert\left(\hat{\tilde{\mathbf{z}}}_j^3\right)^\top \mathbf{x}_{-j}\big\Vert_\infty /n = \tilde{\lambda}_j$. Thus, the first goal in \eqref{eq:goal-alt} is fulfilled using \ref{ass:B-s} and \ref{ass:B-ssj}. Further,
\begin{equation*}
\hat{\mathbf{z}}_j^3 = \mathbf{z}_j^3 + \left(\hat{\mathbf{z}}_j^3-\mathbf{z}_j^3\right) = \mathbf{x}_{-j}\tilde{\boldsymbol{\gamma}}_j + \tilde{\mathbf{z}}_j^3 + \left(\hat{\mathbf{z}}_j^3-\mathbf{z}_j^3\right).
\end{equation*}
From standard Lasso theory (cf.\ \cite{buhlmann2011statistics}), we know that the order of which we should choose the tuning parameter $\tilde{\lambda}_j$ is dependent on the bound for
\begin{align*}
\dfrac{1}{n}\left\Vert \mathbf{x}_{-j} \left(\tilde{\mathbf{z}}_j^3 + \left(\hat{\mathbf{z}}_j^3-\mathbf{z}_j^3\right)\right)\right\Vert_\infty  \leq & \dfrac{1}{n}\left\Vert \mathbf{x}_{-j} \tilde{\mathbf{z}}_j^3 \right\Vert_\infty + \dfrac{1}{n}\left\Vert \mathbf{x}_{-j} \left(\hat{\mathbf{z}}_j^3-\mathbf{z}_j^3\right)\right\Vert_\infty \\ 
\leq & \dfrac{1}{n}\left\Vert \mathbf{x}_{-j} \tilde{\mathbf{z}}_j^3 \right\Vert_\infty + \dfrac{1}{n}\left\Vert \mathbf{x}_{-j} \right\Vert_\infty \left\Vert\hat{\mathbf{z}}_j^3-\mathbf{z}_j^3\right\Vert_1.
\end{align*}
For the first term, 
\begin{align*}
\EE\left[\underset{j}{\text{max}}\left\Vert\left(\tilde{\mathbf{z}}_j^3 \right)^\top\mathbf{x}_{-j}\right\Vert_\infty\right]= &\EE\left[\underset{j,k\neq j}{\text{max}}\left\vert\left(\tilde{\mathbf{z}}_j^3 \right)^\top\mathbf{x}_k\right\vert\right]= \EE\left[\underset{j,k\neq j}{\text{max}}\left\vert\sum_{i=1}^n \tilde{z}_{ij}^3 x_{ik}\right\vert\right] \\
 = & \EE\left[\underset{j,k\neq j}{\text{max}}\left\vert\sum_{i=1}^n \tilde{z}_{ij}^3 x_{ik} -\EE\left[\tilde{z}_{ij}^3 x_{ik}\right]\right\vert\right].
\end{align*}
We maximize over $j$ as well to receive results uniformly in $j$. In the last equality, we use $\EE\left[\tilde{z}_{ij}^3 x_{ik}\right]=\EE\left[\tilde{Z}_{j}^3 X_{k}\right]=0$.

The terms of the type $\tilde{z}_{ij}^3 x_{ik}$ can be viewed as different functions of the vector \\ $\begin{pmatrix}
x_{i1} & \ldots & x_{ip} & \tilde{z}_{i1}^3 & \ldots & \tilde{z}_{ip}^3
\end{pmatrix}^\top$. In total, these are $p\left(p-1\right)$ functions. For these, we can apply the Nemirovski moment inequality from Lemma 14.24 in \cite{buhlmann2011statistics}, which yields
\begin{align*}
& \EE\ \underset{j,k}{\text{max}}\left\vert \sum_{i=1}^n \tilde{z}_{ij}^3 x_{ik}-\EE\left[\tilde{z}_{ij}^3 x_{ik}\right]\right\vert \leq \left(8\log\left(2p\left(p-1\right)\right)\right)^{1/2}\ \EE\left[\underset{j,k}{\text{max}}\sum_{i=1}^n\left(\tilde{z}_{ij}^3 x_{ik}\right)^2\right]^{1/2} \leq \\ & \left(8\log\left(2p\left(p-1\right)\right)\right)^{1/2}\ \EE\left[\sum_{i=1}^n\underset{j,k}{\text{max}} \left(\tilde{z}_{ij}^3\right)^2 x_{ik}^2\right]^{1/2} = \left(8\log\left(2p\left(p-1\right)\right)n\right)^{1/2}\ \EE\left[\underset{j,k}{\text{max}} \left(\tilde{Z}_{j}^3\right)^2 X_{k}^2\right]^{1/2} \\ \leq & \left(8\log\left(2p\left(p-1\right)\right)n\right)^{1/2}\ \EE\left[\underset{k}{\text{max}} X_{k}^8 \right]^{1/2}.
\end{align*}
In the last expression, we simplify the notation and let $k \in \left\{1,\ldots,2p\right\}$ with $X_{p+j}=\left(\tilde{Z}_{j}^3\right)^{1/3}$.
We aim to bound that last expectation term, for which we use the sub-Gaussian assumption.
\begin{align*}
 \EE\left[\underset{k}{\text{max}} X_{k}^8 \right] & = \int_0^{\infty}\mathbb{P}\left(\underset{k}{\text{max}} X_{k}^8 > t \right)dt=\int_0^{\infty}\mathbb{P}\left(\underset{k}{\text{max}} \left\vert X_{k}\right\vert > t^{1/8} \right)dt \\ &
 \leq \int_0^{\infty} \text{min}\left\{1, \sum_k \mathbb{P}\left(\left\vert X_{k}\right\vert > t^{1/8} \right)\right\}dt  \leq \int_0^{\infty} \text{min}\left\{1, 2p\underset{k}{\text{ max}} \mathbb{P}\left(\left\vert X_{k}\right\vert > t^{1/8} \right)\right\}dt \\ &
 \leq \int_0^{\infty} \text{min}\left\{1, 4p\underset{k}{\text{ max}} \ \text{exp}\left(-\dfrac{t^{1/4}}{2\sigma_k^2}\right)\right\} dt  \leq \int_0^{a} 1 dt + 4p \int_a^{\infty} \text{exp}\left(-\dfrac{t^{1/4}}{2\sigma_{\text{max}}^2}\right) dt \\ & 
 = a +p \text{ exp}\left(-\dfrac{a^{1/4}}{2\sigma_{\text{max}}^2}\right)\text{poly}\left(a\right) = a + \text{exp}\left(-\dfrac{a^{1/4}}{2\sigma_{\text{max}}^2} + \log\left(p\right)\right)\text{poly}\left(a\right)
\end{align*}
This holds for any positive integration bound $a$. If we choose $a>16\sigma_{\text{max}}^8 \log\left(p\right)^4$, the second term will vanish as $p \rightarrow \infty$ leading to
\begin{align*}
\EE\left[\underset{k}{\text{max}} X_{k}^8 \right] & \leq \mathcal{O}\left(\log\left(p\right)^4\right) \ \text{such that} \\ \EE\left[\underset{j}{\text{max}}\dfrac{1}{n}\left\Vert\left(\tilde{\mathbf{z}}_j^3 \right)^\top\mathbf{x}_{-j}\right\Vert_\infty\right] & \leq \dfrac{1}{n} \left(8\log\left(2p\left(p-1\right)\right)n\right)^{1/2}\left(\mathcal{O}\left(\log\left(p\right)^4\right)\right)^{1/2} = \mathcal{O}\left(\log\left(p\right)^{5/2}n^{-1/2}\right)\\
\dfrac{1}{n}\left\Vert\left(\tilde{\mathbf{z}}_j^3 \right)^\top\mathbf{x}_{-j}\right\Vert_\infty & = \mathcal{O}_p\left(\log\left(p\right)^{5/2}n^{-1/2}\right) \quad \text{uniformly in }j.
\end{align*}
The last conclusion is a simple application of Markov's inequality. We now turn to the second term to be bounded
\begin{align*}
\dfrac{1}{n}\left\Vert\left(\hat{\mathbf{z}}_j^3 -\mathbf{z}_j^3 \right)^\top \mathbf{x}_{-j}\right\Vert_\infty & \leq \dfrac{1}{n}\left\Vert\hat{\mathbf{z}}_j^3 -\mathbf{z}_j^3 \right\Vert_1\left\Vert\mathbf{x}_{-j}\right\Vert_\infty \\ \left\Vert\mathbf{x}_{-j}\right\Vert_\infty & = \mathcal{O}_p \left(\sqrt{\log \left(p\right)}\right) \quad \text{from the sub-Gaussian assumption.}
\end{align*}
Thus,
\begin{align*}
&\dfrac{1}{n}\left\Vert\hat{\mathbf{z}}_j^3 -\mathbf{z}_j^3 \right\Vert_1  = \dfrac{1}{n}\left\Vert \left(\hat{\mathbf{z}}_j-\mathbf{z}_j\right)^3 + 3\mathbf{z}_j \odot \left(\hat{\mathbf{z}}_j-\mathbf{z}_j\right)^2 + 3\mathbf{z}_j^2 \odot \left(\hat{\mathbf{z}}_j-\mathbf{z}_j\right)\right\Vert_1\\
\leq &\dfrac{1}{n}\left\Vert \left(\hat{\mathbf{z}}_j-\mathbf{z}_j\right)^3\right\Vert_1 + \dfrac{3}{n}\left\Vert \mathbf{z}_j \odot \left(\hat{\mathbf{z}}_j-\mathbf{z}_j\right)^2\right\Vert_1 + \dfrac{3}{n}\left\Vert \mathbf{z}_j^2 \odot \left(\hat{\mathbf{z}}_j-\mathbf{z}_j\right)\right\Vert_1\ \\
 \leq &\dfrac{1}{n}\left\Vert \hat{\mathbf{z}}_j - \mathbf{z}_j \right\Vert_\infty \left\Vert \left(\hat{\mathbf{z}}_j-\mathbf{z}_j\right)^2\right\Vert_1 + \dfrac{3}{n}\left\Vert \mathbf{z}_j \right\Vert_\infty\left\Vert \left(\hat{\mathbf{z}}_j-\mathbf{z}_j\right)^2\right\Vert_1 + \dfrac{3}{n}\left\Vert \mathbf{z}_j^2 \odot \left(\hat{\mathbf{z}}_j-\mathbf{z}_j\right)\right\Vert_1 \\
 \leq &\left\Vert \hat{\mathbf{z}}_j - \mathbf{z}_j \right\Vert_\infty \dfrac{1}{n} \left\Vert \hat{\mathbf{z}}_j - \mathbf{z}_j \right\Vert_2^2 + 3\left\Vert \mathbf{z}_j \right\Vert_\infty \dfrac{1}{n}\left\Vert \hat{\mathbf{z}}_j - \mathbf{z}_j \right\Vert_2^2+ 3\sqrt{\dfrac{1}{n} \left\Vert \mathbf{z}_j ^2 \right\Vert_2^2 \dfrac{1}{n} \left\Vert \hat{\mathbf{z}}_j - \mathbf{z}_j \right\Vert_2^2} \\
& \leq \left\Vert \mathbf{x}_{-j}\right\Vert_\infty \left\Vert \hat{\boldsymbol{\gamma}}_j - \boldsymbol{\gamma}_j \right\Vert_1 \dfrac{1}{n} \left\Vert \hat{\mathbf{z}}_j - \mathbf{z}_j \right\Vert_2^2 + + 3\left\Vert \mathbf{z}_j \right\Vert_\infty \dfrac{1}{n}\left\Vert \hat{\mathbf{z}}_j - \mathbf{z}_j \right\Vert_2^2+ 3\sqrt{\dfrac{1}{n} \left\Vert \mathbf{z}_j ^2 \right\Vert_2^2 \dfrac{1}{n} \left\Vert \hat{\mathbf{z}}_j - \mathbf{z}_j \right\Vert_2^2}\\
& = \mathcal{O}_p \left(\sqrt{\log\left(p\right)}s_j^2 \lambda_j ^3\right) + \mathcal{O}_p \left(\sqrt{\log\left(p\right)} s_j\lambda_j^2\right)+ \mathcal{O}_p \left(\sqrt{s_j\lambda_j^2}\right).
\end{align*}
In summary,
\begin{equation*}
\dfrac{1}{n}\left\Vert \mathbf{x}_{-j} \left(\tilde{\mathbf{z}}_j^3 + \left(\hat{\mathbf{z}}_j^3-\mathbf{z}_j^3\right)\right)\right\Vert_\infty =\mathcal{O}_p\left(\log\left(p\right)^{5/2}n^{-1/2} + \log\left(p\right)s_j^2 \lambda_j ^3 + \log\left(p\right) s_j\lambda_j^2 + \sqrt{\log\left(p\right)s_j\lambda_j^2} \right) .
\end{equation*}
If we choose $\tilde{\lambda}_j$ of this order (as we do in the statement of Lemma \ref{lemm:HD-par-est-conv-alt}), we have 
\begin{equation*}
\left\Vert\tilde{\boldsymbol{\gamma}}_j -\hat{\tilde{\boldsymbol{\gamma}}}_j\right\Vert_1 = \mathcal{O}_p \left(\tilde{s}_j \tilde{\lambda}_j\right) \quad \text{and} \quad \dfrac{1}{n}\left\Vert\mathbf{x}_{-j}\left(\tilde{\boldsymbol{\gamma}}_j -\hat{\tilde{\boldsymbol{\gamma}}}_j\right)\right\Vert_2^2 = \mathcal{O}_p \left(\tilde{s}_j \tilde{\lambda}_j^2\right).
\end{equation*}
For $\left(\hat{\tilde{\mathbf{z}}}_j^3\right)^\top \mathbf{x}_j$, we use the decomposition
\begin{equation*}
\hat{\tilde{\mathbf{z}}}_j^3=\hat{\mathbf{z}}_j^3- \mathbf{x}_{-j}\hat{\tilde{\boldsymbol{\gamma}}}_j=\mathbf{z}_j^3 + \left(\hat{\mathbf{z}}_j^3-\mathbf{z}_j^3\right) - \mathbf{x}_{-j}\tilde{\boldsymbol{\gamma}}_j + \mathbf{x}_{-j}\left(\tilde{\boldsymbol{\gamma}}_j -\hat{\tilde{\boldsymbol{\gamma}}}_j\right) = \tilde{\mathbf{z}}_j^3 + \left(\hat{\mathbf{z}}_j^3-\mathbf{z}_j^3\right) + \mathbf{x}_{-j}\left(\tilde{\boldsymbol{\gamma}}_j -\hat{\tilde{\boldsymbol{\gamma}}}_j\right).
\end{equation*}
Thus,
\begin{align*}
& \dfrac{1}{n}\left\vert \left(\hat{\tilde{\mathbf{z}}}_j^3\right)^\top \mathbf{x}_j - \left(\tilde{\mathbf{z}}_j^3\right)^\top \mathbf{x}_j\right\vert =\dfrac{1}{n}\left\vert \left(\hat{\mathbf{z}}_j^3-\mathbf{z}_j^3\right)^\top \mathbf{x}_j+ \left(\tilde{\boldsymbol{\gamma}}_j -\hat{\tilde{\boldsymbol{\gamma}}}_j\right)^\top \mathbf{x}_{-j}^\top \mathbf{x}_j \right\vert \\
& \leq \dfrac{1}{n}\left\Vert \hat{\mathbf{z}}_j^3-\mathbf{z}_j^3\right\Vert_1 \left\Vert\mathbf{x}_j\right\Vert_\infty +\dfrac{1}{n} \left\Vert\mathbf{x}_{-j}\left(\tilde{\boldsymbol{\gamma}}_j -\hat{\tilde{\boldsymbol{\gamma}}}_j\right)\right\Vert_2 \left\Vert\mathbf{x}_j\right\Vert_2 \\
&=\mathcal{O}_p\left(\log\left(p\right)s_j^2 \lambda_j ^3 + \log\left(p\right) s_j\lambda_j^2 + \sqrt{\log\left(p\right)s_j\lambda_j^2} \right) + \mathcal{O}_p\left(\sqrt{\tilde{s}_j \tilde{\lambda}_j ^2 } \right),
\end{align*}
which is ${\scriptstyle \mathcal{O}}_p\left(1\right)$ by assumption. This leads to
\begin{equation*}
\dfrac{1}{n}\left(\hat{\tilde{\mathbf{z}}}_j^3\right)^\top \mathbf{x}_j = \left(\tilde{\mathbf{z}}_j^3\right)^\top \mathbf{x}_j + { \scriptstyle \mathcal{O}}_p\left(1\right) = \EE\left[\tilde{Z}_j^3 X_j\right] + { \scriptstyle \mathcal{O}}_p\left(1\right).
\end{equation*}
The last equality could be derived using the Nemirovski moment inequality in a very similar fashion. For the expectation, we have
\begin{equation*}
\EE\left[\tilde{Z}_j^3 X_j\right] = \EE\left[\tilde{Z}_j^3 \left(Z_j +\boldsymbol{\gamma}_j ^\top \mathbf{X}_{-j}\right)\right] = \EE\left[\tilde{Z}_j^3 Z_j \right] = \EE\left[\left(Z_j^3 - \tilde{\boldsymbol{\gamma}}_j^\top \mathbf{X}_{-j}\right) Z_j \right] = \EE\left[Z_j^4 \right]
\end{equation*}
such that the second goal in \eqref{eq:goal-alt} is fulfilled as well. As all these derivations hold uniformly in $j$, $\left\vert \Delta_j^{HOLS} \right\vert = { \scriptstyle \mathcal{O}}_p\left(1\right)$ implies $\left\Vert \boldsymbol{\Delta}^{HOLS} \right\Vert = { \scriptstyle \mathcal{O}}_p\left(1\right)$.
\subsection{Proof of Lemma \ref{lemm:HD-var-est-conv-alt}}
We analyse the error term in \eqref{eq:HOdiff-alt}. From the proof of Lemma \ref{lemm:HD-par-est-conv-alt} as well as results in \cite{van2014asymptotically}, we know
\begin{equation*}
\left(\hat{\tilde{\mathbf{z}}}_j^3\right)^\top \mathbf{x}_j / n = \EE\left[Z_j^4\right] + { \scriptstyle \mathcal{O}}_p\left(1\right), \quad \hat{\mathbf{z}}_j^\top\mathbf{x}_j/n = \EE\left[Z_j^2\right] + { \scriptstyle \mathcal{O}}_p\left(1\right) \quad \text{and} \quad \left\Vert \hat{\mathbf{z}}_j\right\Vert_2^2 /n = \EE\left[Z_j^2\right] + { \scriptstyle \mathcal{O}}_p\left(1\right).
\end{equation*}
For the remaining terms in $\hat{\mathbf{v}}_j^\top \hat{\mathbf{v}}_j / n$, we want to ensure
\begin{equation}\label{eq:errgoal-alt}
\dfrac{1}{n}\left\Vert \hat{\tilde{\mathbf{z}}}_j^3\right\Vert _2 ^2 = \EE\left[\tilde{Z}_j^6\right] + { \scriptstyle \mathcal{O}}_p\left(1\right) \quad \text{and} \quad \dfrac{1}{n}\left(\hat{\tilde{\mathbf{z}}}_j^3\right)^\top \hat{\mathbf{z}}_j = \left\Vert \hat{\mathbf{z}}_j^2\right\Vert _2 ^2 /n =\left[Z_j^4\right] + { \scriptstyle \mathcal{O}}_p\left(1\right).
\end{equation}
Using the Nemirovski equation in a similar fashion as before, we know
\begin{equation*}
\underset{j}{\text{max}}\left\vert\dfrac{1}{n}\sum_{i=1}^n z_{ij}^r -\EE\left[Z_j^r\right]\right\vert = \mathcal{O}_p\left(\dfrac{\log\left(p\right)^{\left(r+1\right)/2}}{n^{1/2}}\right).
\end{equation*}
We assume this to be $\mathcal{O}_p \left(1\right) \ \forall r \leq 10$ and even ${\scriptstyle \mathcal{O}}_p \left(1\right) \ \forall r \leq 6$ (which is implied by the first condition). 
We look at some intermediary results. Each difference is ${\scriptstyle \mathcal{O}}_p \left(1\right)$ using the sparsity assumptions.
\begin{align*}
\dfrac{1}{n}\left\Vert \mathbf{z}_j^2 \odot \hat{\mathbf{z}}_j^2 - \mathbf{z}_j^4\right\Vert_1 & = \dfrac{1}{n}\left\Vert \mathbf{z}_j^2 \odot \left(\hat{\mathbf{z}}_j^2- \mathbf{z}_j^2\right)\right\Vert_1 = \dfrac{1}{n}\left\Vert \mathbf{z}_j^2 \odot \left(\left(\hat{\mathbf{z}}_j- \mathbf{z}_j\right)^2 + 2 \mathbf{z}_j \odot \left(\hat{\mathbf{z}}_j- \mathbf{z}_j\right)\right)\right\Vert_1 \\ &
\leq \dfrac{1}{n}\left\Vert\mathbf{z}_j^2\right\Vert_\infty \left\Vert \hat{\mathbf{z}}_j- \mathbf{z}_j \right\Vert_2^2 + \dfrac{2}{n} \left\Vert\mathbf{z}_j^3\right\Vert_2\left\Vert \hat{\mathbf{z}}_j- \mathbf{z}_j \right\Vert_2 \\
& =\mathcal{O}_p\left(\log\left(p\right)s_j\lambda_j^2\right) + \mathcal{O}_p\left(\sqrt{s_j\lambda_j^2}\right).
\end{align*}
This implies $\dfrac{1}{n}\left\Vert \mathbf{z}_j^2 \odot \hat{\mathbf{z}}_j^2\right\Vert_1=\dfrac{1}{n}\left\Vert \mathbf{z}_j \odot \hat{\mathbf{z}}_j\right\Vert_2^2=\mathcal{O}_p\left(1\right)$. With this, we can refine our result in a stepwise fashion:
\begin{align*}
\dfrac{1}{n}\left\Vert \mathbf{z}_j^4 \odot \hat{\mathbf{z}}_j^2 - \mathbf{z}_j^6\right\Vert_1 & = \dfrac{1}{n}\left\Vert \mathbf{z}_j^4 \odot \left(\hat{\mathbf{z}}_j^2- \mathbf{z}_j^2\right)\right\Vert_1 = \dfrac{1}{n}\left\Vert \mathbf{z}_j^4 \odot \left(\left(\hat{\mathbf{z}}_j- \mathbf{z}_j\right)^2 + 2 \mathbf{z}_j \odot \left(\hat{\mathbf{z}}_j- \mathbf{z}_j\right)\right)\right\Vert_1 \\ &
\leq \dfrac{1}{n}\left\Vert\mathbf{z}_j^4\right\Vert_\infty \left\Vert \hat{\mathbf{z}}_j- \mathbf{z}_j \right\Vert_2^2 + \dfrac{2}{n} \left\Vert\mathbf{z}_j^5\right\Vert_2\left\Vert \hat{\mathbf{z}}_j- \mathbf{z}_j \right\Vert_2 \\
& =\mathcal{O}_p\left(\log\left(p\right)^2 s_j\lambda_j^2\right) + \mathcal{O}_p\left(\sqrt{s_j\lambda_j^2}\right) \\
\text{such that} \ & \dfrac{1}{n}\left\Vert \mathbf{z}_j^4 \odot \hat{\mathbf{z}}_j^2\right\Vert_1=\dfrac{1}{n}\left\Vert \mathbf{z}_j^2 \odot \hat{\mathbf{z}}_j\right\Vert_2^2=\mathcal{O}_p\left(1\right).
\end{align*}
\begin{align*}
\dfrac{1}{n}\left\Vert \mathbf{z}_j^2 \odot \hat{\mathbf{z}}_j^4 - \mathbf{z}_j^4 \odot \hat{\mathbf{z}}_j^2 \right\Vert_1 & = \dfrac{1}{n}\left\Vert \mathbf{z}_j^2 \odot \hat{\mathbf{z}}_j^2 \odot \left(\hat{\mathbf{z}}_j^2- \mathbf{z}_j^2\right)\right\Vert_1 \\
& = \dfrac{1}{n}\left\Vert \mathbf{z}_j^2 \odot \hat{\mathbf{z}}_j^2 \odot \left(\left(\hat{\mathbf{z}}_j- \mathbf{z}_j\right)^2 + 2 \mathbf{z}_j \odot \left(\hat{\mathbf{z}}_j- \mathbf{z}_j\right)\right)\right\Vert_1 \\ &
\leq \dfrac{1}{n}\left\Vert\mathbf{z}_j^2\right\Vert_\infty \left\Vert\hat{\mathbf{z}}_j^2\right\Vert_\infty \left\Vert \hat{\mathbf{z}}_j- \mathbf{z}_j \right\Vert_2^2 + \dfrac{2}{n}\left\Vert\mathbf{z}_j\right\Vert_\infty \left\Vert\hat{\mathbf{z}}_j\right\Vert_\infty \left\Vert\mathbf{z}_j^2 \hat{\mathbf{z}}_j\right\Vert_2\left\Vert \hat{\mathbf{z}}_j- \mathbf{z}_j \right\Vert_2 \\
& =\mathcal{O}_p\left(\log\left(p\right)^2 \left(1 + s_j^2 \lambda_j^2\right) s_j\lambda_j^2\right) + \mathcal{O}_p\left(\log\left(p\right) \left(1 + s_j \lambda_j\right)\sqrt{s_j\lambda_j^2}\right) \\
\text{such that} \ & \dfrac{1}{n}\left\Vert \mathbf{z}_j^2 \odot \hat{\mathbf{z}}_j^4\right\Vert_1=\dfrac{1}{n}\left\Vert \mathbf{z}_j \odot \hat{\mathbf{z}}_j^2\right\Vert_2^2=\mathcal{O}_p\left(1\right).
\end{align*}
\begin{align*}
\dfrac{1}{n}\left\Vert \hat{\mathbf{z}}_j^6 - \mathbf{z}_j^2 \odot \hat{\mathbf{z}}_j^4 \right\Vert_1 & = \dfrac{1}{n}\left\Vert \hat{\mathbf{z}}_j^4 \odot \left(\hat{\mathbf{z}}_j^2- \mathbf{z}_j^2\right)\right\Vert_1 = \dfrac{1}{n}\left\Vert \hat{\mathbf{z}}_j^4 \odot \left(\left(\hat{\mathbf{z}}_j- \mathbf{z}_j\right)^2 + 2 \mathbf{z}_j \odot \left(\hat{\mathbf{z}}_j- \mathbf{z}_j\right)\right)\right\Vert_1 \\ &
\leq \dfrac{1}{n} \left\Vert\hat{\mathbf{z}}_j^4\right\Vert_\infty \left\Vert \hat{\mathbf{z}}_j- \mathbf{z}_j \right\Vert_2^2 + \dfrac{2}{n}\left\Vert\hat{\mathbf{z}}_j^2\right\Vert_\infty \left\Vert\mathbf{z}_j \hat{\mathbf{z}}_j^2\right\Vert_2\left\Vert \hat{\mathbf{z}}_j- \mathbf{z}_j \right\Vert_2 \\
& =\mathcal{O}_p\left(\log\left(p\right)^2 \left(1 + s_j^4 \lambda_j^4\right) s_j\lambda_j^2\right) + \mathcal{O}_p\left(\log\left(p\right) \left(1 + s_j ^2\lambda_j^2\right)\sqrt{s_j\lambda_j^2}\right).
\end{align*}
Finally, it follows
\begin{align*}
\dfrac{1}{n}\left\Vert \mathbf{z}_j^3 \odot \hat{\mathbf{z}}_j^3 - \mathbf{z}_j^4 \odot \hat{\mathbf{z}}_j^2 \right\Vert_1 & = \dfrac{1}{n}\left\Vert \mathbf{z}_j^3 \odot \hat{\mathbf{z}}_j^2 \odot \left(\hat{\mathbf{z}}_j- \mathbf{z}_j\right)\right\Vert_1 = \dfrac{1}{n}\left\Vert\mathbf{z}_j^2\right\Vert_\infty \left\Vert\mathbf{z}_j \hat{\mathbf{z}}_j^2\right\Vert_2\left\Vert \hat{\mathbf{z}}_j- \mathbf{z}_j \right\Vert_2 \\
& =\mathcal{O}_p\left(\log\left(p\right)\sqrt{s_j\lambda_j^2}\right) \ \text{such that} \\
 \dfrac{1}{n}\left\Vert \hat{\mathbf{z}}_j^3 -\mathbf{z}_j^3\right\Vert_2^2 & = \dfrac{1}{n}\left\Vert \hat{\mathbf{z}}_j^6 +\mathbf{z}_j^6 - 2 \hat{\mathbf{z}}_j^3\odot\mathbf{z}_j^3\right\Vert_1 \leq \dfrac{1}{n}\left\Vert \hat{\mathbf{z}}_j^6 - \hat{\mathbf{z}}_j^3\odot\mathbf{z}_j^3\right\Vert_1 + \dfrac{1}{n}\left\Vert \mathbf{z}_j^6 - \hat{\mathbf{z}}_j^3\odot\mathbf{z}_j^3\right\Vert_1 \\
& =\mathcal{O}_p\left(\log\left(p\right)^2 \left(1 + s_j^4 \lambda_j^4\right) s_j\lambda_j^2\right) + \mathcal{O}_p\left(\log\left(p\right) \left(1 + s_j ^2\lambda_j^2\right)\sqrt{s_j\lambda_j^2}\right)\\
& = { \scriptstyle \mathcal{O}}_p\left(1\right),
\end{align*}
This can now be applied to find the desired convergence.
\begin{align*}
 & \dfrac{1}{n}\left\vert \left(\hat{\tilde{\mathbf{z}}}_j^3\right)^\top \hat{\mathbf{z}}_j - \left(\tilde{\mathbf{z}}_j^3\right)^\top \mathbf{z}_j \right\vert  =  \dfrac{1}{n}\bigg\vert \left(\tilde{\mathbf{z}}_j^3\right)^\top\left(\hat{\mathbf{z}}_j- \mathbf{z}_j\right) +\left(\hat{\mathbf{z}}_j^3- \mathbf{z}_j^3\right)^\top \mathbf{z}_j +  \\
 &  \left(\hat{\mathbf{z}}_j^3- \mathbf{z}_j^3\right)^\top\left(\hat{\mathbf{z}}_j- \mathbf{z}_j\right) + \left(\tilde{\boldsymbol{\gamma}}_j - \hat{\tilde{\boldsymbol{\gamma}}}_j\right)^\top \mathbf{x}_{-j}^\top \mathbf{z}_j + \left(\tilde{\boldsymbol{\gamma}}_j - \hat{\tilde{\boldsymbol{\gamma}}}_j\right)^\top \mathbf{x}_{-j}^\top\left(\hat{\mathbf{z}}_j- \mathbf{z}_j\right)\bigg\vert \leq \\
 & \dfrac{1}{n} \left\Vert \tilde{\mathbf{z}}_j^3\right\Vert_2 \left\Vert \hat{\mathbf{z}}_j- \mathbf{z}_j\right\Vert_2 + \dfrac{1}{n} \left\Vert \hat{\mathbf{z}}_j^3- \mathbf{z}_j^3\right\Vert_2 \left\Vert \mathbf{z}_j \right\Vert_2 + \dfrac{1}{n} \left\Vert \hat{\mathbf{z}}_j^3- \mathbf{z}_j^3\right\Vert_2 \left\Vert \hat{\mathbf{z}}_j- \mathbf{z}_j\right\Vert_2+ \\
& \dfrac{1}{n} \left\Vert \mathbf{z}_j\right\Vert_2 \left\Vert\mathbf{x}_{-j}\left(\tilde{\boldsymbol{\gamma}}_j - \hat{\tilde{\boldsymbol{\gamma}}}_j\right)\right\Vert_2 +  \dfrac{1}{n} \left\Vert \mathbf{x}_{-j}\left(\tilde{\boldsymbol{\gamma}}_j - \hat{\tilde{\boldsymbol{\gamma}}}_j\right)\right\Vert_2 \left\Vert \hat{\mathbf{z}}_j- \mathbf{z}_j \right\Vert_2 
\end{align*}
All these terms have already been bounded. Thus, we do not need any further assumptions to claim
\begin{equation*}
\dfrac{1}{n}\left(\hat{\tilde{\mathbf{z}}}_j^3\right)^\top \hat{\mathbf{z}}_j = \dfrac{1}{n}\ \left(\tilde{\mathbf{z}}_j^3\right)^\top \mathbf{z}_j + { \scriptstyle \mathcal{O}}_p\left(1\right) = \EE\left[\tilde{Z}_j^3 Z_j \right] + { \scriptstyle \mathcal{O}}_p\left(1\right) = \EE\left[ Z_j^4 \right] + { \scriptstyle \mathcal{O}}_p\left(1\right).
\end{equation*}
We turn to the final term in the error scaling
\begin{align*}
 & \dfrac{1}{n}\left\vert \left(\hat{\tilde{\mathbf{z}}}_j^3\right)^\top \left(\hat{\tilde{\mathbf{z}}}_j^3\right) - \left(\tilde{\mathbf{z}}_j^3\right)^\top \left(\tilde{\mathbf{z}}_j^3\right)\right\vert  = \\
 & \dfrac{1}{n} \bigg\vert 2 \left(\tilde{\mathbf{z}}_j^3\right)^\top \left(\hat{\mathbf{z}}_j^3- \mathbf{z}_j^3\right) + 2 \left(\tilde{\mathbf{z}}_j^3\right)^\top \mathbf{x}_{-j}\left(\tilde{\boldsymbol{\gamma}}_j - \hat{\tilde{\boldsymbol{\gamma}}}_j\right) + \left(\hat{\mathbf{z}}_j^3- \mathbf{z}_j^3\right)^\top \left(\hat{\mathbf{z}}_j^3- \mathbf{z}_j^3\right)+\\
 &  2 \left(\hat{\mathbf{z}}_j^3- \mathbf{z}_j^3\right) \mathbf{x}_{-j}\left(\tilde{\boldsymbol{\gamma}}_j - \hat{\tilde{\boldsymbol{\gamma}}}_j\right) +\left(\tilde{\boldsymbol{\gamma}}_j - \hat{\tilde{\boldsymbol{\gamma}}}_j\right)^\top \mathbf{x}_{-j}^\top \mathbf{x}_{-j}\left(\tilde{\boldsymbol{\gamma}}_j - \hat{\tilde{\boldsymbol{\gamma}}}_j\right) \bigg\vert \leq \\
 & \dfrac{2}{n} \left\Vert\tilde{\mathbf{z}}_j^3 \right\Vert_2 \left\Vert\hat{\mathbf{z}}_j^3- \mathbf{z}_j^3\right\Vert_2 + \dfrac{2}{n} \left\Vert\tilde{\mathbf{z}}_j^3 \right\Vert_2 \left\Vert\mathbf{x}_{-j}\left(\tilde{\boldsymbol{\gamma}}_j - \hat{\tilde{\boldsymbol{\gamma}}}_j\right)\right\Vert_2 + \dfrac{1}{n} \left\Vert\hat{\mathbf{z}}_j^3- \mathbf{z}_j^3\right\Vert_2^2 +\\
 & \dfrac{2}{n} \left\Vert\hat{\mathbf{z}}_j^3- \mathbf{z}_j^3 \right\Vert_2 \left\Vert\mathbf{x}_{-j}\left(\tilde{\boldsymbol{\gamma}}_j - \hat{\tilde{\boldsymbol{\gamma}}}_j\right)\right\Vert_2 +  \dfrac{1}{n} \left\Vert\mathbf{x}_{-j}\left(\tilde{\boldsymbol{\gamma}}_j - \hat{\tilde{\boldsymbol{\gamma}}}_j\right)\right\Vert_2^2.
\end{align*}
Again, these are all terms that we have seen before such that we do not need any additional assumptions to claim
\begin{equation*}
\dfrac{1}{n} \left(\hat{\tilde{\mathbf{z}}}_j^3\right)^\top \left(\hat{\tilde{\mathbf{z}}}_j^3\right) = \dfrac{1}{n}\left(\tilde{\mathbf{z}}_j^3\right)^\top \left(\tilde{\mathbf{z}}_j^3\right) + { \scriptstyle \mathcal{O}}_p\left(1\right) = \EE\left[\left(\tilde{Z}_j^3\right)^2\right] +	{\scriptstyle \mathcal{O}}_p\left(1\right) .
\end{equation*} 
Thus, we have shown convergence for all terms in $\hat{\mathbf{v}}_j^\top \hat{\mathbf{v}}_j / n$. Finally, note that
\begin{align*}
\EE\left[V_j^2\right] & =\EE\left[\left(\dfrac{\tilde{Z}_j^3}{\EE\left[Z_j^4\right]}- \dfrac{Z_j}{\EE\left[Z_j^2\right]}\right)^2\right] = \dfrac{\EE\left[\left(\tilde{Z}_j^3\right)^2\right] }{\EE\left[Z_j^4\right]^2}-\dfrac{2\EE\left[\tilde{Z}_j^3 Z_j\right]}{\EE\left[Z_j^4\right]\EE\left[Z_j^2\right]}+ \dfrac{\EE\left[Z_j^2\right]}{\EE\left[Z_j^2\right]^2}\\
& = \dfrac{\EE\left[\left(\tilde{Z}_j^3\right)^2\right] }{\EE\left[Z_j^4\right]^2} -\dfrac{1}{\EE\left[Z_j^2\right]}
\end{align*}
such that convergence is towards $\EE\left[V_j^2\right] $ as claimed.
\subsection{Proof of Theorem \ref{theo:HD-local-gauss}}
With Lemmata \ref{lemm:HD-par-est-conv-alt} and \ref{lemm:HD-var-est-conv-alt}, we have already established that the bias terms vanish and the denominator converges. It remains to look at $\hat{\mathbf{v}}_j^\top \boldsymbol{\eps}/\sqrt{n}$. 
\begin{align*}
\EE\left[\left(\hat{\mathbf{v}}_j- \mathbf{v}_j\right)^\top  \boldsymbol{\eps}/\sqrt{n}\right] & =\EE\left[\EE\left[\left(\hat{\mathbf{v}}_j- \mathbf{v}_j\right)^\top  \boldsymbol{\eps}/\sqrt{n}\vert\mathbf{x}\right]\right]=\EE\left[\left(\hat{\mathbf{v}}_j- \mathbf{v}_j\right)^\top\EE\left[  \boldsymbol{\eps}/\sqrt{n}\vert\mathbf{x}\right]\right]=0 \quad \text{and}\\
\text{Var}\left(\left(\hat{\mathbf{v}}_j- \mathbf{v}_j\right)^\top  \boldsymbol{\eps}/\sqrt{n}\right)  & = \EE\left[\text{Var}\left(\left(\hat{\mathbf{v}}_j- \mathbf{v}_j\right)^\top  \boldsymbol{\eps}/\sqrt{n}\vert \mathbf{x}\right) \right] + \text{Var}\left(\EE\left[\left(\hat{\mathbf{v}}_j- \mathbf{v}_j\right)^\top  \boldsymbol{\eps}/\sqrt{n}\vert \mathbf{x}\right]\right) \\
& = \sigma^2\EE\left[\left\Vert \hat{\mathbf{v}}_j- \mathbf{v}_j \right\Vert_2^2/n 
\right] + 0 = {\scriptstyle\mathcal{O}}\left(1\right)
\end{align*}
The last equality uses the convergence rates from the proof of Lemma \ref{lemm:HD-var-est-conv-alt}. By Chebyshev's inequality and the CLT
\begin{equation*}
\hat{\mathbf{v}}_j^\top \boldsymbol{\eps}/\sqrt{n}  \overset{\mathbb{P}}{\to} \mathbf{v}_j^\top \boldsymbol{\eps}/\sqrt{n} \overset{\mathbb{D}}{\to} \mathcal{N}\left(0, \sigma^2 \EE\left[V_j^2\right]\right)
\end{equation*}
By Slutsky's theorem, we can replace $\sigma^2$ with a consistent estimate such that the theorem's statement follows.
\end{document}